\documentclass[a4paper,notitlepage,12pt]{article}%

\usepackage{appendix}

\usepackage{amsfonts}
\usepackage{amsmath}
\usepackage{amssymb}
\usepackage{mathptmx}
\usepackage{array,multirow,makecell} 

\usepackage{rotating}
\usepackage{setspace}
\usepackage{lscape}
\usepackage{pdflscape}
\usepackage{caption}

\usepackage{fancyhdr}
\usepackage{subfiles}
\usepackage{float}
\usepackage{makecell}

\usepackage{url}
\usepackage{placeins}
\usepackage{longtable}
\usepackage{supertabular}
\usepackage{tabularx}

\usepackage[UKenglish]{babel}
\usepackage{eurosym}
\usepackage[utf8]{inputenc}
\usepackage{csquotes}
\usepackage[T1]{fontenc}

\usepackage[justification=centering,font={bf}]{caption}
\usepackage{graphicx, subfigure}

\usepackage{comment}
\usepackage[usenames]{color}
\usepackage{colortbl}
\usepackage{xcolor}
\usepackage{textcomp}

\usepackage{dcolumn}
\usepackage{placeins}
\usepackage{chngcntr}
\usepackage{afterpage}

\usepackage{geometry}
\usepackage{tikz}
\usepackage{apptools}
\usepackage{arydshln}

\usepackage{footnote}

\usepackage[backend=biber,style=authoryear,maxcitenames=2,mincitenames=1,block=space,date=year,isbn=false,url=false,eprint=false,clearlang=false]{biblatex}
\DeclareLanguageMapping{british}{british-apa}
\AtEveryBibitem{
 \clearlist{address}
 \clearfield{eprint}
 \clearfield{isbn}
 \clearfield{issn}
 \clearlist{location}
 \clearfield{month}
 \clearfield{series}
 \clearfield{note}
 \clearlist{language}
 
 \ifentrytype{book}{}{
  \clearlist{publisher}
  \clearname{editor}
 }
}
\usepackage{xpatch}
\xpretobibmacro{bib+doi+url}
{\iffieldundef{doi}
	{}
	{\clearfield{url}}}
{}{}

\xpretobibmacro{cite+doi+url}
{\iffieldundef{doi}
	{}
	{\clearfield{url}}}
{}{}
\AtAppendix{\counterwithin{equation}{section}}
\providecommand{\U}[1]{\protect\rule{.1in}{.1in}}

\evensidemargin\oddsidemargin
\begin{document}

    \begin{titlepage}

	\newgeometry{top=0.5 in, bottom=1 in, left=1 in, right= 1 in}


\title{Minimum pricing or volumetric taxation? \\ Quantity, quality and competition effects of price regulations in alcohol markets$^{\ast}$\thanks{\scriptsize We thank the Institut National du Cancer (INCa, AAP-SHS-E-SP-2015), the Mission interministérielle de lutte contre les drogues et les conduites addictives (MILDECA), the Agence nationale de la recherche (ANR-19-CE21-0004 PRIMOFOOD), Toulouse School of Economics and ANR (ANR-17-EURE-0010), Paris School of Economics and ANR (ANR-17-EURE-0001), and the AXA Research Fund (AXA Awards 2015, Fabrice Etilé) and . This project was part of the “Impact épidémiologique et économique des politiques de prix de l’alcool sur les cancers” (Epidemiological and economic impact of alcohol pricing policies on cancer) research project, coordinated by Fabrice Etilé. INCa, MILDECA, ANR, AXA, INRAE, PSE, TSE, and BSE were not involved in the design of the study and are not responsible for its implementation, results and interpretations. We thank Christine Boizot-Szantai and Val\'{e}rie Orozco, for collaboration on data acquisition and preparation, and Marine Friant-Perrot, Amandine Garde, Chantal Julia and Mathilde Touvier for advices on the epidemiological and legal aspects of alcohol markets. We thank seminar participants at Paris School of Economics, UMR GAEL (Grenoble), UMR SMART (Rennes), UMR BETA (Nancy), 2nd WAP conference (Bordeaux) for comments.}}
\author{C\'{e}line Bonnet\thanks{{\scriptsize Toulouse School of Economics and INRAE, }},  
        Fabrice Etil\'{e}\thanks{{\scriptsize Paris School of Economics and INRAE,  Corresponding author: UMR INRAE 1393 Paris-Jourdan Sciences Economiques and Ecole d'Economie de Paris, 48 Boulevard Jourdan, 75014 Paris. fabrice.etile@inrae.fr }}, 
        S\'{e}bastien Lecocq\thanks{{\scriptsize INRAE, University of Bordeaux, CNRS, INRAE, BSE.}}
        }
\date{{\small June 2025}}
\maketitle
\vspace{-1cm}
\thispagestyle{empty} 
    
\begin{abstract}
    \setstretch{1}
    {\small Reforming alcohol price regulations in wine-producing countries is challenging, as current price regulations reflect the alignment of cultural preferences with economic interests rather than public health concerns. We evaluate and compare the impact of counterfactual alcohol pricing policies on consumer behaviors, firms, and markets in France. We develop a micro-founded partial equilibrium model that accounts for consumer preferences over purchase volumes across alcohol categories and over product quality within categories, and for firms' strategic price-setting. After calibration on household scanner data, we compare the impacts of replacing current taxes by ethanol-based volumetric taxes with a minimum unit price (MUP) policy of €0.50 per standard drink. The results show that the MUP in addition to the current tax outperforms a tax reform in reducing ethanol purchases (-15\% vs. -10\% for progressive taxation), especially among heavy drinking households (-17\%). The MUP increases the profits of small and medium wine firms (+39\%) while decreasing the profits of large manufacturers and retailers (-39\%) and maintaining tax revenues stable. The results support the MUP as a targeted strategy to reduce harmful consumption while benefiting small and medium wine producers. This study provides ex-ante evidence that is crucial for alcohol pricing policies in wine-producing countries.}
\end{abstract}

{\footnotesize 

}

\end{titlepage}
    
    \restoregeometry 
    
    \pagenumbering{gobble} 
    
    \thispagestyle{plain} 
    
    \pagenumbering{roman} 
    
    \pagenumbering{arabic}

\newpage
    \section{Introduction}
        Alcohol consumption is a major contributor to the global burden of disease, with risks including cancer, cardiovascular disease, car accidents, injuries, mental health problems, and cognitive decline. It represented 5.3\% of all deaths worldwide in 2016 \parencite{shield_national_2020}. In France, it remains a major cause of health and social problems, although consumption has been halved since the Second World War. Seven per cent of adult deaths (41,000) were attributable to alcohol in 2015 \parencite{bonaldi_mortalite_2019,shield_new_2018,gbd_global_2018,praud_cancer_2016}. Pricing policies are considered by international public health experts as part of any national comprehensive strategy to reduce the burden of alcohol-related disease \parencite{oms_global_2010,inserm_reduction_2021}, with calls to target the ethanol content of beverages through minimum unit pricing (MUP) or volumetric taxes \parencite{cheatley_policies_2021,cecchini_addressing_2021}. Implementing these recommendations would completely overhaul the regulatory framework for alcohol prices, in France as in other wine-producing Mediterranean or Central European countries, where alcohol taxation is historically geared more towards protecting domestic wine producers than towards public health objectives.\footnote{France is the world's 2nd largest wine producer (ahead of Spain, but behind Italy) and 1st largest exporter. The wine sector represents almost 550,000 jobs, including 250,000 direct jobs \parencite{palle_poids_2013}.} The specific tax rates on wines are close to zero while they represent approximately 55\% of the total volumes of ethanol purchased by households.\footnote{\textcite{angus_comparing_2019} compares the 2018 tax rates (excise duties) on wine, beer, and spirits in European countries. The UK, Ireland, Nordic, and Baltic countries have higher rates than other EU countries, with no exemptions for wines. Wines are not subject to specific duties in the Mediterranean and Central European countries. Unsurprisingly then, France, Spain, Italy, Portugal and Bulgaria tried to block the Scottish Minimum Unit Price policy after its introduction in 2012.} Therefore, it is not surprising that, despite the international consensus, attempts to reform alcohol price regulation policies have faced strong opposition from political representatives. In this paper, we estimate and compare the impacts of minimum pricing and volumetric taxes on alcohol purchases, equilibrium prices, consumer welfare, and firms' profits. Our objective is to understand whether there exists a price policy that could benefit public health while being acceptable by most wine producers.
\bigskip

We develop a model that account both for consumer choices over quantity and quality and for manufacturers' and retailers' price-setting decisions. Modeling the supply-side of the market is crucial for two reasons. First, strategic price reactions on the supply-side can limit the effectiveness of behavioral health policies \parencite{bonnet_tax_2013,allais_mandatory_2015,dubois_effects_2018}. Two ex-post evaluation studies of the UK alcohol market found that tax increases are under-transmitted to the prices of basic products and over-transmitted to the prices of premium products, for all alcohol categories \parencite{wilson_effect_2021,ally_alcohol_2014}. This under-transmission reduces the effectiveness of taxation policies if heavy drinkers prefer to buy basic brands.
Second, whether or not a government will be able to reform price regulations will depend on how much different reforms can affect the profits of the different alcohol producers, from large companies to small wine farmers, who have a lot of political influence in France. It is therefore important to document the potential impacts of any reform on firms' profits.
On the demand-side, we consider a multi-level demand system setting in which households can purchase various volumes of alcohol in six distinct categories, ciders, beers, aperitifs, spirits, still wines, and sparkling wines. Each category represents a specific market in which consumers can choose from a variety of products with different ethanol content. This separability assumption allows us to neatly disentangle the quality and quantity effects of price policies. As the heterogeneity in the ethanol content between alcohol categories is large, substitutions in volumes between categories can have large health impacts. Quality effects within categories, i.e. substitutions between product varieties, may have smaller impacts, but they drive the strategic price responses of firms. We calibrate the demand functions using longitudinal homescan data on the alcohol purchases of a representative panel of French households. These data contain detailed information on prices and product attributes, and the panel dimension allows us to account for household preferences for ethanol by conditioning all the estimations on purchase volumes observed during the previous year. On the supply-side, manufacturers face an oligopolistic price competition and contract with retailers to sell their products to the consumers. Retailers also sell their own brands (Private Labels). We use the market-specific demand functions to calibrate the supply-side models, and we are able to recover margins and marginal costs for each product variety sold in the six markets. It renders possible the simulation of firms' strategic price reactions to changes in price regulation policies and consumer behavior. We eventually gain access to evaluations of the impact of pricing policies on firms' profits, and we can identify the losers and winners from various counterfactual pricing policies.  
\bigskip

We consider several reforms of price regulation policies. In the current tax system, wines are almost exempt from specific taxes, taxes on beers and ciders are low, and spirits bear most of the burden. We test replacing current taxes by two alternative volumetric tax systems as in \textcite{griffith_tax_2019}: a uniform tax rate based on the ethanol content of products; a progressive tax rate that increases with the ethanol content. The progressive tax maintains a relative disadvantage for spirits, which consists of a large share of total ethanol purchases observed for high-risk households. 
These two taxes are calibrated on one of the two following objectives, assuming that there are no market reactions. The first objective is neutrality in terms of fiscal revenues, while the second objective is neutrality of alcohol-specific taxes for public finances.\footnote{The total social cost in 2010 was €120 billion, of which €66 billion was due to mortality and €48 billion to lost welfare and productivity. The cost to public finances, net of tax revenues and pensions saved due to premature deaths, was estimated at around €3 billion \parencite{kopp_cout_2015}. Tax revenues covered only 42\% of public health costs. These figures justify the implementation of new policies aimed at reducing the population's alcohol consumption in order to reduce alcohol-related health and social harm.} Fiscal neutrality leads to a lower tax rate (around 7.2 cts/g of ethanol) than neutrality for public finances (14.6 cts/g).

As an alternative to these four tax reform scenarios, we consider introducing a minimum unit price policy on top of the current tax system. The introduction of a minimum price may make it possible to better target abusive consumers if they disproportionately buy cheap products with a high alcohol content. In fact, the results of ex-post evaluations of the minimum pricing policy implemented in Scotland and Wales show that it led to the substitution of high-alcohol beers and ciders for lower-alcohol products and that the impact was concentrated in 20\% of households with the highest per capita alcohol consumption, regardless of income category \parencite{llopis_impact_2021,odonnell_immediate_2019}. \textcite{griffith_price_2022} also use their reduced-form empirical model of the UK market to show that generalizing the Scottish minimum price throughout the UK would be preferable to a uniform tax with the same impact on total consumption, because it affects the heavy drinkers relatively more. In a theoretical paper, \textcite{calcott_minimum_2019} shows that the two instruments can be complementary if quality and quantity of alcohol are substitutes and if the harm is caused by consumers of cheaper alcohol. We therefore examine a sixth scenario, where the minimum price is combined with the replacement of the current tax system by a progressive tax. 
\bigskip

We find that the addition of a Minimum Unit price set at 0.5\euro /standard drink to the current tax system outperforms the four tax reforms. A minimum price policy would reduce average purchases (in g. of ethanol) by 15.0\%, compared to -10.3\% for a high progressive ethanol tax. Supply-side price responses would only slightly increase the effectiveness of the minimum price policy (-13.7\% without response). Low tax rates will have the unintended effect of increasing purchases (in g. of ethanol), by favoring substitutions from still wines to spirits as mechanic consequences of rising taxes on the former and falling taxes on the latter.

The superiority of the minimum price policy is essentially due to the quantity effect, as it reduces purchases in all alcohol categories. We also find larger effects for heavy-drinking households (-17.2\%). Importantly, a minimum price policy would significantly increase the profits of small and medium wine manufacturers (+39.3\%) to the detriment of large manufacturing firms and retailers (-38.7\%). In fact, it would increase the price competitiveness and market shares of products positioned in the middle-quality segment, to the detriment of entry-level products currently produced and marketed by large firms. As the minimum price mechanically increases price-cost margins, lower quality products make higher unit profits, but their market share drops dramatically, so that their total profit falls. Finally, tax revenues remain virtually unaffected (-2.3\%), because the increase in VAT revenues offsets the contraction of the market.   
\bigskip

 Our results complement studies that have provided ex-post evidence that minimum prices and volumetric taxes are effective in increasing consumer prices, reducing purchases and consumption, and improving health \parencite{wagenaar_effects_2010,roodman2020,elder_effectiveness_2010,stockwell_raising_2012,zhao_impacts_2017,anderson_impact_2021,robinson_evaluating_2021,xhurxhi_early_2020,griffith_price_2022}.\footnote{Our results are in line with ex-post evaluation studies of the Scottish Minimum Unit Price that have also exploited Kantar homescan data. In particular, \textcite{anderson_impact_2021}, using a difference-in-difference ex-post evaluation design with England as the control group, find that the Scottish MUP resulted in a -7.7\% drop in purchase volumes (-8.6\% for the Welsh MUP). \textcite{griffith_price_2022} find an average reduction of 11\% in alcohol units purchased per adult per week.} This ex-post evidence is mainly available for English-speaking countries, which historically have a drinking culture of binge drinking, beer and spirits, as opposed to wine-producing countries where wine consumption is integrated into everyday social occasions such as meals \parencite{gordon_rethinking_2012}. 
 
 Here, we provide ex-ante evidence on the structural impacts of price policy reforms on consumers, firms, and markets for a wine-drinking country. As in \textcite{griffith_price_2022} for the UK, we compare minimum pricing with progressive volumetric taxation. Beyond the many methodological differences, they also find that minimum pricing and ethanol-based progressive taxation with high tax rates can achieve the same impacts on consumption (with taxation slightly outperforming minimum pricing in their case). However, our study goes further in the identification of losers and winners on the supply-side because we are interested in the acceptability of pricing policies, due to the strong lobbying of the wine industry \parencite{millot_lobbying_2022}. Here, the industrial organization of the market is a key difference between Anglo-Saxon countries, and France and other Mediterranean countries. The deep-rooted cultural preservation of the wine sector in Mediterranean countries and the resulting current alcohol tax policies explain why minimum pricing is more effective in reducing ethanol purchases. Our article sheds light on the public debate in France on how to regulate alcohol more effectively to reduce its consumption and the burden it places on society.
 
 We finally contribute to the literature on empirical IO modeling, with a structural model of demand for quality and quantity that capture patterns of substitution between alcohol products within each alcohol category, and substitutability or complementarity between alcohol categories. Our approach is essentially based on the sequential estimation of demand for quality and then demand for quantity, the latter being conditioned on the former through a structural household-specific quality index for each alcohol category that measures household preferences for product differentiation within the category. Most studies to date have used continuous choice demand models that treat alcohol or alcohol categories as homogeneous \parencite{sharma_effect_2016,srivastava_disaggregated_2015,ciderova_estimation_2022}, or discrete choice models that better account for product differentiation but force substitutability between categories \parencite{griffith_tax_2019, miravete_market_2018}. Our model goes beyond this division between the quantity and quality dimensions of household choices to account for richer patterns of behavioral reactions on the demand side.

The remainder of the paper is structured as follows. In the next section, we present our data and discuss some stylized facts about the French alcohol market. This helps us to motivate our pricing policy reform scenarios. In Section 3, we detail the structural model, which is calibrated in Section 4. In Section 5, we present the key results, detail the main mechanisms at play, and report additional heterogeneity analyzes on households and firms. Section 6 concludes with a discussion of the limits of the study.

    \section{Data, facts and price regulation scenarios}
        
This section presents the data and stylized facts regarding the structure of alcohol markets in France. In line with these, we define scenarios of policy reform regarding minimum unit pricing and ethanol-based volumetric taxation.
  
\subsection{Data}

\subsubsection{Households}

This study is based on the 2014 Kantar Worldpanel (KWP) homescan data. The data cover all purchases for consumption at home made by a representative panel of French households (excluding overseas territories).\footnote{While the KWP data covers well the population and purchases for consumption at home, as benchmarked against Family Budget Surveys \parencite{lecocq_alcohol_2023}, alcohol consumption away from home is not observed. We come back to this point in the discussion of the results.} In 2014, 28,375 households were enrolled in KWP, of which 24,177 were considered actively reporting their purchases of food and beverages for at least one four-week period. They report their purchases of barcoded products using handheld scanning devices. Approximately half of them (N = 13,395) also belong to a representative subpanel in which participants must report their purchases of products without barcodes.\footnote{Information on products with barcodes is directly retrieved from a product database, by using the barcode. For products without barcodes, KWP asks participants to report a number of standardized attributes, e.g. commercial segment, brand, packaging, etc.} Wines, ciders and beers produced by small manufacturers do not always have barcodes because they are sold directly in the vineyard or brewery, in open-air markets, or in cellars or specialty stores. We work with this subpanel and select the 12,170 households that were active in at least three periods of four weeks out of 13 periods that the year contains. 

We further restrict the sample to those households that purchased alcoholic beverages at least once in 2014 (N=11,463), under the reasonable assumption that a reform of alcohol price regulations will be justified by a public health objective and will not have the effect of encouraging abstinent households to become consumers.  

In all of our analyses and estimations, we use the four-week sample weights provided by KWP for each household and each household-purchase act, to make our analysis representative of the French market. Appendix Section \ref{App:Subsamples} provides more details on sample selection and sample weights. Appendix Table \ref{AppData:S1S2S3S4_Var_HH_indiv} presents some descriptive statistics on household characteristics.   

\subsubsection{Markets and products}

We choose to group products into six categories: ciders, beers, aperitifs, spirits, still wines, and sparkling wines. This definition of alcohol categories closely follows the SIRENE classification of manufacturers used by the French National Statistics Office (INSEE).\footnote{The SIRENE classification downloaded from \url{https://www.data.gouv.fr/fr/datasets/base-sirene-des-entreprises-et-de-leurs-etablissements-siren-siret/} distinguishes spirits (1101Z: Production of distilled alcohol beverages), sparkling wines (1102A: Production of sparkling wines), still wines (1102B: Production of still wines), ciders and fruits wines (1103Z: Production of ciders and fruit wines), aperitifs (essentially the 1104Z class - Production of other non-distilled fermented beverages, and some products in other classes), and beers (1105Z and 1106Z: Production of beers, malt production)} The SIRENE open data were completed by a careful reading of the professional press on the organization of these six markets to understand their structure and what the largest manufacturers were in 2014. It appeared that each of these six alcohol categories constitutes a specific alcohol market with its own supply and demand characteristics. In supply side modeling, we treat them as six disjoint alcohol markets. We again used insights from the professional press to divide each alcohol category into subcategories that map key quality differences in the product space. For example, beers are divided into three subcategories, depending on their alcohol content: Alcohol-free,  Bock/Premium, Special. We distinguish sweet ciders from raw ciders. Sparkling wines include champagne and other sparkling wines. For still wines, we use the well-known distinction between table wines (Vins de table), local wines (Vins de Pays), and wines with a designation of origin (Vins d'Appellation). Spirits include five subcategories (Rum, Whisky, Aniseed, Liquors, Other spirits), and this is also the case for Aperitifs (Cocktails/Punches, Liquor wines, Natural Sweet Wines, Amer/Gentiane/Vermouth, Other aperitifs). 

We added information on brand manufacturers using the SIRENE open data and manual search of information. These information are matched with our purchase data to construct a nomenclature that makes a distinction between the largest market players, firms with national brands (NB), and retailers with private label brands (PL), and an aggregate of smaller manufacturers.\footnote{The term 'National Brand' designates the brands that are distributed on the whole French territory, by all retailers (e.g. Ricard is distributed by all retailers). The term 'Private label' designates products produced by the retailer itself (for example, a specific brand produced by \textit{Intermarché} and sold only in \textit{Intermarché}).} Within manufacturing firms, other smaller brands are aggregated.
\newline

We eventually consider manufacturers and retailers as additional characteristics because they play a critical role in pricing and marketing (e.g., shelf positioning, price discounts). Vertical relationships between manufacturers and retailers are likely to modify supply-side adjustments to price regulations, and therefore it is important to consider brands sold at different retailers as distinct competing choice alternatives. The retailers were grouped into seven distribution channels, corresponding to the five largest retailers, an aggregate of hard-discounters, and an aggregate of other shopping places (see Appendix Section \ref{AppData:Nomenclature} for details). 

By interacting categories, subcategories, brands, manufacturers, and retailers, we obtain 1,662 different product varieties. For each variety and four-week period, we calculate the purchase volumes and the expenditure at national level (adjusting them for sampling weights), and finally the average unit price in €/litre. Appendix Section \ref{AppData:Alcohol} provides details on the computation of key variables, such as alcohol content, product prices, purchase volumes.

 KWP Homescan data offer the advantage of providing detailed information on purchase volumes, expenditures, and product characteristics: brand, key differentiating attributes, and alcohol content.\footnote{Appendix Section \ref{AppData:CompareINSEE} show that our data match well with the expenditure statistics provided by the 2017 Family Budget Survey.} Available information on alcohol content is crucial for the analysis of alcohol taxation, which consists of excise duties (ie, based on the volume purchased, not the value), which can vary depending on the category of alcohol, the size of the manufacturing firm, and the alcohol content of products. We constructed the excise and volumetric taxes faced by each variety in 2014 by combining fiscal information on tax rates applied in 2014 with information on the alcohol categories and subcategories, the size of manufacturing firms and the alcohol content. Appendix Section \ref{AppData:CurrentTax} presents the excise taxes applied in 2014 and 2024, showing that the tax regulations did not change much over 10 years.   

\subsection{Facts}

\subsubsection{Markets}

Table \ref{tab:marketstats} presents some market statistics for each alcohol category. Still wines represent 45.34\% of purchase acts, followed by beers (21.89\%) and spirits (15.38\%).\footnote{A purchase act is the purchase of a product variety in a given day at a given retailer by a given household.} Aperitifs, sparkling wines, and ciders represent less than 10\% of purchase acts each. The second line shows that ciders, beers and still wines are the cheapest alcohol categories on average, whereas spirits form the most expensive category. 
\bigskip 

\renewcommand{\arraystretch}{1.2}
\begin{table}[]
    \centering
    \caption{Market statistics}\label{tab:marketstats}
    \footnotesize
    \hspace*{-0.75cm}
    \begin{tabular}{l| c c c c c c c c}
        \hline \hline
          & \textbf{Ciders} & \textbf{Beers} & \textbf{Aperitifs} & \textbf{Spirits}  & \makecell{\textbf{Still} \\ \textbf{wine}}  & \makecell{\textbf{Sparkling} \\ \textbf{wines}}  \\
        \hline
         Share purchase volumes ($\%$) &  2.22  &  31.56 & 5.13 & 7.81  & 48.28    & 4.99   \\
        Average price (\euro/L) &  2.43 (0.67)  &  2.36 (3.52)& 6.07 (2.82)& 16.38 (5.72) & 3.78 (1.34)    & 10.41 (8.86)   \\
       
        \hline
        Number of subcategories &  2  &  3 & 5 & 5 & 3     & 2  \\
        Number of manufacturers  &  6 &  25 & 29 & 14 & 12    & 35   \\
        Number of brands &  8  &  52 & 32 & 48 & 26  & 38   \\
         Number of product varieties  & 75 &  395 & 336 & 404 & 230    & 222  \\
        \hline 
        $CR_{4}$ & 36.07 &  61.37  & 35.20 &  51.11  & 11.78    & 21.60  \\
        HHI index ($\in ]0;10,000[$) & 1473 &  1388  & 608 & 986 & 342     & 391   \\
        \hline 
        Market shares, in volumes (\%) & & & & & & \\
        \quad Main national brands  &  36.12 &   73.17     &   47.26 & 62.58 &   13.09 &   43.89             \\
        \quad Private labels  &  54.50 &  21.02  & 33.20  & 28.99 & 38.26    & 32.34   \\
        \quad Aggregate of other small brands&  9.38 & 5.81 & 19.54 & 8.43 & 48.65   & 23.16 \\
        \hline
        Prices (\euro/L) & & & & & &  \\
        
        \quad Main national brands&  2.95 (0.70)   &  2.93 (1.22)  &  7.09 (3.57)&    18.24 (6.84)  & 3.81 (1.89)    &   14.58 (14.46)             \\
       \quad Private labels & 2.04 (0.24) &   1.34 (0.51)  &   5.08 (2.15) &   11.51 (5.18)  &  2.86 (1.04)      &  11.74 (7.51)          \\
        \quad Aggregate of other small brands&  3.23 (0.74)  & 3.45 (1.10)& 6.75 (4.12) & 16.93 (8.28)& 3.26 (1.37)     & 12.42 (7.23)  \\
 \hline\hline
    \end{tabular}
    \begin{center}
		\begin{minipage}{1\textwidth}
			{\footnotesize \textbf{Notes}: Market statistics computed over N=1,662 brands observed over 13 periods of 4-weeks, and purchase acts made by a sample of N=11,463 households. A purchase act is the purchase of a given product variety in a given day at a given retailer by a given household. Standard deviations in parentheses describe the variability of prices across periods and product varieties. Market shares are calculated from household purchase volumes (in L) adjusted for household and purchase period-specific sample weights. }
		\end{minipage}
	\end{center}
\end{table}

Table \ref{tab:marketstats} statistics also show that the six markets differ considerably in terms of structure and competition. The number of varieties is quite heterogeneous in the six markets, from 75 for ciders to 404 for spirits. For ciders, only six manufacturers (including the aggregate of "other manufacturers") compete by selling eight brands. For spirits, 14 manufacturers compete through 48 brands.  To better measure the level of competition in each market, we compute the concentration ratio ($CR_{4}$) and the Herfindahl-Hirschman Index (HHI).\footnote{The $CR_{4}$ is defined as the sum of the market shares of the four largest national firms in an industry. The HHI is defined as the sum of the squares of the market shares of the national firms within the industry. Hence, $CR_{4}$ represents the market power of the largest firms, whereas HHI measures more globally the concentration of all national large firms within the industry}
Their computation is based on the market shares of firms apart from private labels and the aggregate of other small brands, meaning that these two measures represent a lower limit of concentration. These two categories can be thought of as a competitive fringe offering product varieties that are cheaper or with specific characteristics.

The beer and spirits markets can be perceived as oligopolies, since their $CR_{4}$ are equal to 61 and 51\%, respectively. In contrast, there are no such large firms in the market of still wines, since $CR_{4}$ is equal to 12\%. Except for ciders, beers, and spirits where the HHI of the main national manufacturers is at a moderate level of competition, the three other markets do not seem to have a high level of concentration at the manufacturer, given their HHI below 1000. This is partly explained by a large private label market share and/or a large market share of small national brands. In fact, those three markets are largely made up of small producers, and it is particularly important for wines, 23.2$\%$ for sparkling wine and 48.6$\%$ for still wines, and to a lesser extent 19.5\% for aperitifs. This share is much lower for the other markets: 5.7\% for beers, 9.4\% for ciders, 8.4\% for spirits, and 5.8\% for beers. This emphasizes a specificity of the French wine market: compared to the US, Australian or Chilean markets, the French wine sector was and is still characterized by the important role of small and medium manufacturers as independent winemakers and cooperatives \parencite{cubertafond_entreprendre_2015}.

Although all national brands are always more expensive than private label products, the ranking between the main national brands and the other small national brands varies between markets. Although small national brands for cider and beer are on average more expensive than the main national brands, the reverse is true for the other four markets. Additional descriptive statistics in Table \ref{Tab:DescStatClass} in the Appendix further illustrate the heterogeneity of the product offer between the subcategories within the six markets.

\subsubsection{Household choices over quantity and quality}

Households buy on average 0.83 liters of ethanol per week per household member aged 16 or older. Taking into account the alcohol content of the varieties, this translates to 79.4g of pure ethanol per week, or nearly 8 standard drinks, while the French recommendation guidelines set the maximum at 10 standard drinks per week. 

The first panel of Table \ref{tab:hholdstats} shows the breakdown of purchases in terms of volume, in liters, and standard drinks. Still wines top the list, with 48.3\% of volume in liters and 49.9\% in standard drinks. The second category with the highest purchase, beer, accounts for 31.6\% of volumes but only 13.2\% of standard drinks. The figures for spirits are 7.8\% and 25.7\%, respectively, showing their important contribution to alcohol intake. 

The second panel of Table \ref{tab:hholdstats} describes the distribution of the unit values of purchases. The price in \euro\ per liter paid by households is the highest for spirits, ahead of sparkling wines and aperitifs.\footnote{The large dispersion of unit values for sparkling wines is explained by the vertical differentiation between the subcategory of champagnes and that of other sparking wines.} With the exception of wines, variations in median unit prices between categories are positively correlated with the median ethanol content of the alcohol category. Unit prices for still wines show little price differential with beer, contrary to what is often observed in countries with more of a brewing tradition than a wine-growing one. 

\renewcommand{\arraystretch}{1.2}
\begin{table}[]
    \centering
   \caption{Household purchase volumes, prices and taxes across the six alcohol markets}
    \label{tab:hholdstats}
    \footnotesize
    \hspace*{-0.5cm}
    \begin{tabular}{l| c c c c c c c c}
        \hline \hline
          & \textbf{Ciders} & \textbf{Beers} & \textbf{Aperitifs} & \textbf{Spirits}  & \makecell{\textbf{Still} \\ \textbf{wine}}  & \makecell{\textbf{Sparkling} \\ \textbf{wines}}  \\
        \hline
        \textbf{Consumption} &&&&&&\\
        Purchase volumes (L/year) &   1.69 & 24.05 & 3.91 & 5.95 & 36.79 & 3.80 \\
        Share purchase volumes ($\%$) &  2.22  &  31.56 & 5.13 & 7.81  & 48.28    & 4.99   \\
        Standard drinks/week &   0.09 & 1.80 & 0.74 & 3.51 & 6.80 & 0.69\\
         Share standard drinks  ($\%$) &  0.69 & 13.17 & 5.51 & 25.72 & 49.87 & 5.05    \\
                 \hline
         \textbf{Alcohol content} &&&&&&\\     
                         Median alcohol degree & 4.4 & 5.8 & 15.0 & 40.0 & 12.0 & 12.0 \\
        \hline
         \textbf{Prices per liter} &&&&&&\\       
      Unit value - First quartile (\euro/L) &  2.17  &  1.98 & 4.22 & 14.33 & 2.44    & 5.68   \\
        Median unit value (\euro/L) &  2.74  &  2.86 & 6.64 & 16.93 & 3.22    & 8.06   \\
        Unit value - Third quartile  (\euro/L) &  3.33  &  3.53 & 9.15 & 20.07 & 4.37    & 24.38   \\

                \hline 
         \textbf{Taxes} &&&&&&\\   
        Average unit value (\euro/L) - all tax included   & 2.88 & 2.53 & 6.40 & 20.31 & 4.10 & 13.28 \\
         Average unit value (\euro/L) - VAT excluded & 2.40 & 2.11 & 5.37 & 16.93 & 3.41 & 11.07 
        \\
        Average unit value (\euro/L) - all taxes excluded & 2.39 & 1.76 & 4.37 & 8.54 & 3.38 & 10.98 \\
        Implicit tax rate (\%) & 20.79 & 50.40 & 62.83 & 219.42 & 21.56 & 21.14 \\
        \hline
                 \textbf{Prices per standard drink} &&&&&&\\   
         Average unit value (\euro/ standard drink) - all tax included  & 1.00 & 0.65 & 0.64 & 0.66 & 0.43 & 1.41 \\
        Average unit value (\euro/ standard drink) - VAT excluded & 0.84 & 0.54 & 0.54 & 0.55 & 0.35 & 1.18 \\
        Average unit value (\euro/ standard drink) - all taxes excluded & 0.83 & 0.45 & 0.44 & 0.28 & 0.35 & 1.17 \\
         \hline\hline
                  \textbf{Policy scenarios} &\multicolumn{6}{c}{Expected $\%$ change in average unit price (\euro/L)}\\   
                &\multicolumn{6}{c}{without market reactions}\\ 
     \hline
        Low uniform tax & +10.4 & -0.4 & -1.7 & -33.2 & +24.4 & +6.9 \\
        High uniform tax & +21.2 & +16.6 & +15.5 & -16.6 & +50.0 & +14.7 \\
        Low progressive tax & +4.9 & -7.5 &  -1.4 & -19.4 & +21.5 & +6.0 \\
        High progressive tax & +9.7 & +0.4 & +13.0 & +5.8 & +40.2 & +11.7 \\
        Minimum price & +1.4 & +6.3 & +8.3 & +3.2 & +40.0 & +3.2 \\
        Minimum price + Low prog. tax & +5.9 & +1.6 & +7.2 & -8.0 & +46.6 & +8.1 \\
        \hline\hline
\end{tabular}
    \begin{center}
		\begin{minipage}{\textwidth}
			{\footnotesize \textbf{Notes}: Statistics computed over N=11,463 households. Statistics weighted by household and purchase period-specific sample weights - see Appendix Section \ref{AppData:KWPFormula} for the exact formula.  Implicit tax rate = 100 x tax revenue / (sales - tax revenue).}
		\end{minipage}
	\end{center}
\end{table}

\subsection{Motivations for price policy reforms}

In general, the current price regulation system does not align with public health objectives. The third panel of Table \ref{tab:hholdstats} also shows that the different alcohol categories are subject to very different tax burden due to the differences in the specific volumetric taxes currently applied. The implicit tax rate, corresponding to excise duties and other taxes including VAT \parencite{ruiz_caractere_2008}, ranges from a rate close to VAT (nearly 21\%) for the three least taxed categories, ciders, still and sparkling wines, to almost 220\% for spirits and 50\% and 63\% for beers and aperitifs, respectively. The fourth panel of Table \ref{tab:hholdstats} reports the average tax levels per standard drinks. Clearly, the current tax design is much more favorable to ethanol in wines than to the ethanol in spirits.  

In addition, the current price regulation system is regressive, as shown by \textcite{lecocq_alcohol_2023} in a descriptive study based on the same data. They find that the household tax effort rates (the ratio of taxes excluding VAT paid to disposable income) is 0.45\% for low-income households compared to 0. 11\% for high-income households.\footnote{The latter is likely to reflect both social differences in circumstantial exposure to addiction risks, and differences in individual self-control efforts. It is clearly difficult to measure the relative importance of circumstances and efforts in these inequalities.} Consistent with the gap in consumption patterns and effort rates between income groups, the implicit tax rate (the ratio of tax burden to pretax expenditure) is 53.2\% for low-income households versus 41.7\% for high-income households. They show that the observed regressivity of the current tax system can be attributed to higher overall volumes of alcohol per capita purchases and higher share of spirits relative to wines in low-income households, where income is adjusted for household OECD consumption units (see Appendix Tables \ref{Tab:DescStatHholdLiters} and \ref{Tab:DescStatHholdDrinks}). From an externality-correction perspective, this raises the question of horizontal equity, since a euro spent on pure alcohol is not taxed in the same way depending on household income. This can be an additional legal motivation for a tax reform as discussed in \textcite{sassi_taxation_2014}.

These statistics finally illustrate the specific economic importance of wine in France. It is the most consumed and least taxed alcohol category due to its historical role in gastronomic culture and agricultural production. Given the economic and cultural concerns about the still wine market, it is important to clarify its market segmentation. In our data, most still wine purchases fall in the price range that defines the low-end quality level according to market professionals \parencite{cubertafond_entreprendre_2015}: more than 80\% of wine volumes are purchased at less than 5\euro/L. Wines purchased at less than 3\euro/L account for 35\% of purchases and almost 50\% of volume, which is explained by the fall in unit prices for wines in bag-in-box packaging, which is used primarily for low-quality product varieties. These descriptive statistics highlight a fact that has been overlooked in the public debate on alcohol regulation in France: a significant proportion of the volume of wine placed on the market is of poor quality and has a low price. However, the economic importance of the sector implies that we must pay particular attention to the designs of potential policy reforms, and we must be able to document their effects on producers and consumers.

\subsection{Policy scenarios}

We are interested in scenarios of price policy reforms that can promote public health objectives while still complying with legal constraints. From a legal perspective, any price regulation reform must be calibrated to match either current fiscal revenues or the monetary externalities on public finances. Regarding the first objective, Article 40 of the French Constitution prohibits any creation or aggravation of a public burden and only authorizes the reduction of fiscal revenues if this is offset by an increase in another revenue. The second objective, internalizing externalities (at least partially), is a classical rationale for public policies.\footnote{We abstract here from considerations regarding an optimal tax design that would specifically account for the private externalities (e.g. productivity losses) and the internalities generated by the tax, as in Allcott et al. (2019) for instance. Their approach requires identifying a priori those consumers who face internalities, and in the specific case of wine in France, the argument is not well accepted by stakeholders and the general public.} Beyond fiscal neutrality, a second legal constraint is that Article 110 of the Fundamental Treaty of the European Union stipulates that direct or indirect taxation must be the same for similar goods moving freely within the Union and must not serve as indirect protection for other goods.\footnote{Article 110: "No Member State shall impose, directly or indirectly, on the products of other Member States any internal taxation of any kind in excess of that imposed directly or indirectly on similar domestic products. Furthermore, no Member State shall impose on the products of other Member States any internal taxation of such a nature as to afford indirect protection to other products"}  This article severely limits the scope for reform, as has been shown in several cases in the past.\footnote{Article 110 raises the question of the similarity of products within a category and of possible substitutions between categories. In case 243/84 John Walker (1986), it was ruled that whisky and liqueur wines are not like goods. In case 106/84 Commission vs. Denmark (1986), it was held that grape wines and other wines are similar. In case 170/78, Commission vs. United Kingdom (1980), the introduction of a tax on wine five times higher than that on beer was refused on the grounds of substitutions between the two product categories.} Only a reform motivated by a legitimate public health objective and proportionate to that objective (i.e. sufficiently effective compared to less stringent policies such as information campaigns) can be deemed to comply with Article 110 \parencite{sassi_taxation_2014}. It will therefore necessarily have to target the ethanol content of products, whatever their category. 

Given these constraints, we will first simulate the replacement of all alcohol-specific taxes by a single volumetric tax, either uniform or increasing progressively with the ethanol content of the product. Under the uniform volumetric tax, one gram of ethanol has the same impact, whether it comes from a glass of cider, wine, or spirit. In contrast, the progressive tax scheme is based on the premise that each additional gram of ethanol is associated with an increase in marginal risks to health. Hence, the marginal tax rate on each additional degree is increasing. This progressive volumetric tax is similar in spirit to the sugar taxes on soft-drinks that have recently been implemented in the UK, Chile, or France. In practice, we will consider a multi-tiered tax scheme with six intervals for the alcohol degree: [0;5[, [5;10[, [10;15[, [15;25[, [25;45[ and [45;100].\footnote{For 100mL of alcohol, one alcohol degree is 1mL of pure ethanol, or $0.8$g. In France, a standard drink is 10g of ethanol.} We will assume that marginal taxation is twice higher in the second interval than in the first, three times higher in the third than in the first, etc. 

We calibrate the uniform and progressive tax schemes to achieve one or another of the following two objectives under the assumption that there are no reactions on the demand and supply sides of the market: (1) fiscal neutrality (2) neutrality for public finances. Under fiscal neutrality, all alcohol tax revenues (including VAT revenues) are left unchanged by the reform. Under neutrality for public finances, the alcohol tax revenues (excluding VAT revenues) must cover the costs of alcohol for public finances (health cares and enforcement of the law, minus retirement pensions). Public finance neutrality always leads to higher calibrated tax rates as we exclude VAT revenues from the accounting equation. We thus end with four scenarios: Low uniform tax (fiscal neutrality), High uniform tax (public finance neutrality), Low progressive tax (fiscal neutrality), High progressive tax (public finance neutrality). 

We will also evaluate the implementation of a minimum unit price on top of the current tax system. The minimum unit price paid by the consumer (VAT included) will be set to 0.5\euro$ $ per standard drink (10g of ethanol), which is similar to the MUP initially implemented in Scotland in 2012. This is our fifth scenario, where existing taxes are left unaltered. Finally, as a final scenario, we will evaluate the implementation of the minimum price combined with the replacement of existing excises and duties by the low progressive tax. 

Appendix \ref{AppTaxDesign} provides additional details on the design and calibration of all policy scenarios, while the lower panel of Table \ref{tab:hholdstats} shows their impact on unit prices in the absence of consumer and market adjustments. Interestingly, given the distortions of current tax schemes, the average price of still wines would increase by 20\% to 50\% depending on the scenario, while the average price of spirits may fall by up to -33\% under a flat tax. The flat tax is also unfavorable for ciders and beers. This is not the case with progressive taxes, which benefit beers, penalize wines less and are less favorable to spirits. Only a high progressive tax, aimed at cost neutrality for public finances, would prevent a fall in the price of spirits and aperitifs. To a lesser extent, a minimum price would also prevent such a drop. However, in all cases, the relative price of wine would increase sharply, suggesting substitution towards other alcohols that become relatively cheaper. Only by modeling household substitution behavior can we make predictions about these substitutions and their consequences for pure alcohol consumption.
\FloatBarrier

    \section{Model}
In this section, we build a tractable quantitative model of alcohol markets that is informed by our descriptive analysis of markets and household behaviors. Our goal is to capture the behavioral mechanisms that are the most relevant to explain changes in household purchase patterns and market equilibrium: first, substitutions between product varieties of different degrees of alcohol, and substitutions \textit{and} complementarities between alcohol categories; second, the optimal price-setting strategies of retailers and manufacturers. Finally, we discuss the assumptions of the model.

\subsection{Consumer demand for quality and quantity}\label{Sec:Demand_Model}

We consider alcohol consumption to be separable from the rest of consumption expenditures. The set of
products is partitioned into $A$ mutually exclusive product
categories. In each period $t$, the household $h$ first allocates a predetermined alcohol budget $Y_{t}^{h}$ for at-home consumption. Then, it will have a number of
consumption occasions that will exhaust its budget $Y_{t}^{ah}$ for category
$a$. Each consumption occasion consists of the choice of either consuming one
product $j$ among the choice set $J_{t}^{a}$,
or not consuming in category $a$ and keeping the money for expenditures in
other categories (their outside option). At the beginning of the period,
households have perfect anticipations regarding the availability and price of
products during the period. They allocate their budget between categories according to the expected utility of consuming a product variety in each of
these categories. 

 Define $\mathbf{q}_{t}^{a}$ the quantity vector of products
$j$ in category $a$ in period $t$, with the related price vector $\mathbf{p}_{t}^{a}$. We assume that preferences over quantity
are weakly separable, so that each household solves at each period
the following maximization program:

\begin{gather}
Max_{\mathbf{q}_{t}^{a},\forall a=1...A}F^{h}\left(  U_{t}^{1h}(\mathbf{q}_{t}
^{1}),...,U_{t}^{Ah}(\mathbf{q}_{t}^{A})\right) \\
u.c.~\sum_{a=1}^{A}\mathbf{p}_{t}^{a}{}^{\prime}\mathbf{q}_{t}^{a}=Y_{t}^{h} \nonumber
\end{gather}

\noindent where $F^{h}$ is a regular utility function, and $U_{t}^{ah}$ is the expected utility of consuming quantities $\mathbf{q}_{t}^{a}$ of products in category $a$ at various occasions of consumption over period $t$. 

We define a consumption occasion $o$ as the consumption
of one unit of a product $j$ in any alcohol category (whatever the unit: the goods are perfectly
divisible). The number of consumption occasions in $t$ for an alcohol category is therefore
equal to the aggregate physical quantity $Q_{t}^{ah}=\sum _{j\in J_{t}^{a}}q_{jt}^{ah}$. At each consumption occasion, conditional on choosing a product
$j$ in category $a$, the household gets some
consumption utility that is noted $v_{jto}^{ah}$. Note that this utility
should be the same for all products $j$ in $a$ if the household had no taste
for variety or if the products were homogeneous. Thus, within-category product
differentiation is an extra-source of utility, beyond the consumption of
quantity $Q_{t}^{ah}$.

When deciding to allocate its budget across alcohol categories and planning
consumption occasions, the household shopper has in mind the expected utility
that could be obtained from consuming a product variety in each product
category conditional on the purchase of at least one product variety in
this category:

\begin{equation}
u_{to}^{ah}\sim\mathbb{E}_{j\in J_{t}^{a}}\left(  v_{jto}^{ah}|Q_{t}
^{ah}>0\right)
\end{equation}

The conditional distribution of utility $v_{jto}^{ah}$ is identified by
the observation of household purchase behaviors. When shopping, the household
decides which variety to purchase according to a standard trade-off between
money and the average utility of consuming the product on any consumption occasion. Specifically, we assume
that, at the point of purchase, the consumption utility enters the purchase utility $V_{jto}^{ah}$, through the following relationships:

\begin{equation}\label{Eq:utility_product}
V_{jto}^{ah}=-\alpha^{ah}p_{jt}^{a}+v_{jto}^{ah} 
\end{equation}

\noindent \textit{where }$\alpha^{ah}>0$\textit{\ is the household marginal utility of expenditure
in category }$a$. The consumption utility itself,
$v_{jto}^{ah}$, is specified so that purchase decisions can be described by a
random utility model, where the stochastic term captures some taste for
variety plus other idiosyncratic factors that affect the expected value of $j$
at consumption occasion $o$. Households also have the choice not to purchase a product in category $a$ and keep their money for purchasing any product in any other category on the same shopping trip: this is their outside option.

Purchase decisions are finally related
to consumption utility through a Random Utility Model, with the purchase
utility associated with option $j$ being equal to:

\begin{align}
\forall j  &  \in J_{t}^{a}, \forall o, V_{jto}^{ah}=-\alpha^{ah}
p_{jt}^{a}+\underset{v_{jto}^{ah}}{\underbrace{\beta^{ah}Z_{jt}^{a}
+\varepsilon_{jto}^{ah}}} \nonumber\\
&  =V_{jt}^{ah}+\varepsilon_{jto}^{ah} \label{Eq:RUM}
\end{align}

\noindent where the $\varepsilon_{jto}^{ah}$ have i.i.d. Gumbel
distributions, and $Z_{jt}^{a}$ are sets of
product attributes.
In Appendix \ref{AppModel}, we show that this setup implies that the expected utility of a consumption occasion is:

\begin{align}
    u_{to}^{ah}=(1+b_{t}^{ah}(\mathbf{p}_{t}^{a}))
\end{align}

\noindent where $b_{t}^{ah}(\mathbf{p}_{t}^{a})$ is the consumer surplus from
differentiation between products \emph{within} category $a$. We suppose that the anticipated utility over the period $t$ of product purchases in category $a$ is equal to the linear sum of expected occasion utilities:

\begin{equation}\label{eq:adjusted_quantity}
U_{t}^{ah}(\mathbf{q}_{t}^{a})=\sum_{o}u_{to}^{ah}=Q_{t}^{ah}(1+b_{t}%
^{ah}(\mathbf{p}_{t}^{a}))
\end{equation}

\bigskip

\noindent We relabel the anticipated utility level $U_{t}^{ah}(\mathbf{q}_{t}^{a})$ as a quantity index $\mathbb{Q}_{t}^{ah}$. Then, we derive a corresponding price
index $\mathbb{P}_{t}^{ah}$ that satisfies the budget constraint for alcohol category $a$:

\begin{equation}\label{eq:Adjusted_price}
\mathbb{P}_{t}^{ah}\mathbb{Q}_{t}^{ah}=Y_{t}^{ah}\Rightarrow\mathbb{P}%
_{t}^{ah}=\frac{Y_{t}^{ah}}{Q_{t}^{ah}(1+b_{t}^{ah}(\mathbf{p}_{t}^{a}))}%
\end{equation}

\noindent This price index equals the unit value of purchases in category $a$ ---
$Y_{t}^{ah}/Q_{t}^{ah}$ --- adjusted for the index of the expected welfare
surplus from quality differentiation within category $a$, $b_{t}^{ah}(\mathbf{p}_{t}^{a})$. If the products in $a$ were not differentiated, then all would have the same price,
which in turn would be equal to the price index since $b_{t}^{ah}(\mathbf{p}_{t}^{a})=0$. Only differentiation between
product categories, as captured by the first-stage budgeting decision, would
matter.
\bigskip

Finally, we rewrite the first-stage budgeting decision with the quality-adjusted price
and quantity indices as follows:

\begin{gather}
Max_{\mathbb{Q}_{t}^{ah},\forall a=1...A}F^{h}\left(  \mathbb{Q}_{t}
^{1h},...,\mathbb{Q}_{t}^{Ah}\right) \\
u.c.~\sum_{a=1}^{A}\mathbb{P}_{t}^{ah}\mathbb{Q}_{t}^{ah}=Y_{t}^{h}
\end{gather}

\noindent A standard demand system can be estimated to recover the price elasticity
of the quantities. The econometric calibration of the demand models is eventually performed in two steps. First, we estimate a discrete random utility choice model for each product category. Second, we use these estimates to construct the quality indices $b_{t}^{ah}(\mathbf{p}_{t}^{a})$ and fit the first-stage demand system for quantities.
    
\subsection{Supply-side price-setting decisions}

To simulate the impact of price regulations on markets, we need a model of how manufacturers and retailers set product prices. We do not observe all the determinants of these decisions \emph{directly}, especially the production costs and technologies and the nature of vertical contracts. We therefore need structural assumptions regarding the supply-side of markets to infer this information from the observed data on purchase behaviors. 

Each alcohol category in the demand model corresponds to a market. For each market $a$, we assume that manufacturers and retailers engage in two part-tariff contracts, whereby manufacturers propose to retailers a wholesale price and a fixed franchise fee. In addition, manufacturers implicitly use resale price maintenance (RPM), and therefore set the retail prices of national brand products (NB) while retailers set the prices of their private label products (PL).\footnote{The two-part tariffs with RPM has been largely used in the empirical literature on the French retail market. \textcite{bonnet_inference_2010} proposed an empirical test of various vertical relationships models on the French market for bottled water. They demonstrated that the two-part tariff model with RPM provides the best performance in terms of statistical fit. Following \textcite{rey_resale_2010}, this model may account for situations where manufacturers have all the bargaining power for a product and make take-it or leave-it offers, as well as for situations where retailers have all the bargaining power.} Profits are shared between manufacturers and retailers via a set of product-specific fixed fees. This model of vertical relationship helps to take into account the specific role played by PL products in the price-setting strategies of retailers, especially in low-quality market segments.

Formally, we define $R$ the number of retailers, and $M^a$ the number of manufacturers in each separate market $a$ in each time period $t$. For notational convenience, we drop the subscripts $a$ and $t$.

The profit of retailer $r$ is given by:
\begin{equation}
	\Pi ^{r}=\sum\limits_{j\in S_{r}} \left[ Q\left( \frac{p_{j}}{1+\tau}-w_{j}-c_{j}\right) s_{j}(\mathbf{p})-F_{j} \right], \forall r=1,... R \label{Eq:RetProfit}
\end{equation}

\noindent where $S_{r}$ is the set of products that retailer $r$ sells, $Q$ is the size of the market, $\tau$ is the VAT rate, $p_j$ is the retail price of product $j$, $w_j$ is the wholesale price of product $j$, $c_{j}$ is the constant marginal cost to distribute product $j$, $s_{j}(\mathbf{p})$ is the market share of product $j$, and $F_j$ if the franchise fee that the retailer pays to the manufacturer.\footnote{As a product is a combination of a brand and a retailer, one can define a fixed fee $F_{j}$ for each product and separate the contracts that manufacturers have with each retailer. $F_{j}$ is then the fixed fee received by a given manufacturer (if positive) from a given retailer.}  In the specific case of PLs, we assume that the products are sold to retailers at a wholesale price $w_{j}$ equal to the marginal cost of the manufacturing firms noted $\mu_{j}$.\footnote{A retailer defines the characteristics of his own PL, organizes competition among manufacturers for delegating the production of these PL products. This should be interpreted as a price competition with homogeneous products that leads to a wholesale price equal to the marginal costs.}

Each manufacturer $m$ maximizes its profit :
\begin{equation}
	\Pi ^{m}=\sum \limits_{j\in G_{m}} \left[ Q(w_{j}-\mu _{j}-T_{j}^{0})s_{j}(\mathbf{p})+F_{j} \right], \forall m=1,..., M^a \label{Eq:FirmProfit}
\end{equation}

\noindent subject to the participation constraint of each retailer $\Pi ^{r}\geq \widetilde{\Pi }^{r}$, $\forall r=1,...,R$. $G_{m}$ is the set of products that are sold by manufacturer $m$,  $w_{j}$ is the wholesale price of product $j$, $\mu_{j}$ is the marginal production cost of $j$, $T_{j}^{0}$ is the excise tax, and $\widetilde{\Pi }^{r}$ is a fixed reservation utility that we can normalize to zero without loss of generality.

Following \textcite{rey_resale_2010}, under RPM, the participation constraints are binding and manufacturers choose to set retail margins to zero.\footnote{The intuition is that RPM produces a situation where prices are set as if the retailer and the manufacturer were integrated, and this increases total profit. The fixed fees are used to share the profits between manufacturers and retailers.} Therefore, the prices of NB products sold by the manufacturer are given by the following first order conditions for the manufacturer $m$ and for all products $j\in G_{m}$:
\begin{gather}
	\sum \nolimits_{k\in G_{m}}\left( \frac{p_{k}}{1+\tau}-\mu _{k}-c_{k}-T_{k}^{0}\right) \frac{\partial s_{k}(\mathbf{p})}{\partial p_{jt}}+s_{j}(\mathbf{p})+\sum_{r=1}^R \sum \nolimits_{k \in \widetilde{S}_{r}}\left( \frac{p_{k}}{1+\tau}-\mu_{k}-c_{k}-T_{k}^{0}\right)\frac{\partial s_{k}(\mathbf{p})}{\partial p_{jt}}=0  \label{eq1}
\end{gather}

\noindent where $\widetilde{S}_{r}$ is the set of PL products that retailer $r$ sells.\footnote{The marginal retail profit from NB products is zero, but they can receive fees, that do not depend on prices, to share the total profit on NB products. Hence, the manufacturer $m$ has just to care about the profit made by the retailer with its PL brands, and the competition of the latter with its own brands.} These first-order conditions implies that the manufacturer's pricing policy depends on two factors. First, it has to optimize the competition between products in its portfolio (the two leading terms). Second, it has to account for the competition of retailer brands, the so-called common agency problem (the third term).

On the retailer side, the prices of PLs in $\widetilde{S}_{r}$ are given by the first-order condition:

\begin{equation}
	\sum \nolimits_{k\in \widetilde{S}_{r}}\left(\frac{p_{k}}{1+\tau}-\mu _{k}-c_{k}-T_{k}^{0}\right)\frac{\partial s_{k}(\mathbf{p})}{\partial p_{jt}}+s_{j}(\mathbf{p})=0, \forall j \in \widetilde{S}_{r}.
	\label{eq2}
\end{equation}

Let $C_{j}$ be the total marginal costs for each product $j$, where $C_{j}=\mu_{j}+c_{j}$ is the sum of the marginal costs of production and distribution. Knowing the demand functions $s_{j}(\mathbf{p})$, a vector of marginal costs for all products in $J_{t}^{a}$, $\widehat{C}_{t}=\left(\widehat{C}_{1t},..,\widehat{C}_{jt},..,\widehat{C}_{Jt}\right)$, can be calibrated by solving equations (\ref{eq1}) and (\ref{eq2}) simultaneously for all manufacturers and retailers \parencite[see, for details,][Appendix]{bonnet_inference_2010}.

\subsection{Discussion of assumptions}

Our modeling strategy relies on three assumptions. First, to identify the expected utility of a consumption occasion within a period, we conflate purchase acts with consumption occasions. However, each purchase of a product observed in the data could be associated with one or several consumption occasions (depending on the number of units purchased) that are given the same utility. We thus assume that within each alcohol category, anticipated variations in purchase volumes across purchase acts are not correlated with consumers' intrinsic valuations of products on point of purchase. In their first-stage decision, consumers anticipate that purchase volumes will vary randomly across purchase acts within a period, around an average that only depends on the purchase volume decided for the period in the first stage of decision, and the number of purchase acts in the category, which is exogenously determined. This rules out optimal inventory management within the period. This also excludes the possibility that, conditional on unit price, the number of units purchased is systematically higher or lower for products associated with a higher consumption utility.\footnote{For instance, consumers do not systematically purchase one bottle of a Vin d'Appellation vs. two bottles of a Vin de Pays when they opt for the former rather than for the latter and the two products have the same unit price.}

Second, we assume that each alcohol category is one market (cider, beers, aperitifs, spirits, still wines, sparkling wines) and that the six markets are disjoint. Hence, we impose that changes in one market do not affect competition in the other markets, but only the total volume sold in other markets. As a consequence, neither manufacturers nor retailers are allowed to develop a common pricing strategy across several markets.\footnote{National brand manufacturers were mostly different in the six alcohol markets. See the list in Appendix Table \ref{AppTab:manufacturers}} Specifying separate models for each market allows us to implement a detailed modeling of substitutions between product varieties while keeping the dimensionality of the household choice set reasonable. In addition, we can apply a supply-side model that accounts for strategic price-setting decisions by manufacturers and retailers, and for the vertical relationships between them.  

Third, consistent with the assumption of disjoint markets, we separate household choices of quantity and quality by assuming that households follow a two-stage budgeting procedure, as for instance in \textcite{hausman_utility-consistent_1995}. We want to focus on substitutions in quality between product varieties within alcohol categories and substitutions or complementarities in quantity between alcohol categories. We thus ignore other possibilities of substitutions. In particular, we do account neither for substitutions toward on-trade consumption nor for intertemporal substitutions. We could not model substitutions toward on-trade due to a lack of appropriate data. We will discuss this limitation in our conclusion. Regarding stockpiling, inventory behaviors may be of some importance in understanding the seasonality of the wine market. Intertemporal substitutions could be driven by the stockpiling of products during price promotion periods. The KWP data do not provide precise information on product price promotions, and we do not observe promotions that the consumer may face on other products.
 
However, following \textcite{griffith_tax_2019}, we test for habit formation and stockpiling by examining the associations between past purchases and steady-state inventory levels and current purchases. While stockpiling can produce intertemporal substitutions, habit formation would conversely produce intertemporal complementarities. Our reduced form estimation results in Appendix \ref{AppStockpiling} show very small effects of past purchase volumes on current purchases in regressions that control for household and time fixed effects. We also find positive rather than negative associations between purchase behaviors and a steady-state inventory level measure constructed under the assumption that households manage their inventory to smooth consumption \parencite[see][]{hendel_sales_2006}. Hence, although we cannot categorically reject the existence of some form of short-term intertemporal dependence in consumer behaviors, these dynamics are not of primary importance once household preference heterogeneity is taken into account. To this end, we systematically control for the average purchase volume (in standard drinks per household adult) observed over 2013, when estimating the models on the 2014 purchase data.    
\bigskip

    \section{Calibration}
        \subsection{Market models}

\subsubsection{Specification of demand functions}

For each alcohol category, inference on household purchase behaviors is based on a structural demand system that is derived from the Random Utility Model in equation (\ref{Eq:RUM}). The econometric counterpart of the household utility function is a \textit{random coefficient logit model} that allows for estimating flexible household substitution patterns. 

Specifically, we let price disutility vary with the characteristics of households and between alcohol categories: $\alpha^{ah}= \alpha^{a} + \delta^{a} D^{h} + \sigma^{a}\zeta^{ah}$, where $D_{h}$ is a vector of household characteristics, and $\zeta^{ah}$ is a random term distributed normally with standard deviation $\sigma^{a}$ that captures the unobserved taste of the household $h$ for the alcohol category $a$. Household characteristics include the household income category, the age category of the main shopper, and a measure of the household preferences for alcohol. The latter is constructed from average household purchases observed in 2013, converted into standard drinks per adult household member per day, categorized into three categories of alcohol habits: less than one standard drink, between one and two, or two and more.

We control for product-specific household utility ($\beta^{ah}Z^{a}_{jt}$ in equation (\ref{Eq:RUM})) through a rich set of fixed-effects: (1) retailer fixed-effects that account for differences in marketing policies (assortments, promotions) across retailers; (2) brand fixed-effects, that account for differences in marketing efforts (especially advertising), beyond mere differences in consumer tastes for brands; (3) fixed-effects specific to product subcategories and sociodemographic groups, that capture sociodemographic differences in consumer tastes for product subcategories; (4) an alcohol-free times brand fixed-effects. The set of selected fixed-effects varies from one market to another, based on a descriptive analysis of household heterogeneity in product choices (see Appendix \ref{AppInference:fixedeffects}). 

The choice set does not vary across consumers. At each observed purchase occasion, the household can choose not to purchase a product in the category $a$ and keep its budget for any outside alcohol product. The utility of the outside option is normalized to zero, so that $\alpha^{ah}$ is the marginal dis-utility of spending one more euro in $a$ rather than keeping it for a purchase in any other alcohol category. Without this outside option, a homogeneous increase in the price of all products in $a$ would not affect the size of the market $a$ and therefore the reactions of manufacturers and retailers to changes in overall purchase volumes. The estimates of the first-stage demand system presented thereafter shows that none of the compensated cross-price elasticities between the six categories is negative. This preference structure implies that, for each category, the aggregate 'all products in other alcohol categories' can effectively be considered a substitute. 
 
\subsubsection{Identification of price elasticities}

Identification of price effects leverages price variations over time and across brands and retailers because we define product varieties by interacting brands with retailers. Prices may be endogenous due to unobserved product characteristics that vary over time and are correlated with both price and demand. For example, we do not observe neither the advertising efforts made by firms for their brands nor the display on shelves in stores. These unobserved marketing efforts are likely affecting the demand and correlated with prices, making them endogenous even if we control for retailer fixed-effects. We therefore apply a control function approach as in \textcite{petrin_control_2010}. In a first stage, we regress the prices on a set of instrumental variables as well as the exogenous variables of the baseline utility function.  The estimated error term of the first stage is a proxy measure of the unobserved time-varying product characteristics that are correlated with prices. We include this residual as a covariate in the estimation of logit models as a control function for the impact of unobserved product characteristics on prices and demand (see Appendix \ref{App:firststage} for details). 

The identification of the effect of the control function (the residual) relies on instrumental variables that are excluded from the demand model. We tested three types of instrument that have been widely used in the Empirical Industrial Organisation literature. First, taxes represent a large part of the final price in categories such as aperitifs and spirits. As the alcohol content of these products varies greatly across subcategories and even within brands, we can leverage tax variations across alternatives. In the spirits market, the tax level is only 0.08\euro/L for alcohol-free aniseed spirits, and raises up to 13.50\euro/L for some liquors. In the aperitifs market, the tax level varies from 0.04\euro/L for some wine-based aperitifs to 12.8\euro/L for some cocktails. Second, we also exploit Berry-Levinsohn-Pakes (\emph{BLP}) instruments such as the number of \textit{other} competing product varieties within a given subcategory at each period. These competing varieties are likely to impact markups and therefore prices, but should not have a direct impact on the household's utility from a specific alternative \parencite{berry_automobile_1995}. We sometimes interact these variables with manufacturer fixed-effects to form instruments that capture the manufacturer-specific market power. Finally, we can approximate the variation in \emph{costs shifters} across products by considering the mean price of same brand or retailer product varieties in other periods \parencite{hausman_competitive_1994}. Here, the identifying assumption is that the month-variety-specific shocks on prices are uncorrelated across periods. Hence, the mean retailer and brand prices observed in other periods captures variations in costs that are orthogonal at period-variety-specific shocks.

We selected for each market (alcohol category) the instruments that are significant in the first stage regression (F-test statistics to eliminate weak instruments). Appendix Table \ref{Tab:firststage} displays the estimation results for the first-stage price equations.

\subsubsection{Estimation results}

The six demand models are estimated by simulated maximum likelihood (see details in \ref{App:maxlik}) on the sample of households that have purchased alcohol in 2014 and were also observed in 2013 so that we can compute an average number of standard drinks per adult household member in 2013 (sample S2, see Appendix Section \ref{App:Subsamples}). 

Table \ref{Tab:RCL} presents the detailed results of the estimations of the six logit random coefficient models, where price endogeneity is controlled by the introduction of the estimated error terms of the first-stage instrumental variable equations. The coefficient on this control variable is always positive, which means that unobserved time-varying product characteristics, such as marketing efforts, simultaneously increase prices and consumer utility. 

The price coefficients reflect the willingness-to-pay of households for the various alcohol categories. For instance, as compared to the other income categories, upper-income households (the reference category) have a stronger preference for beers, sparkling wines, and still wines, but a weaker preference for aperitifs and spirits. Older households express a stronger preference for wines, aperitifs, and ciders than young households, who have a stronger preference for beers and spirits. Heavy drinking households have a stronger preference for spirits and, to a lesser extent, for still wines and aperitifs. The estimated standard deviation ($\sigma^{a}$) of the distribution of the price coefficient is nearly zero for ciders and still wines (and close to zero for sparkling wines), meaning that age, income, and the overall preference for alcohol capture most of the household taste heterogeneity in these alcohol categories.

In Appendix \ref{App:elasticity}, Table \ref{Tab:Elastclassif}, we report the average estimated own-price elasticities, by alcohol category and subcategory. They reflect the willingness to substitute to any alternative product (including the outside option). While the average elasticities are the largest for beers (-5.02) and still wines (-4.43), the standard deviation associated with the distribution of elasticities within these categories is also greater (+1.6 for beers and +2.15 for still wines). This reflects the greater heterogeneity of these alcohol categories, in terms of brand and quality segmentation. For still wines, the elasticities are larger in the lower-quality segments (\enquote{de Table} and \enquote{de Pays}), which implies that competition is more intense between products within these categories. 

Finally, Table \ref{Tab:marginsclassif} shows the average margins (in \% of the consumer price) that were calculated after calibrating the supply-side models. The higher the margin, the higher is the profit per unit. As expected, margins are generally higher for private labels than for national labels. They are also higher for lower quality wines than for higher quality ones (\enquote{de Table} vs. \enquote{d'Appellation}).   

 \renewcommand{\arraystretch}{1.3}
	\begin{center}
		\begin{table}
			\caption{Random coefficients logit models - Estimation results}
			\label{Tab:RCL}
			\def\sym#1{\ifmmode^{#1}\else\(^{#1}\)\fi}
			\centering
			\footnotesize
		\begin{tabular}{l| c| c| c| c| c| c}
			\hline \hline
			& \textbf{Ciders}& \textbf{Beers} & \textbf{Aperitifs} & \textbf{Spirits} & \makecell{\textbf{Still} \\ \textbf{wines}} & \makecell{\textbf{Sparkling} \\ \textbf{wines}} \\
			\hline
			Price &&&&&& \\
			\quad Mean ($\alpha^{a}$) & -1.700*** & -1.991***   & -0.643*** & -0.408*** & -1.572*** & -0.381***    \\
			\qquad x $D^{Income}_{Upper-middle}$  & 0.025*** & -0.021***  & 0.016*** & 0.041***  & -0.065***   & -0.012***        \\
			\qquad x $D^{Income}_{Lower-middle}$& 0.013***     & -0.105***  & 0.022***  & 0.020***    & -0.183***     & -0.016***   \\
			\qquad x $D^{Income}_{Lower}$  & -0.054***  & -0.092***  & 0.031***  & 0.006***  & -0.275***   & -0.034**        \\
			\qquad x $D^{Age}_{[35;55]}$  & 0.454*** & -0.165***   & 0.028***  & -0.006***  & 0.052*** & 0.083***       \\
			\qquad x $D^{Age}_{> 55}$  & 0.539*** & -0.374***  & 0.056*** & -0.037***   &  0.133***   & 0.054***     \\
			\qquad x $D^{Habit}_{]0;1]}$ & -0.237*** & -0.138***  & 0.051***  & 0.026***   & 0.042***  &  -0.009***      \\
			\qquad x $D^{Habit}_{>2}$  & -0.547*** & -0.187***  & 0.010***   & 0.057***  & 0.012***   & -0.025***    \\
			\qquad x $\delta^{Champagne}$  &  &  &  & &   &0.188***  \\
			\quad Standard Deviation ($\sigma^{a}$) & 0.000  & 0.411***  & 0.109*** & 0.150***   & 0.000 & 0.005***    \\
            Control function  & 1.153***    & 1.562***   & 0.369 ***  & 0.092 *** & 1.499*** & 0.153***     \\
            \hline
			Fixed Effects & &&&&&\\
			\quad Retailer    & Yes & Yes & Yes & Yes & Yes & Yes \\
			\quad Subcategory    & Yes & No & Yes & Yes & Yes  & Yes \\
			\qquad Subcategory x $D^{Income}_h$  & - & - & - & Yes & Yes& - \\ 
			\qquad Subcategory x $D^{Age}_h$  & Yes & - & - & - & - & Yes\\ 
			\qquad Subcategory x $D^{Habit}_h$  & - & - & Yes & - & - & -\\ 
			\quad Brand    &Yes & Yes & Yes & Yes & No & Yes \\
			\qquad Brand x Alcohol-free     &    No &    Yes &    No  &   No  &    No  &    No  \\
			\quad Alcohol-free & No & No & Yes & Yes & No & No \\
			\hline\hline
			\end{tabular}
			\medskip			\begin{center}
				\begin{minipage}{\textwidth}
					{\footnotesize \textbf{Notes}: N=266,460 household-purchase observations. $^{*} p < 0.10$, $^{**} p < 0.05$, $^{***} p < 0.01$. $D^{Income}_{.}$ are dummies for the household income categories (Upper, Upper-middle, Lower-middle, Lower) and Upper is the reference; $D^{Age}_{.}$ are dummies for the three age of the household head categories (18-35, 35-55, >55 years old) and 18-35 is the reference; $D^{Habit}_{.}$ are dummies for average household purchases observed in 2013 ($\le 1$, $[1;2]$, $>2$ standard drinks/adult/day) and $\le 1$ is the reference; $\delta^{Champagne}$ is a dummy distinguishing champaigns from lower quality sparkling wines.}
				\end{minipage}
			\end{center}
		\end{table}
	\end{center}

\begin{landscape}
 \renewcommand{\arraystretch}{1.3}
	\begin{table}
		\caption{Estimated own-price elasticities and calibrated margins}
		\label{Tab:marginsclassif}
		\centering
		\footnotesize
	   \begin{tabular}{l| c c c c c c}
		\hline \hline
		&\textbf{Ciders}& \textbf{Beers}& \textbf{Aperitifs}& \textbf{Spirits} & \textbf{Still wines}  & \textbf{Sparkling wines}    \\

		\hline
        Own-price elasticity & -3.40 (0.84)& -5.02 (1.60) & -3.04 (1.20) & -3.55 (0.20) & -4.43 (2.15)   & -2.71 (1.09)   \\

         \hline
		Margins  & 32.73 (6.73)& 23.67 (18.23)& 38.59 (19.54)& 33.80 (13.13) & 24.42 (7.76)   & 42.82 (18.34)  \\
        
		&  \textit{Sweet} &  \textit{Alcohol-free}  & \makecell{\textit{Cocktails} \\ \textit{and Punch}}& \textit{Rum}  &  \textit{de Table}   &  \textit{Champagne}                \\
		\quad National Brands &  25.92 (0.00)&  36.82 (6.82) &  37.13 (12.18) &  31.78 (3.05) & 38.70 (5.03)  &  26.52 (5.10)     \\
		\quad Private Labels &  37.90 (3.88)&  48.28 (5.30) &  57.88 (18.64) &  32.02 (0.82)& 34.21 (1.59)  &  31.49 (1.83)   \\
		&  \textit{Raw} &  \textit{Bock/premium} &   \textit{Liquor wines} & \textit{Whisky} &  \textit{de Pays  }    &  \textit{Other sparkling wines}             \\
		\quad National Brands  & 28.30 (7.03)&  23.07 (6.84)& 22.23 (3.18)&  30.18 (6.81) & 25.51 (6.46)  & 76.16 (23.70)   \\
		\quad Private Labels &  36.09 (3.57) & 36.22 (0.75) & 24.60 (1.65)  &  29.76 (0.81) & 26.98 (1.65) &  68.27 (8.12)    \\
		& & \textit{Special} &   \textit{Natural sweet wines}  & \textit{Aniseed}  &  \textit{d'Appellation}      &                 \\
		\quad National Brands & &  20.36 (3.69)& 37.53 (5.66)&  32.09 (0.62)& 15.57 (2.51)\\
        \quad Private Labels & &  26.49 (0.00)& 40.68 (2.69) &  30.24 (0.80)& 16.29 (1.59)  & \\
		&       &       &  \makecell{\textit{Amer, gentiane} \\ \textit{and vermouth}}    &  \textit{Liquors}      &       &    \\
		\quad National Brands & & & 36.07 (15.37)&38.07 (7.93) &  &  \\
		\quad Private Labels & & & 49.86 (4.13)&41.61 (1.66) & &  \\
		&       &       &  \textit{Other aperitifs}     &  \textit{Other spirits}     &      &     \\
		\quad National Brands & & &54.16 (10.40) & 31.26 (2.17) &  & \\
		\quad Private Labels & & & 48.05 (6.66)&29.92 (0.78) &  &  \\
		\hline \hline
		\end{tabular}
		\begin{center}
			\begin{minipage}{1.2\textwidth}
				{\footnotesize \textbf{Notes}: N=1,662 product varieties. In parenthesis, the standard deviation of the distribution of calibrated margins in the subcategory or category. Margins are expressed as a percentage of the consumer price.}
			\end{minipage}
		\end{center}
	\end{table}
\end{landscape}

\FloatBarrier
\newpage

\subsection{Demand for quantity}

\subsubsection{A pseudo-panel approach}

We use a pseudo-panel approach to estimate the quantity demand model, which allows us to avoid the issue of dealing with zero purchases in certain alcohol categories. The pseudo-panel is constructed by grouping households into clusters.
We define 120 clusters by combining the following variables: whether the household consists of
single male or single female; couple without children; or couple with at least one child under 16.
, whether the head of the household is under 35, between 35 and 54, or over 54 years old.
than 35, between 35 and 54, or more than 54 years old; whether the household
belongs to one of the following four Kantar income categories: rich, mid-rich,
mid-poor or poor; and whether the average number of standard drinks consumed per adult household member per day in 2013 was lower than or equal to 1, between 1 and 2, or greater than 2. The latter characteristic control for heterogeneity in household habits. The Rich and mid-rich, and poor and mid-poor income categories are grouped when the size of the cluster is too small. Appendix \ref{App:pseudopanel} provides details on the construction of the pseudo-panel.

Table \ref{Tab:DescQUAIDSdata} reports some descriptive statistics on budget shares, quality indices and quality-adjusted and unadjusted prices and quantities in the estimation sample (grouping all periods of 2013 and 2014).  

 As a result of the demand for quality models, the quality indices are larger in categories that are more differentiated in terms of classifications, such as aperitifs, spirits, and sparkling wines (with the distinction between mousseux and champagnes), and for which there is a larger heterogeneity in unobserved consumer taste measured by the standard deviation of the marginal utility of price in Table \ref{Tab:RCL}. Larger quality indices tend to produce lower adjusted prices (third column) and higher adjusted quantity indices (i.e. larger utility flows for consumers). This is especially true for spirits, whose small consumption volume in L (0.58 on average vs. 1.14 for wines) is associated with a very large utility flow (10.74 vs. 3.72 for wines). Adjusting the prices and physical quantities by quality reveals, to some extent, the preferences of consumers between the various alcohol categories. The low quality index for still wines specifically confirms that French households do not particularly value product differentiation on the mass market of wines. This shows the disconnection between the discourse of lobbyists, experts, and marketers of the French wine sector, which emphasizes its richness, excellence, and diversity, and the reality of the mass market. French often drink wine without paying much attention to quality. The last column shows that, despite clustering, there is still a significant number of zero expenditures for each category. Out of 3,120 observations, we have only 1,624 observations for which expenditures on all 6 categories are strictly positive. Considering that this results from finite sampling rather than from variations in cluster-specific taste, we keep all observations for estimating the model.  
\bigskip

\begin{center}
    \renewcommand{\arraystretch}{1.3}
	\begin{table}[]
		\caption{Prices, quantities and budget shares}\label{Tab:DescQUAIDSdata}
		\footnotesize
		\centering
		\medskip
        \hspace*{-0.7cm}
		\begin{tabular}[c]{lcccccccc}\hline\hline
			& \multicolumn{2}{c}{Prices (\euro/L)} &  &
			\multicolumn{2}{c}{Quantities (L/4-weeks)} & Quality & Shares & Obs. \\\cline{3-4}%
			\cline{6-7}
			$Q>0$&  Unadjusted, $P$ & Adjusted, $\mathbb{P}$ &  & Unadjusted, $Q$ & Adjusted,
			$\mathbb{Q}$ & $b$ & $w$ &\\\hline
			Ciders &  \hspace{0.07in}2.87 (0.70) & \hspace{0.07in}0.86 (0.25) &  &
			\hspace{0.07in}0.21 (0.17) & \hspace{0.07in}0.71 (0.60) &
			\multicolumn{1}{r}{\hspace{0.07in}2.37 (0.35)} & 0.03 (0.03) &1,853 \\
			Beers  & \hspace{0.07in}2.57 (0.54) & \hspace{0.07in}0.86 (0.18) &  &
			\hspace{0.07in}2.52 (1.63) & \hspace{0.07in}7.57 (4.94) &
			\multicolumn{1}{r}{\hspace{0.07in}1.99 (0.15)} & 0.24 (0.12) & 2,330\\
			Aperitifs  & \hspace{0.07in}6.69 (1.68) & \hspace{0.07in}0.73 (0.18) &
			& \hspace{0.07in}0.37 (0.24) & \hspace{0.07in}3.45 (2.30) &
			\multicolumn{1}{r}{\hspace{0.07in}8.22 (0.36)} & 0.09 (0.05) & 2,223\\
			Spirits  & 19.89 (3.38) & \hspace{0.07in}1.18 (0.27) &  &
			\hspace{0.07in}0.58 (0.53) & 10.74 (11.23) & \multicolumn{1}{r}{16.16 (2.24)}
			& 0.36 (0.12) & 2,242\\
			Still wines  & \hspace{0.07in}5.00 (1.75) & \hspace{0.07in}1.55 (0.54) &  &
			\hspace{0.07in}1.14 (1.19) & \hspace{0.07in}3.72 (3.92) &
			\multicolumn{1}{r}{\hspace{0.07in}2.23 (0.08)} & 0.17 (0.09) & 2,267\\
			Sparkling wines & 11.84 (6.48) & \hspace{0.07in}0.82 (0.44) &  &
			\hspace{0.07in}0.32 (0.30) & \hspace{0.07in}4.56 (4.16) &
			\multicolumn{1}{r}{13.48 (1.04)} & 0.13 (0.10) & 2,046 \\\hline\hline
		\end{tabular}
		\begin{center}
			\begin{minipage}{\textwidth}
				{\footnotesize \textbf{Notes}: The first column report the number of cluster-period observations with positive aggregate purchase volumes. Prices are expressed in Euro/Liter when unadjusted and in Euro/Utility
					unit when adjusted (cf equation \ref{eq:Adjusted_price}), quantities in Liter/4-weeks when unadjusted  and Utility unit/4-weeks when adjusted (cf equation \ref{eq:adjusted_quantity}) . The quality index is defined in equation \ref{Eq:quality_index} and its construction is detailed in Appendix Section \ref{App:construction_price_indices} All observations are weighted by the weight of the cluster in the total sample. Standard deviations in parenthesis show the variability across clusters and over time.}
			\end{minipage}
		\end{center}
	\end{table}
\end{center}

\subsubsection{Specification and estimation}

The quantity demand functions are derived from a Quadratic Almost Ideal Demand System \parencite[QUAIDS:][]{banks_quadratic_1997}, whereby the budget share $w_{t}^{ac}$ for
alcohol category $a$ and cluster $c$ in period $t$ is modeled as a function of log total
expenditure $\ln Y_{t}^{c}=\ln\left( \sum_{a=1}^{A}Y_{t}^{ac}\right)=\ln\left(\sum_{a=1}^{A}\mathbb{P}_{t}^{ac}\mathbb{Q}
_{t}^{ac}\right)$, the vector of log prices, ln$\mathbf{P}_{t}^{c}=(\ln
\mathbb{P}_{t}^{1c},...,\ln\mathbb{P}_{t}^{Ac})^{\prime}$, and
cluster-averages of socio-demographic control variables.\footnote{\textcite{banks_quadratic_1997} show that the Engel curves for alcohol are
	non-linear, as opposed to other food items. This justifies the use of a QUAIDS instead of the simpler AIDS originally proposed by \textcite{deaton_almost_1980}.} Appendix \ref{App:quantity_spe} provides more details about the specification.

The demand system in the Appendix Equation \ref{App:quantity_spe} is estimated using an iterative reweighted least-squares procedure to deal with non-linearities, imposing homogeneity and symmetry constraints \parencite[see][]{lecocq_estimating_2015}. In the estimation, each cluster observation is weighted by the proportion of French households
represented by the cluster. This therefore limits the influence of less populated clusters on the estimates. Although we cluster observations and include additional observed characteristics, some residual cluster-level demand side variable may be correlated with the total alcohol budget $Y_{t}^{c}$. We follow the usual practice of instrumenting the latter by log household income (cluster average).

\subsubsection{Identification}

 The price indices may be endogenous because they are constructed as the ratio of the average price of product varieties weighted by the probability of purchase (average unit value for the category) to the quality index. For identification purposes, we first assume that any shocks to consumer preferences for a category of alcohol do not affect relative preferences for certain products over others within that category. This assumption is realistic in so far as unobservable shocks to the specific utility of a variety can be decomposed additively into an average shock specific to the alcohol category and a deviation from this average shock specific to the product. The average shocks to the alcohol category therefore have no effect on the relative preferences of consumers for the different varieties within the category. They do not affect market shares or the quality index.
 
However, preference shocks specific to the category as a whole may be correlated with the prices of the varieties. We include the average sociodemographic characteristics of the clusters in the regressions to control for differences in time-invariant preferences that may simultaneously affect households' intra- and inter-category trade-offs. We also include period fixed-effects to take into account increased demand from all households at certain times, such as the end of the year holidays or the wine fairs period. The use of aggregate household data mitigates potential issues with regional/local shocks on demand that could be correlated with prices. Finally, our identification strategy assumes that the prices of the varieties of a category of alcohol are not affected by transitory variations in the preferences of certain clusters of consumers for that category of alcohol. This assumption excludes, for example, a transitory increased demand for sparkling wine from wealthy households, which would drive up prices.

Nevertheless, as we cannot rule out the possibility that omitted demand-side factors be correlated with prices, we propose various robustness checks and sensitivity analyses in the next section. We first estimate the model with period effects interacted with income categories and with alcohol habit categories. We also replace the quality-adjusted price indices by more exogenous Laspeyres price indices.\footnote{These Laspeyres price indices are constructed for each category $a$ and each household cluster $c$ as $P^{ac}_{L,t}=\sum_{j \in J^{a}_{t}} S^{c}_{j} p_{jt}$, where $S^{c}_{j}$ is the cluster-average share of product $j$ in purchases. We also tried an instrumental variable strategy by instrumenting our quality-adjusted price indices with these Laspeyres indices (a form of shift-share instrumentation). This approach did not pass a weak instrument test, likely because the sources of price variations (e.g. cost shocks) are correlated across alcohol categories.}


  \subsubsection{Estimation results}

 Table \ref{Tab:ElastQuaids} reports the estimated elasticities averaged over all cluster-period observations with cluster weights in the population. The first line shows that all categories except ciders have very significant income elasticities. These are clearly lower than one for ciders and still and sparkling wines, which means that an increase 1\% in the total alcohol budget increases expenditures in these categories less than proportionately. Expenditures on beers and spirits are more elastic to alcohol expenditure (+1.2\% and +1.4\%). Higher prices are also associated with a higher budget ($\mathbf{E}_{PY}$).
 The price effect is not significant for ciders. We also find that an increase in household income increases total alcohol expenditures, with an elasticity of +0.05\% (not reported here).

 The compensated cross-price elasticities reported off-diagonal in the middle panel of the Table measure consumer preferences in terms of willingness to substitute one alcohol category for another. Consumers show a strong preference for sparkling wines, whose cross-price elasticities are never significant. They have milder preferences for cider and beers. We find nine cross-price effects significant at the level of 5\%, eight of which are symmetrical: beers and
 spirits are substitutes --- consumers are willing to substitute beers for spirits when the price of beers increases, and vice versa -- as well as beers and still wines, aperitifs and spirits, spirits and still wines. Households are willing to substitute ciders for beers if the price of the latter increases, but this is not reciprocal. The absence of negative compensated cross-price elasticities validates our definition of the outside option as "all products in other alcohol categories" in our modeling of demand for quality.
 
 The uncompensated own-price elasticities are reported in the diagonal of the lower panel of Table \ref{Tab:ElastQuaids}. They are important for understanding the potential impacts of price policies, as they summarize both consumer preferences for substitutions and the income effect of price variations. The own-price elasticities are lower than one for all categories, larger for spirits and beers (-0.8) and very low for sparkling wines (-0.2). For spirits and beers, a large part of the large own-price effect is explained by the large budget elasticity of these categories. As shown in the middle panel, when compensated for the income loss produced by the price increase, the own-price elasticity for spirits falls to -0.4\%, a compensated own-price elasticity that is similar to what is observed for the other categories. All 14 significant uncompensated cross-price elasticities are negative. Hence, even though they are Hicksian substitutes, they are gross complements due to income effects. As expected, the latter are rather large for sparkling wines, with significant cross-price elasticities (except with still wines) that are on par with its own-price elasticity. It is noteworthy that an increase in the prices of still and sparkling wines is likely to generate non-negligible reports on spirits, with cross-price elasticities close to $+0.2$. We will see in the next section that these price effects are going to be an important factor driving our results.  

 Table \ref{AppTab:SensitivityQuaids} in Appendix Section \ref{AppResults} compares these results with alternative estimates from a specification with income-period and alcohol habit-period fixed-effects to control for variations in prices that may be related to shocks specific to certain consumer segments, depending on their income and their habitual consumption level. The results are unchanged with own-price uncompensated elasticities that barely vary. We also replace the quality-adjusted price indices by Laspeyres price indices because residual price endogeneity might still be a concern. We then find significantly higher own-price elasticities for wines: $-0.7$ instead of $-0.4$ for still wines, and $-0.7$ instead of $-0.2$ for sparkling wines. Given this slight discrepancy, we will also test the sensitivity of our simulation results to the replacement of quality-adjusted prices by Laspeyres indices.

\begin{landscape}
\begin{center}
    \renewcommand{\arraystretch}{1.3}
	\begin{table}
		\caption{Average income, budget and price elasticities}\label{Tab:ElastQuaids}
		\centering
        \footnotesize
		\begin{tabular}[c]{llcccccc}
			\hline\hline
			&  & {\textbf{Ciders}} & {\textbf{Beers}} & {\textbf{Aperitifs}} & {\textbf{Spirits}} & \makecell{\textbf{Still} \\ \textbf{wines}} & \makecell{\textbf{Sparkling} \\ \textbf{wines}}\\\hline
			\multicolumn{2}{l}{{\small Income elasticity}} &
			\multicolumn{1}{l}{{\small \ 0.503 (0.302)*}} &
			\multicolumn{1}{l}{{\small \ 1.210 (0.112)***}} &
			\multicolumn{1}{l}{{\small \ 1.017 (0.154)***}} &
			\multicolumn{1}{l}{{\small \ 1.434 (0.083)***}} &
			\multicolumn{1}{l}{{\small \ 0.349 (0.126)***}} &
			\multicolumn{1}{l}{{\small \ 0.385 (0.187)**}}\\
			\multicolumn{2}{l}{{\small Budget price elasticity}} &
			\multicolumn{1}{l}{{\small \ 0.007 (0.022)}} &
			\multicolumn{1}{l}{{\small \ 0.098 (0.031)***}} &
			\multicolumn{1}{l}{{\small \ 0.070 (0.021)***}} &
			\multicolumn{1}{l}{{\small \ 0.071 (0.027)***}} &
			\multicolumn{1}{l}{{\small \ 0.059 (0.020)***}} &
			\multicolumn{1}{l}{{\small \ 0.122 (0.011)***}}\\\hline
		
			\multicolumn{8}{l}{{\small Compensated price elasticities}} \\
			& {\small Ciders} & \multicolumn{1}{l}{{\small -0.491 (0.084)***}} &
			\multicolumn{1}{l}{{\small \ 0.217 (0.110)**}} &
			\multicolumn{1}{l}{{\small \ 0.044 (0.083)}} &
			\multicolumn{1}{l}{{\small \ 0.183 (0.104)*}} &
			\multicolumn{1}{l}{{\small \ 0.109 (0.083)}} &
			\multicolumn{1}{l}{{\small -0.063 (0.043)}}\\
			& {\small Beers} & \multicolumn{1}{l}{{\small \ 0.024 (0.031)}} &
			\multicolumn{1}{l}{{\small -0.542 (0.041)***}} &
			\multicolumn{1}{l}{{\small \ 0.039 (0.031)}} &
			\multicolumn{1}{l}{{\small \ 0.349 (0.039)***}} &
			\multicolumn{1}{l}{{\small \ 0.084 (0.031)***}} &
			\multicolumn{1}{l}{{\small \ 0.045 (0.017)}}\\
			& {\small Aperitifs} & \multicolumn{1}{l}{{\small \ 0.012 (0.043)}} &
			\multicolumn{1}{l}{{\small \ 0.100 (0.055)*}} &
			\multicolumn{1}{l}{{\small -0.438 (0.042)***}} &
			\multicolumn{1}{l}{{\small \ 0.242 (0.054)***}} &
			\multicolumn{1}{l}{{\small \ 0.077 (0.042)*}} &
			\multicolumn{1}{l}{{\small \ 0.008 (0.023)}}\\
			& {\small Spirits} & \multicolumn{1}{l}{{\small \ 0.014 (0.023)}} &
			\multicolumn{1}{l}{{\small \ 0.237 (0.030)***}} &
			\multicolumn{1}{l}{{\small \ 0.063 (0.023)***}} &
			\multicolumn{1}{l}{{\small -0.381 (0.030)***}} &
			\multicolumn{1}{l}{{\small \ 0.066 (0.023)***}} &
			\multicolumn{1}{l}{{\small \ 0.000 (0.010)}}\\
			& {\small Still wines} & \multicolumn{1}{l}{{\small \ 0.017 (0.037)}} &
			\multicolumn{1}{l}{{\small \ 0.117 (0.049)**}} &
			\multicolumn{1}{l}{{\small \ 0.041 (0.037)}} &
			\multicolumn{1}{l}{{\small \ 0.136 (0.047)***}} &
			\multicolumn{1}{l}{{\small -0.337 (0.037)***}} &
			\multicolumn{1}{l}{{\small \ 0.026 (0.019)}}\\
			& {\small Sparkling wines} & \multicolumn{1}{l}{{\small -0.013 (0.058)}} &
			\multicolumn{1}{l}{{\small \ 0.084 (0.076)}} &
			\multicolumn{1}{l}{{\small \ 0.005 (0.057)}} &
			\multicolumn{1}{l}{{\small \ 0.001 (0.071)}} &
			\multicolumn{1}{l}{{\small \ 0.034 (0.057)}} &
			\multicolumn{1}{l}{{\small -0.112 (0.032)***}}\\\hline
            \multicolumn{8}{l}{{\small Uncompensated price elasticities }}   \\
			& {\small Ciders} & \multicolumn{1}{l}{{\small -0.504 (0.085)***}} &
			\multicolumn{1}{l}{{\small \ 0.098 (0.133)}} &
			\multicolumn{1}{l}{{\small -0.002 (0.093)}} &
			\multicolumn{1}{l}{{\small \ 0.008 (0.114)}} &
			\multicolumn{1}{l}{{\small \ 0.024 (0.106)}} &
			\multicolumn{1}{l}{{\small -0.127 (0.058)**}}\\
			& {\small Beers} & \multicolumn{1}{l}{{\small -0.008 (0.032)}} &
			\multicolumn{1}{l}{{\small -0.828 (0.047)***}} &
			\multicolumn{1}{l}{{\small -0.072 (0.034)**}} &
			\multicolumn{1}{l}{{\small -0.072 (0.043)*}} &
			\multicolumn{1}{l}{{\small -0.121 (0.040)***}} &
			\multicolumn{1}{l}{{\small -0.109 (0.022)***}}\\
			& {\small Aperitifs} & \multicolumn{1}{l}{{\small -0.014 (0.043)}} &
			\multicolumn{1}{l}{{\small -0.140 (0.068)**}} &
			\multicolumn{1}{l}{{\small -0.531 (0.045)***}} &
			\multicolumn{1}{l}{{\small -0.112 (0.058)*}} &
			\multicolumn{1}{l}{{\small -0.096 (0.054)*}} &
			\multicolumn{1}{l}{{\small -0.122 (0.030)***}}\\
			& {\small Spirits} & \multicolumn{1}{l}{{\small -0.024 (0.024)}} &
			\multicolumn{1}{l}{{\small -0.102 (0.038)***}} &
			\multicolumn{1}{l}{{\small -0.068 (0.025)***}} &
			\multicolumn{1}{l}{{\small -0.881 (0.030)***}} &
			\multicolumn{1}{l}{{\small -0.178 (0.030)***}} &
			\multicolumn{1}{l}{{\small -0.183 (0.015)***}}\\
			& {\small Still wines} & \multicolumn{1}{l}{{\small \ 0.008 (0.038)}} &
			\multicolumn{1}{l}{{\small \ 0.035 (0.058)}} &
			\multicolumn{1}{l}{{\small \ 0.009 (0.040)}} &
			\multicolumn{1}{l}{{\small \ 0.014 (0.049)}} &
			\multicolumn{1}{l}{{\small -0.396 (0.046)***}} &
			\multicolumn{1}{l}{{\small -0.019 (0.025)}}\\
			& {\small Sparkling wines} & \multicolumn{1}{l}{{\small -0.023 (0.059)}} &
			\multicolumn{1}{l}{{\small -0.007 (0.089)}} &
			\multicolumn{1}{l}{{\small -0.030 (0.062)}} &
			\multicolumn{1}{l}{{\small -0.133 (0.078)*}} &
			\multicolumn{1}{l}{{\small -0.031 (0.072)}} &
			\multicolumn{1}{l}{{\small -0.161 (0.033)***}}\\\hline\hline
		\end{tabular}
		\begin{center}
			\begin{minipage}{1.4\textwidth}
				{\scriptsize \textbf{Notes}: average elasticities where each cluster observation is weighted by the proportion of French
					households represented by the cluster. Standard errors in parentheses. ***,**
					and * Significant at the 1\%, 5\% and 10\% level. The first line display Income elasticities, i.e. the \% change in expenditures on a given alcohol category when total alcohol expenditures increase by 1\%. The second line displays the \% change in total alcohol expenditures when the price of a given alcohol category (columns) increase by 1\%.}
			\end{minipage}
		\end{center}
	\end{table}
	\end{center}
\end{landscape}
\FloatBarrier
\newpage

    \section{Minimum Unit Pricing vs. Volumetric Taxation: main results}
        
We now present the main results from our counterfactual simulations. The simulation of the impact of a counterfactual price policy on the alcohol markets (household purchases and equilibrium prices) proceeds in two steps. First, for each market/alcohol category, we use the demand and supply functions to simulate the market equilibrium that would result from the implementation of the policy. Second, we use the new equilibrium prices to construct new price indices, and variations in quantities are predicted using the estimated demand functions for quantities. We provide further details in Appendix Section \ref{AppSimulation}.

We first compare the aggregate impact of the six counterfactual price policy scenarios on household purchases of pure alcohol. We then detail the market mechanisms explaining these results: quality, quantity and competition effects.

\subsection{Minimum Unit Price has the largest impacts on household purchases}

Table \ref{Tab:hholdpurchasespure} presents the simulation results for the impact of the six policies on average household purchase volumes converted to pure alcohol (upper panel).\footnote{The estimates are not very different when we adjust for the number of adult household members, or when we individualize household consumptions.} We report the average impacts, with 95\% confidence intervals that have been computed by Monte-Carlo simulations of the model (1000 replications). We obtain four important results.

Firstly, a tax reform calibrated for the objective of fiscal neutrality (low tax rate) unexpectedly leads to an increase in the purchase and consumption of pure alcohol. This increase is observed for both a uniform and a progressive tax. It reaches +14.0\% (CI95\%[11.6\%;18.7\%]) in the Low uniform tax scenario when the price reactions of manufacturers and retailers are taken into account.

Secondly, tax reforms calibrated to achieve public finance neutrality result in a significant reduction in consumption, up to -10.3\% in the High progressive tax scenario (CI95\%[-12.7\%;-7.8\%]).

Third, the Minimum Unit Price would yield the greatest benefits for public health, with household purchase decreasing by -15.0\% (CI95\%: [-16.4\% ;-13.5\%]). The aggregate impacts are virtually the same whether or not a tax reform is implemented at the same time (Minimum Unit Price + Low progressive tax). It is important to note that the MUP scenario dominates the High progressive tax scenario, with non-overlapping confidence intervals when we account for supply-side reactions (Columns 3 and 4).

Fourth, price reactions from producers and retailers have impacts that are substantial in the Low uniform and Low progressive tax scenarios. Supply-side reactions are predicted to amplify expected consumption increases (first and third lines). In contrast, supply-side reactions amplify the expected benefits of minimum price policies, with, for example, an additional drop of -1.3 percentage points in the MUP scenario (-13.7\% without price reaction vs. -15\% with price reaction). 

In the lower panel of Table \ref{Tab:hholdpurchasespure}, we report the impacts on purchase volumes in L, i.e., not accounting for the variability in alcohol content across products. The MUP and high-tax scenarios negatively impact both purchase volumes and volumes of pure alcohol. Interestingly, the Low uniform and Low progressive tax scenarios have almost no impact on purchase volumes expressed in L. Hence, their positive impacts on purchases of pure alcohol (once we account for the ethanol content of products) essentially reflect substitutions between products, with households choosing varieties with higher alcohol content. This result emphasizes the interest in modeling household purchase behavior along both the quantity and the quality dimension.

As announced in the previous section, to alleviate concerns about the potential endogeneity of the quality-adjusted price indices, we also produce simulation results with quantity elasticities obtained from the estimation of a QUAIDS where prices are measured by Laspeyres indices. Appendix Table \ref{AppTab:Laspeyresresults} shows that this tends to shift downwards all effects without affecting much the ranking of scenario. The impacts of Low uniform and progressive taxes on household purchases expressed in pure alcohol are still positive, but closer to zero. The Minimum Unit Price decreases household purchases by -23.1\% vs. -18.8\% for the High progressive tax, and -26.1\% for a MUP associated with a Low progressive tax. These simulated impacts are larger than those we obtain with quality-adjusted prices. This is essentially due to larger price elasticities, as reported  in  Table \ref{AppTab:SensitivityQuaids}. Since the estimates obtained with quality-adjusted prices are finally more conservative, we focus on these in the rest of this article. 

\begin{center}
    \renewcommand{\arraystretch}{1.5}
    \begin{table}
        \caption{Impacts on household purchases of pure alcohol}
    	\label{Tab:hholdpurchasespure}
    	\centering
    	\medskip
    	\footnotesize
        \begin{tabular}{l c c c c c}
            \hline\hline
             & \multicolumn{2}{c}{\textbf{Without supply-side reactions}} && \multicolumn{2}{c}{\textbf{With supply-side reactions}} \\
            \textbf{Scenario}& Avg. impact& CI95\% && Avg. impact& CI95\% \\
            \hline
            \multicolumn{6}{l}{\textit{Volumes of pure alcohol (Std. Drinks/hhold)}} \\
            Low uniform tax & +10.1\%  & [+7.2,+13.1]   && +14.1\%  & [+10.6,+17.7]  \\
            High uniform tax & -8.2\% & [-11.8,-4.7]  && -5.9\% & [-9.6,-2.2] \\
            Low progressive tax & +4.2\%  & [+2.2,+6.2]   && +6.8\%  & [+4.5,+9.1]  \\
            High progressive tax & -10.3\% & [-12.9,-7.8] && -10.3\% & [-12.8,-7.8] \\
            Minimum Unit Price & -13.7\% & [-14.7,-12.6] && -15.0\% & [-16.4,-13.5] \\
            MUP + Low prog. tax & -12.4\% & [-13.7,-11.1] && -14.7\% & [-16.8,-12.5] \\
            \hline
            \multicolumn{6}{l}{\textit{Purchase volumes (L/hhold)}} \\
            Low uniform tax & +0.3\%  & [-2.7,+3.3]   && +1.7\%  & [-2.0,+5.4]  \\
            High uniform tax & -10.3\% & [-14.2,-6.5]  && -9.3\% & [-13.2,-5.3] \\
            Low progressive tax & +0.2\%  & [-1.8,+2.2]   && +1.3\%  & [-0.9,+3.5]  \\
            High progressive tax & -7.2\% & [-9.9,-4.6] && -6.8\% & [-9.4,-4.3] \\
            Minimum Unit Price & -12.7\% & [-13.7,-11.6] && -13.5\% & [-15.0,-11.9] \\
            MUP + Low prog. tax & -12.2\% & [-13.5,-10.8] && -13.9\% & [-16.1,-11.6] \\
            \hline\hline
        \end{tabular}
        \begin{center}
            \begin{minipage}{0.8\textwidth}
                {\scriptsize \textbf{Notes}: Mean simulated relative impacts on household purchase volumes, in standard drinks (upper panel) and purchase volumes (lower panel); 95\% Confidence Intervals in brackets.}
            \end{minipage}
        \end{center}
    \end{table}
\end{center}
\FloatBarrier

\subsection{Understanding the mechanisms: quantity, quality and competition}

The estimated average impacts result from the combination of quantity, quality, and competition effects. In each of the six markets, price variations induce changes in choices of product variety, generating variations in the average alcohol content of a purchase. Beyond this quality effect within each market, price variations can produce changes in the total quantities purchased on each market, as households make substitutions between alcohol categories.
The upper panel of Table \ref{Tab:hholdpurchasesvol} shows the effect of the different policies on purchase volumes (in L/household/year), after taking into account the supply-side reactions.\footnote{For the sake of readability, we report the average simulated impacts without their confidence intervals in the table.} Examination of the results by alcohol category reveals fairly uneven variations between scenarios, particularly for spirits, where consumption increases sharply in the first three scenarios. In contrast, the consumption of still wines decreases systematically and very sharply in the Minimum Unit Price scenario (-23\%, CI95\%:[-25.2,-20.6]). 

\begin{center}
    \renewcommand{\arraystretch}{1.5}
    \begin{table}
        \caption{Household purchases: quantity effects}
    	\label{Tab:hholdpurchasesvol}
    	\medskip
    	\footnotesize
        \hspace*{-0.5cm}
        \centering
        \begin{tabular}{l c c c c c c}
            \hline\hline
             & \textbf{Ciders} & \textbf{Beers} & \textbf{Aperitifs} & \textbf{Spirits}  & \makecell{\textbf{Still} \\ \textbf{wine}}  & \makecell{\textbf{Sparkling} \\ \textbf{wines}}\\
            \hline
            Initial volumes (L/year) & 1.7 & 23.9 & 4.0 & 6.2 & 37.6 & 3.8 \\
            \hline
            \multicolumn{7}{c}{\textbf{Simulated \% change in average household purchase volumes (L)}}\\
            Low uniform tax & -6.8 & -1.6 & +4.9  & +63.6  & -10.7 & +0.0  \\
            High uniform tax & -7.0 & -11.0 & -4.1 & +22.3  & -17.7 & -7.1 \\
            Low progressive tax & -4.4 & +5.3  & +6.4  & +30.6  & -9.2 & -1.5 \\
            Low progressive tax & -3.1 & +0.4  & +0.1  & -8.0 & -14.0 & -8.5\\
            Minimum Unit Price & +2.7  & -9.5 & -8.2 & -7.3 & -23.0 & -3.6 \\
            MUP + Low prog. tax & -0.2 & -9.0 & -5.8 & -0.8 & -25.9 & -5.0 \\
            \hline
            \multicolumn{7}{c}{\textbf{Simulated \% pts change in average unit prices (\euro/L) of alcohol categories}} \\ \multicolumn{7}{c}{In parenthesis: \% pts change produced by supply-side reactions}\\
            Low uniform tax & +11.4  (+0.4) & +0.1 (+0.4) & -6.8 (-0.2) & -56.3 (-9.1) & +24.8 (-1.1) & +13.5 (+0.3) \\
            High uniform tax & +21.9 (+0.4) & +12.9 (+0.3) & +7.5 (-0.2) & -28.3 (-6.4) & +51 (-2.2) & +28.1 (+0.1)  \\
            Low progressive tax & +5.5 (+0.1) & -8.6 (-0.2) & -9.2 (-0.6) & -33.2 (-6.8) & +21.3 (-1.6) & +11.6  (+0.1) \\
            High progressive tax & +10.4 (+0.2)& -1.8 (-0.1) & +0.6 (-0.4) & +5.1 (+1.0)& +40.8 (-2.0) & +22.4 (+0.2)  \\
            Minimum Unit Price & +0.2 (0.0)& +11.5 (-1.4) & +11.2 (-0.3) & +6.9 (-0.6) & +48.8 (+11.5)& +6.3 (+0.0)  \\
            MUP + Low prog. tax & +5.5 (+0.1)& +9.1 (+1.8)& +6.3 (-0.4) & -2.8 (-1.8) & +62.6 (+18.3) & +13.9 (+0.0)   \\
            \hline
            \multicolumn{7}{c}{\textbf{Simulated shares of alcohol categories (\%) in total household purchase volumes (L)}}\\
            \textit{Initial shares} & \textit{2.2} & \textit{31.0} & \textit{5.2} & \textit{8.0} & \textit{48.6} & \textit{4.9} \\
            Low uniform tax & 2.1 & 31.0 & 5.5 & 13.4 & 44.2 & 5.0  \\
            High uniform tax & 2.3 & 31.3 & 5.6 & 11.1 & 45.5 & 5.1  \\
            Low progressive tax & 2.1 & 33.0 & 5.6 & 10.6 & 44.7 & 4.9  \\
            High progressive tax & 2.3 & 33.9 & 5.7 & 8.1 & 45.7 & 4.9  \\
            Minimum Unit Price & 2.6 & 32.7 & 5.6 & 8.7 & 43.7 & 5.5  \\
            MUP + Low prog. tax  & 2.6 & 32.7 & 5.7 & 9.2 & 41.9 & 5.4  \\
            \hline\hline
         \end{tabular}
         \begin{center}
            \begin{minipage}{\textwidth}
                {\scriptsize \textbf{Notes:} All estimates are based on model simulations and average across households weighted by their sample weights. Simulated changes in average household alcohol purchases in each category/market (in L/household/year). Simulated variations in household average unit prices, where the latter is calculated using the unit prices and the predicted household-level choice probabilities in the observed or counterfactual situations. Simulated shares of alcohol categories in total household purchase volumes (in \% of volumes in L) are average across households (weighted by their sample weights) of the ratio of predicted purchase volumes in each category on total household alcohol purchases. The \% pts change produced by supply-side reactions are calculated as the difference between the \% pts change in average unit prices with and without supply-side reactions.}
            \end{minipage}
        \end{center}
       
    \end{table}
\end{center}

The differences between scenarios can be explained by changes in the market size of the overall alcohol market and by substitutions between alcohol categories, and thus transfers from one market to another. The current tax system is highly favorable to wines (see Table \ref{tab:hholdstats}). Any reform based on the alcohol content of products will therefore tend to reduce the price competitiveness of wines as compared to the other alcohol categories, especially spirits. The middle panel of Table \ref{Tab:hholdpurchasesvol} shows the variations in the average unit prices of purchases in the different alcohol categories after price adjustments by manufacturers and retailers.\footnote{The average unit prices at baseline and in the counterfactual scenario are computed using the product-level unit prices weighted by the predicted choice probabilities. This contrasts to the lower panel of Table \ref{tab:hholdstats}, where we reported expected variations in observed unit values (total expenditures divided by fixed total quantities) without supply side reactions.} The difference induced by supply-side reactions is reported in parentheses (in percentage points). We observe sharp increases in the unit price of still wines in the two Minimum Unit Price scenarios (+48.8\% and +62.6\%), which is explained by very low initial prices in view of their alcohol content. The unit value of spirits decreases significantly in three of the four tax reform scenarios, with price reductions of up to -56.3\% in the Low uniform tax scenario. A progressive tax hardly limits the differential variation in unit prices between still wines and spirits, even though the marginal tax increases with the alcohol content of the product.  The variations in unit prices for ciders, beers, and aperitifs are smaller than those for wines and spirits. 
\bigskip

The supply-side reactions amplify in some cases the immediate impact of policies on prices, as shown by the difference with the price variation without supply-side reactions (reported in parentheses). This is especially the case for spirits in the first three scenarios. In the Low uniform tax scenario, the average unit value is 9.1 percentage points lower than expected in the absence of supply-side reactions. This is caused by heightened competition in the low-price product range, resulting in further price decreases and, therefore, lower unit prices than expected. For wines, the average unit price under the Minimum Unit Price scenario is 11.5 percentage points higher than without supply-side reactions. 

The introduction of a minimum price leads, in fact, to very substantial price increases for low-quality wines, decreasing their competitiveness. The results by wine quality segment reported in Appendix Table \ref{AppTab:wineresults} reveal that, before price adjustment,  the average unit price of still wines initially sold at lower than 3\euro/L increases by 107.8\%, vs. +19.1\% for the wines initially priced between 3\euro/L and 5\euro/L, and +0.8\% for wines priced at 5\euro/L s or more. After supply-side price adjustments, the increases are, respectively, +108. 0\%, +20. 3\%, and +3. 3\%. Hence, strategic price adjustments at the product segment level are little significant on average, even though, as expected, the higher-quality products that face the largest demand increase slightly more their prices. Higher-quality and highly demanded wines adjust only slightly their prices upward (and therefore their margins). They essentially exploit the benefits of enhanced price competitiveness (the price difference with low-quality wines still diminishes) to increase their market share.

These price variations, calculated separately for each market, have direct effects on the size of each market, as well as indirect effects of substitution between markets. The lower panel of Table \ref{Tab:hholdpurchasesvol} illustrates these substitutions, comparing the initial market shares in the overall purchase volumes with the market shares simulated in the different scenarios. In line with the simulated price changes, the market share of still wines falls, while that of spirits rises (except in the High progressive tax scenario). For other categories, there are few variations. This is particularly the case for sparkling wines, for which our estimates show a fairly low direct price elasticity and a rather high budget elasticity (see Table \ref{Tab:ElastQuaids}).
\bigskip

Although quantity effects (direct price effects and substitutions between alcohol categories) explain an essential part of the aggregate results presented in Table \ref{Tab:hholdpurchasespure}, substitutions between product varieties within alcohol categories can also attenuate or amplify the effect of public policies. The upper panel of Table \ref{Tab:hholdpurchasesalccontent} shows that these substitutions are observed for categories that are more heterogeneous in terms of alcohol content: ciders, beers, aperitifs, and spirits. Substitutions vary widely from one scenario to another and specifically show the importance of the tax design. For instance, considering beers, we might think that a progressive tax would reduce more largely the average alcohol content of purchases than a uniform tax, since it would bear proportionally more on strong beers. In contrast, the decrease is greater in the High uniform tax scenario (-11.0\%) than in the High progressive tax scenario (-2.2\%). This is explained by the tax design. In the High uniform tax scenario, the prices of all beers increase, whereas in the High progressive tax scenario their prices slightly decrease with only a small relative advantage for lighter beers. 

Finally, the combination of substitution effects and competition effects explains the changes observed in the contribution of each alcohol category to total household purchase volumes (in g of ethanol), displayed in the lower panel of Table \ref{Tab:hholdpurchasesalccontent}. Unsurprisingly, the share of spirits increases sharply in the first three scenarios due to cumulative quantity and quality effects. Households would purchase larger volumes of stronger varieties of spirits. In the last three scenarios, the most interesting from a public health point of view, there are limited substitution effects from still wines to spirits and, to a lesser extent, to beers, aperitifs and sparkling wines. The large decrease in household purchases of pure alcohol is then not explained by substantial substitutions between alcohol categories, but rather by a drop in total purchase volumes (Table \ref{Tab:hholdpurchasespure}), and a fall in the average alcohol content of varieties for spirits, aperitifs, and beers. In the Minimum Unit Price scenario, the average alcohol content drops by 3.0\% for beers, 1.7\% for aperitifs, and 3.5\% for spirits.

    \begin{center}
        \renewcommand{\arraystretch}{1.5}
        \begin{table}
            \caption{Household purchases: quality effects}
        	\label{Tab:hholdpurchasesalccontent}
        	\medskip
            \footnotesize
            \centering
            \begin{tabular}{l c c c c c c}
                \hline\hline
                & \textbf{Ciders} & \textbf{Beers} & \textbf{Aperitifs} & \textbf{Spirits}  & \makecell{\textbf{Still} \\ \textbf{wine}}  & \makecell{\textbf{Sparkling} \\ \textbf{wines}}\\
                 \hline
                Initial average content (g/L) & 3,6 & 5,2 & 13,0 & 37,5 & 12,0 & 11,7 \\
                \hline
                 \multicolumn{7}{c}{\textbf{Simulated \% change in average alcohol content}} \\
                Low uniform tax & -4.5 & +0.1  & +3.9  & +8.9  & +0.0  & -0.2 \\
                High uniform tax & -9.0 & -11.0 & -5.1 & +5.7  & +0.0  & -0.6 \\
                Low progressive tax & -2.2 & +2.8  & +1.4  & +4.6  & +0.0  & -0.2 \\
                High progressive tax & -4.1 & -2.2 & -7.4 & -5.0 & +0.0  & -0.4 \\
                Minimum Unit Price & -0.1 & -3.0 & -1.7 & -3.5 & +0.2  & -0.1 \\
                Minimum Unit Price + Low prog. tax & -2.2 & -0.7 & -1.2 & -5.5 & +0.1  & -0.2 \\
                \hline
                \multicolumn{7}{c}{\textbf{Simulated share (in \%) of alcohol categories in household purchases (g of ethanol)}} \\
                \textit{Initial shares} & \textit{0.7} & \textit{12.8} & \textit{5.6} & \textit{26.2} & \textit{49.8} & \textit{4.9} \\
                Low uniform tax & 0.5 & 11.1 & 5.3 & 41.0 & 38.9 & 4.3 \\
                High uniform tax & 0.6 & 10.8 & 5.4 & 36.0 & 43.5 & 4.8 \\
                Low progressive tax & 0.6 & 13.0 & 5.6 & 33.6 & 42.3 & 4.5 \\
                High progressive tax & 0.7 & 14.0 & 5.8 & 25.6 & 47.7 & 5.0 \\
                Minimum Unit Price & 0.8 & 13.2 & 5.9 & 27.6 & 45.2 & 5.6 \\
                Minimum Unit Price + Low prog & 0.8 & 13.6 & 6.1 & 28.8 & 43.3 & 5.5 \\
                \hline\hline
            \end{tabular}
            \begin{center}
            \begin{minipage}{\textwidth}
                {\scriptsize \textbf{Notes}: All estimates are based on model simulations and average across households weighted by their sample weights. Simulated changes in average alcohol content are averages across households (weighted by their sample weights) of changes in the sum of product alcohol contents weighted by predicted household-level choice probabilities. Simulated shares of alcohol categories in total household purchase volumes (in \% of volumes in L) are averages across households (weighted by their sample weights) of the ratio of predicted purchase volumes of ethanol in each category on total household ethanol purchases. All formula for household-level predicted changes are derived in Appendix Section \ref{AppSimulation}}
            \end{minipage}
            \end{center}
        \end{table}
    \end{center}

    \section{Heterogeneity of policy impacts}
        
\subsection{Heterogeneity across household category}

The effects of policy reforms can differ in terms of distributional incidence and effectiveness in targeting high-risk households. Table \ref{Tab:hholdheterogeneity} thus presents the simulation results for different household categories: low vs. high income households on the one hand; households with low vs. high risk on the other.\footnote{Kantar WorldPanel classifies households in four income categories, from high to low, using self-reported household monthly income adjusted for the number of consumption units (OECD equivalence scale). High-risk households are defined as those with habitual consumption greater than 2 standard drinks/household adult member/day in 2013. Moderate risk households have habitual consumption levels lower than one standard drink/household member//day in 2013. This classification masks a very significant gap in risk exposure between these two categories: pure alcohol purchases are on average 6.7 times higher for high-risk households compared to low-risk households. } These two dimensions, standard of living and risk level, overlap to some extent, since lower-income households consume more alcohol on average.

The upper panel of Table \ref{Tab:hholdheterogeneity} shows that the impacts on household purchases (in ethanol volumes) do not differ between income categories, except for the Low uniform and Low progressive tax scenarios (but the 95\% confidence intervals overlap).\footnote{+12.5\%, CI95\%[9.1,16.1] for the high-income vs. +16.5\%, CI95\%[13.2,20.0] for the low-income.} Interestingly, the High progressive tax scenario, the Minimum Unit Price and the combined MUP and low progressive tax scenarios yield larger benefits for high-risk households. For instance, Minimum Unit Price achieves a reduction of -14.3\% (CI95\%[-15.7,-12.8]) for the low-risk households as against -17.2\% (CI95\%[-18.8,-15.6]) for the high-risk ones.
 Thus, minimum price policies benefit high-risk households slightly more (+17\% with respect to low-risk households). They would see their consumption fall more sharply in absolute terms, with associated health benefits also greater under the hypothesis of a positive dose-effect relationship. Therefore, the minimum price policy would be the option that best responds to concerns about reducing health inequalities. 

The regressive effects of the different scenarios can be measured by the variations in consumer welfare generated in the short term by price movements and the resulting variations in purchasing power. We are not interested here in the variation in household spending on alcohol, but in the variation in utility resulting both from the implementation of a public policy and from the reaction of households to price changes. This variation in utility can be measured directly, as our structural model enables an exact measurement of utility. It can also be converted into money metrics by computing an equivalent variation representing the income variation that would ex-ante be equivalent to the variation in consumer welfare generated by the counterfactual policy scenario. The equivalent variation integrates together the expected variation in indirect utility and the negative income effects caused by higher prices (see Appendix \ref{App:EVformula}).\footnote{Since we are primarily conducting an ex ante evaluation exercise and we are interested in the ex-ante political acceptability of the reform, we focus on the equivalent variation rather than on the compensating variation.} 

The middle panel of Table \ref{Tab:hholdheterogeneity} reports the impacts on the overall utility households derive from alcohol consumption, while the lower panel of Table \ref{Tab:hholdheterogeneity} shows the expected welfare losses in \euro. Interestingly, the relative utility loss from Minimum Unit Price policies is on average larger for high-income and low-risk households (around -5.0\% for both groups). High-risk households would need to increase their total expenditures by 76.8\euro / year on average to maintain their consumption level and therefore their utility. This explains why their utility loss is lower than that of low-risk households (-1.7\% vs -5.0\%). But this expenditure increase corresponds to an opportunity cost that is measured (ex-ante) by the equivalent variation. The Minimum Unit Price scenario would indeed generate a welfare loss equivalent to around 88.9€/year for high-risk households (CI 95\% [-90.2,-87.0]), as opposed to 115€/year in the High progressive tax scenario (CI95\%[-117.6,-113.3]).
In the three scenarios that are the most attractive from a public health perspective, losses would be rather limited for most households, except those with high consumption. Yet, these variations in consumer welfare do not take into account the beneficial effects on physical and mental health of a reduction in consumption, in the medium or long term.

    \begin{center}
        \renewcommand{\arraystretch}{1.5}
        \begin{table}
            \caption{Household heterogeneity}
            \label{Tab:hholdheterogeneity}
            \medskip
            \centering
            \footnotesize
            \begin{tabular}{l c c c c c} 
                \hline\hline
                 & \multicolumn{2}{c}{\textbf{Income category}} & &\multicolumn{2}{c}{\textbf{Risk category}} \\
                 & \textbf{Low} & \textbf{High} & & \textbf{Low} & \textbf{High} \\
                 \hline
                 \multicolumn{6}{l}{Baseline  } \\
                 Purchase volumes (drinks/household/week) & 14.9 & 13.4 & & 6.3 & 42.5 \\
                 Average expenditure (\euro/year) & 400.0 & 354.2 & & 178.3 & 1081.4 \\
                 \hline
                \multicolumn{6}{c}{\textbf{Impact on household purchases of pure alcohol  (\%)}} \\
                Low uniform tax & +16.5 & +12.5 & & +12.8 & +17.8 \\
                High uniform tax & -5.0 & -6.7 & & -6.5 & -4.2 \\
                Low progressive tax & +8.1 & +5.7 & & +6.8 & +6.8 \\
                High progressive tax & -9.9 & -10.6 & & -9.8 & -12.0 \\
                Minimum Unit Price & -15.1 & -14.8 &  &-14.3 & -17.2 \\
                Minimum Unit Price + Low prog. tax & -14.3 & -15.0 & & -13.8 & -17.6 \\
                \hline
                 \multicolumn{6}{c}{\textbf{Impact on utility derived from alcohol (\%)}}  \\
                Low uniform tax & +7.0 & +5.7 & & +1.5 & +20.5 \\
                High uniform tax & -4.5 & -4.9 & & -8.0 & 5.0 \\
                Low progressive tax & +3.7 & +2.8 & & +1.0 & +9.9 \\
                High progressive tax & -5.9 & -6.7 & & -7.6 & -2.3\\
                Minimum Unit Price & -3.6 & -4.9 & & -5.0 & -1.7 \\
                Minimum Unit Price + Low prog. tax & -3.8 & -6.1 & & -6.2 & -0.6 \\
                \hline
                \multicolumn{6}{c}{\textbf{Equivalent variation in consumer welfare (€/household/year)}} \\
                Low uniform tax & +89.0 & +70.9 & & +37.2 & +248.4 \\
                High uniform tax & -11.7 & -19.0 & & -7.2 & -37.3 \\
                Low progressive tax & +40.4 & +30.4 & & +19.7 & +97.7 \\
                High progressive tax & -35.0 & -42.9 & & -18.7 & -115.4 \\
                Minimum Unit Price & -25.6 & -37.3 & & -15.6 & -88.9 \\
                Minimum Unit Price + Low prog. tax & -22.7 & -39.0 & & -14.5 & -90.9 \\
                \hline\hline
            \end{tabular}
            \begin{center}
                \begin{minipage}{0.8\textwidth}
                    {\scriptsize \textbf{Notes}: Low-risk households: average purchases <1 standard drink/household adult member/day in 2013; High-risk households: average purchases $\ge$2 standard drinks/household adult member/day in 2013.}
                \end{minipage}
            \end{center}
        \end{table}
    \end{center}        
\FloatBarrier

\subsection{Industry profits: losers and winners.}

The different scenarios have heterogeneous consequences for sectors and firms (producers and retailers), depending on their impact on total market size, substitutions between markets, and within markets. Our supply-side models enable us to specify the effects on unit profits in the industry. Although the supply-side model cannot give us insights into profit-sharing between producers and retailers, we constructed our dataset so as to make a distinction between national brands produced by small manufacturers on the one hand (small firms), and national brands produced by large manufacturers or private label brands produced by retailers on the other hand (large firms).

Table \ref{Tab:firmsprofits} details the variations in industry profits by market/alcohol category for all scenarios, and by market and firm size for the three scenarios that yield the largest health benefits. Two results stand out. Firstly, as shown in the upper panel of the table, \textit{with the exception of the Minimum Unit Price scenario}, the profit of the still wine sector falls, with a drop reaching -34.6\% in the Minimum Unit Price scenario with Low progressive tax (CI95\%[-37.9,-31.3] (last line of the upper panel). This drop is due to the implementation of the progressive tax, as the implementation of the Minimum Unit Price alone leaves profits virtually unchanged (+2.3\%, CI95\%[-0.8,+5.5]). Secondly, in the Minimum Unit Price scenario, profits increase slightly in all markets, and for the entire alcohol sector as a whole (+3.7\%, CI95\%[+1.9,+5.5]). At the market level, the increase is significant at the 95\% confidence level for spirits only (+6.9\%, CI95\%[+4.2,+9.7]).

Variations in total profit can be further decomposed into a quantity effect that reflects the size of the market, a quality effect that reflects changes in variety market shares, and a price effect that reflects changes in product-specific margins. When we conduct this decomposition exercise for the Minimum Unit Price scenario (Appendix Table \ref{AppTab:taxprofitsresults}), we find significant quantity effects that are negative for all markets, which is expected since prices increase. The quantity effect is the largest for still wines, representing -30.4\% of initial profits vs -8.5\% for spirits. The quality effects are all negative, except for sparkling wines, and especially large for still wines (-139.7\% of initial profits). This reflects the large substitutions from low-quality wines to higher-quality ones. Finally, the price effect is positive because unit prices mechanically increase while marginal costs remain constant: margins are higher, which offsets the profit loss induced by the contraction of the market. The price component of the change in profit is very large for still wines, representing +172.5\% of the initial profits. Hence, this large price effect offsets the quantity and quality effects, which explain why the Minimum Unit Price has no impact on the overall profit of the wine sector. The same mechanism is at play in the other markets, although the magnitude of the transfer from small to large firms is smaller. The MUP increases the profits of small and medium wine firms (+39\%) while decreasing the profits of large manufacturers and retailers (- 39\%). These results support the MUP as a targeted strategy to curb harmful consumption while benefiting small and medium wine producers.

Finally, Appendix Table \ref{AppTab:taxprofitsresults} also shows that the MUP maintains stable tax revenues. VAT revenues increase with prices (+7\%). This offsets the loss of fiscal revenues produced by the contraction of the market (-10.3\%) and, finally, tax revenues are estimated to drop by 2.3\% only.

\begin{center}
        \renewcommand{\arraystretch}{1.5}
        \begin{table}
            \caption{Variations in profits}
        	\label{Tab:firmsprofits}
        	\medskip
            \centering
            \footnotesize
            \begin{tabular}{l  c c c c c c}
                \hline\hline
                 &Ciders & Beers & Aperitifs & Spirits & \makecell{Still \\ Wines} & \makecell{Sparkling \\  Wines} \\
                 \hline
                \multicolumn{7}{c}{Variation in profit by market (\%)} \\
                Low uniform tax  & -6.1 & -1.3 & +5.7 & +22.7 & -16.5 & +2.4 \\
                High uniform tax & -5.9 & -10.6 & -2.7 & +1.9 & -24.9 & -1.3 \\
                Low progressive tax O1 & -4.1 & +5.7 & +6.8 & +6.5 & -14.7 & +0.4 \\
                High progressive tax O2 & -2.7 & +0.5 & +0.9 & -3.1 & -21.3 & -3.7 \\
                Minimum Unit Price & +3.4 & +1.1 & +3.4 & +6.9 & +2.3 & +1.8 \\
                Minimum Unit Price + Low prog. tax & +0.2 & +10.7 & +8.6 & +45.5 & -34.6 & -1.3 \\
                \hline
                \multicolumn{7}{c}{Variation in profits by market and company size (\%)} \\
                High progressive tax   & & & & & & \\
                \makecell[r]{\textit{Small firms} } & -6.3 & -4.8 & -8.3 & +7.9 & -20.7 & -1.1 \\
                \makecell[r]{\textit{Large firms} } & -2.4 & +0.6 & +3.6 & -4.1 & -22.0 & -5.1 \\
                Minimum Unit Price    & & & & & & \\
                \makecell[r]{\textit{Small firms} } & +3.4 & +7.6 & +6.7 & +5.8 & +39.3 & +2.7 \\
                \makecell[r]{\textit{Large firms} } & +3.4 & +1.0 & +2.5 & +7.0 & -38.7 & +1.3 \\
                Minimum Unit Price + Low prog    & & & & & &  \\
               \makecell[r]{\textit{Small firms} } & -1.8 & +24.7 & -7.8 & +50.8 & -21.5 & +0.3 \\
                \makecell[r]{\textit{Large firms} } & +0.4 & +10.3 & +13.4 & +45.1 & -49.1 & -2.1 \\
                \hline\hline
            \end{tabular}
            \begin{center}
            \begin{minipage}{1.2\textwidth}
                {\scriptsize \textbf{Notes}: Mean simulated impacts on household purchases by alcohol category (L/year).}
            \end{minipage}
            \end{center}
        \end{table}
    \end{center}
\FloatBarrier

    \section{Conclusion}
        Our results clearly outline the superiority of a Minimum Unit Price policy over the alternative tax reforms, especially in the French context where the wine sector has always opposed any reform driven by public health motives. The introduction of a minimum price leads to very substantial price increases for low-quality wines, produced mainly by large manufacturers and retailers. In competition with higher-quality wines, their market shares would fall. The size of the market shrinks as a consequence of an income effect. These results can be explained by the increase in selling prices at constant production cost: the implementation of a minimum price increases producers' unit margins. This largely compensates for the drop in purchase volumes. In a sense, the minimum unit price leads to a transfer of surplus from consumers to producers, with the former subsidizing the latter. The introduction of a minimum price has the advantage of affecting only the largest operators in this sector, while the profits of other operators would increase sharply despite the drop in quantities purchased on this market. On the other hand, all operators would again be affected if a progressive tax were added to the minimum price, replacing existing duties: the extra margin mechanically generated by the introduction of the minimum price would then be taken up by taxation. The MUP entails only a small loss for public finances. It will be largely offset by the expected decrease in health care costs, especially since minimum pricing has higher impacts on high-risk households. These conclusions contrast somewhat with previous empirical evidence that supported the advantage of progressive volumetric taxation over minimum pricing \parencite{griffith_price_2022}. Importantly, these evidence were based on analyses of the UK market and did not really focus on the heterogeneity between supply-side operators. That our estimates lead to different conclusions demonstrate the importance of accounting for country-specific consumption patterns, market organization, and public regulations when analyzing and designing price regulation policies.

Like all modeling exercises, our study has some limits. In particular, we were not able to model on-trade consumption in restaurants, bars, night clubs, or consumption away from home (by friends, etc.). Yet, on-trade consumption is generally more expensive than off-trade consumption, and the policy reform would apply to all consumption occasions. This limits concerns regarding substitutions from off-trade to on-trade. We also ignore the production capacity constraints. These will certainly have important consequences for our predictions regarding the wine sector. In the MUP scenario, smaller wine producers will not be able to meet in the short term an increase in demand volumes. As a consequence, they may increase their prices above the predicted price changes of our model. This will even more reduce overall purchases. In the long term, the structure of the market could be affected by shifting to producing higher quality and fewer quantities. Despite these limits, our modeling framework is well suited to the ex ante evaluation of behavioral price policies that are likely to affect consumption through three first-order mechanisms: adjustments in consumer quantity and quality, and strategic price adjustments by firms and retailers.




    \singlespacing
    \printbibliography	

\pagebreak

\appendix

    {\begin{center}
        \Large \textbf{Supplementary Appendix:\\
    Minimum pricing or volumetric taxation? \\ 
    Quantity, quality and competition effects of price regulations in alcohol markets.}
    \end{center}}

    \counterwithin{figure}{section}
    \counterwithin{table}{section}
    \begin{refsection}

    \section{Data: details}\label{AppData}
        
Our models were developed and calibrated on the 2014 Kantar Worldpanel (KWP) household panel data. It is important to know the main features of the data in order to understand some of our modeling choices. This appendix provides additional details on the Kantar WorldPanel household scanner survey, and on how we prepared the data for the modeling work.

\subsection{Data preparation}\label{App:Subsamples}	
	
\subsubsection{Sample selection}
	
The panel is representative of the French population. Households may be followed for several years. In the KWP data, the calendar year is divided into 13 periods of 4 weeks. The reporting activity of each household is evaluated for each of these periods. Households are awarded points added to an account, depending on their activity during the four-week period, and can redeem their points for an award of their choice.\footnote{See the awards on \url{https://www.boutiquepaneliste.com/decouvrir_cadeaux}}

KWP classifies a household as inactive if it has not reported enough purchases compared to a minimum expected level (based on its past reporting behavior and its sociodemographic characteristics). In 2014, 28,375 households were enrolled in the panel, of which 24,177 were considered to actively report their food and beverage purchases over at least one four-week period. 

A period-specific sample weight is calculated for each active household in order to make the sample of active households representative of the French population for the period. KWP uses marginal calibration based on the following household characteristics: age of the household head (4 categories), income class (4 categories), age of the household head (4 categories) interacted with family structure (5 categories), profession of the household head (7 categories), region of residence (8) interacted with type of residence area (4 categories). The minimum time step for analyzing household purchases is therefore a 4-week period, and inactive households must be dropped from the analysis.   
\bigskip

All participating households must report their purchases of barcoded products: they form the GC panel (GC: Gen-Coded = barcoded, N = 24,177 households). Approximately half of them (N = 13,395) are also members of a second panel, called PF (PF: Produits Frais = Fresh products). PF households also have to report their purchases of products without barcodes. Most alcohol beverages sold in stores have a barcode. However, some wines, ciders, and beers produced by small local manufacturers are only present in the PF panel. We therefore focus on the PF panel. 

For the computation of market statistics, we decided to keep only households that are active for at least 3 periods out of 13 in order to avoid introducing too much noise in the estimates (e.g., households being active only in Christmas, when purchases are more likely). This is sample S0 (N = 12,170 households). Within this sample, we only keep households that purchased alcoholic beverages at least once in 2014, because nonpurchasers are unlikely to start purchasing if a health-oriented price regulation is implemented. This is sample S1 (N = 11,463 households) that is used to produce the market and household statistics in Section 2.
\bigskip

 To control consumption habits in the modeling of demand, we use purchases observed in 2013. The mixed multinomial logit models of quality demand are thus estimated in the subsample S2 of households that are in S1, and were also active at least 3 periods out of 13 in PF 2013. The S2 subsample includes 9,114 households.

Finally, in modeling the substitutions in quantities between broad alcohol categories (e.g., wine for beer), we use a standard Quadratic Almost Ideal Demand System model. To minimize troubles with zero consumptions in some categories - most households do not purchase in all alcohol categories -, we aggregate households into clusters that are based, in part, on consumption habits observed in 2013. To maximize the number of households used in the aggregation, we add to S1 those households that are observed in 2014 not in S1, but were active at least 3 periods out of 13 in PF 2013 \textit{AND} had at least one alcohol purchase in 2013. This is our sample S3. Figure \ref{fig:sampleselection} summarizes the sample selection process.

\usetikzlibrary{positioning}
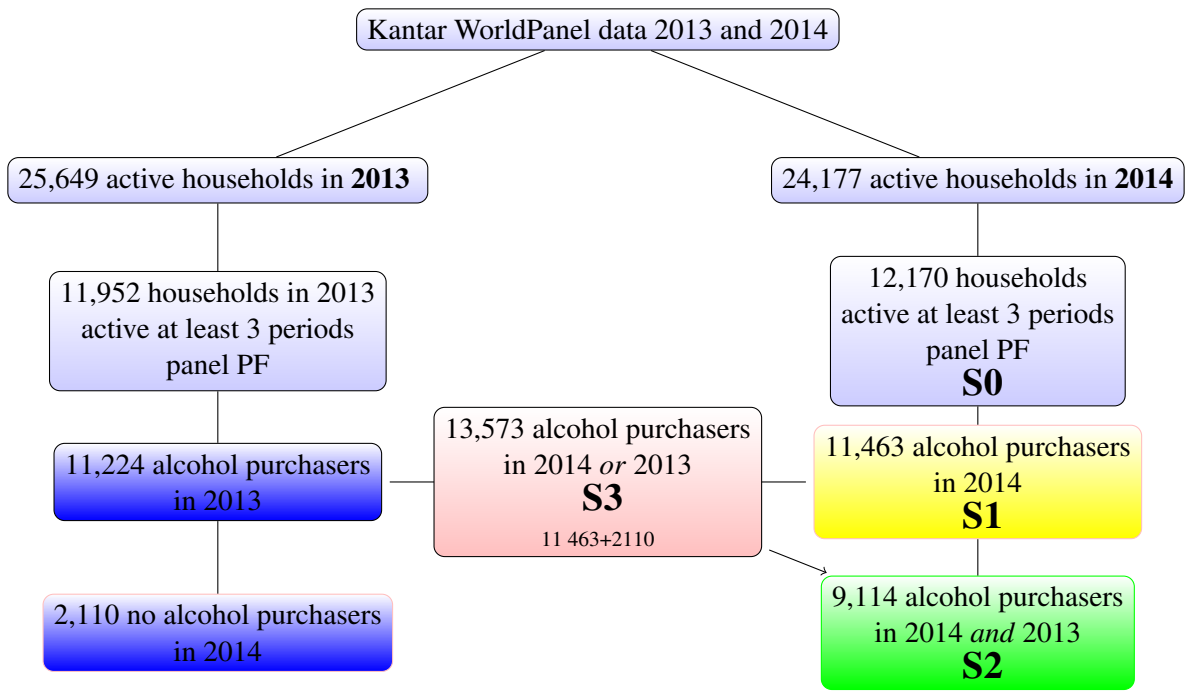
\begin{figure}
	\centering
	\caption{Selection of households and samples construction}\label{fig:sampleselection}
	\begin{tikzpicture}[level 1/.style={sibling distance=10cm},
		level 2/.style={sibling distance=7cm},
		level 3/.style={sibling distance=6cm},
		level 4/.style={sibling distance=4cm},
		level 5/.style={sibling distance=5cm},
		level distance=2cm,
		every node/.style = {shape=rectangle, rounded corners,draw, align=center,font=\small,top color=white, bottom color=blue!20}]
		\node (1) {Kantar WorldPanel data 2013 and 2014}
		child {node {25,649 active households in \textbf{2013} } 
			child {node {11,952 households in 2013 \\ active at least 3 periods \\ panel PF}                         
				child  {node (B) [bottom color=blue] {11,224 alcohol purchasers\\ in 2013       }
					child   {node (G) [draw=pink, bottom color=blue] {2,110 no alcohol purchasers \\ in 2014        }}
				}
			}
		}
		child {node {24,177 active households in \textbf{2014}} 
			child {node {12,170 households\\active at least 3 periods \\ panel PF \\ \begin{large} \textbf{S0}             \end{large}}                            
				child {node (F) [bottom color=yellow, draw=pink]  {11,463 alcohol purchasers \\ in 2014 \\ \begin{large} \textbf{S1}             \end{large}}                            
					child {node [draw=green, bottom color=green] (C) {9,114 alcohol purchasers \\ in 2014 \textit{and} 2013  \\\begin{large} \textbf{S2}               \end{large}}                
					}               
				}
			}
		}
	;
		\draw[-,shorten >=1mm,shorten <=1mm]
		(F) -- node (H) [bottom color=pink]{13,573 alcohol purchasers \\ in 2014 \textit{or} 2013 \\ \begin{large} \textbf{S3} \end{large} \\{\scriptsize 11 463+2110}}   (B)    ;            \draw[->,shorten >=1mm,shorten <=1mm]
		(H) -- (C)                 
		    ;
	\end{tikzpicture}	
\end{figure}

\subsubsection{Sample weights and sample comparison}\label{AppData:weightnorm}

KWP provides sample weights for all household-year observations in the permanent PF subpanel, and all household-period observations in the PF panel. Period-specific weights are non-null only for those households that were active during the period. We normalized the household sample weights so that they sum to the same number $H^{\text{S}}=12,000$ at each period. Let $\omega_{h,t}$ be the sample weight for household $h$, period $t$, we normalized it for sample $S$ so that: 

\begin{equation}
\omega_{h,t}^{\text{S}}  = H^{\text{S}} \frac{\omega_{h,t}}{\displaystyle\sum_{h \in S} \omega_{h,t}}
\end{equation}
\bigskip

Table \ref{AppData:S1S2S3S4_Var_HH_indiv} compares the average characteristics of households and adult household members in the three samples. The three samples are very similar.

\begin{table}[htbp]
	\centering
	\caption{Households and individuals characteristics \label{AppData:S1S2S3S4_Var_HH_indiv}}
	\begin{tabular}{l*{3}{c}}
		\hline\hline \textbf{Sample:}
		&          \textbf{S1}&         \textbf{ S2}&          \textbf{S3}\\
		\hline
		\textit{Nb~of~households}&       11,463&        9,114&       13,573\\
		\textit{Nb~of~adults}&       22,728&       18,197&       26,768\\
		\textit{Nb~of~individuals}&       30,077&       23,456&       35,707\\\hline
		\multicolumn{4}{c}{\textbf{Households~characteristics (\%)}} \\
		\textit{Not~in~the~2013~panel} &        19.2&           0.0&        16.9\\
		\textit{City~size}&           .&           .&           .\\
		More~20000  &        57.6&        57.8&        57.9\\
		Less~20000  &        42.4&        42.2&        42.1\\
		\textit{Nb~of~children}&           .&           .&           .\\
		0           &        72.3&        74.8&        71.2\\
		1           &        13.0&        11.7&        13.5\\
		2           &        10.9&        10.1&        11.4\\
		{[}3+{[}        &         3.8&         3.4&         3.9\\
		\textit{Income categories}&           .&           .&           .\\
		High &        15.7&        15.8&        15.7\\
		High-middle &        29.7&        30.0&        29.2\\
		Low-middle &        40.2&        40.8&        40.2\\
		Low &        14.4&        13.4&        14.9\\
		\textit{Standard~drinks/day/household member in 2013}&           .&           .&           .\\
	
		{[}0,1{]}       &        66.5&        66.5&        68.9\\
		{]}1,2{]}       &        16.5&        16.5&        15.7\\
		{]}2,$+\inf${[}        &          17.0&          17.0&        15.4\\
		\hline\hline
	\end{tabular}
	\begin{center}
		\begin{minipage}{0.7\textwidth}
			{\footnotesize \textbf{Notes}: Percent for \textit{Standard drinks/day/household adult members} are calculated on households that are present in 2013.
				All statistics are produced using the households sample-specific weights (average of all period weights over 2013-2014) - see \ref{AppData:weightnorm} for their computation. The four income categories are produced by Kantar WorldPanel from self-reported household monthly income adjusted for the number of consumption units (OECD equivalence scale). }
		\end{minipage}
	\end{center}
\end{table}

\subsection{Definition of product nomenclature}\label{AppData:Nomenclature}

In the markets for beers, ciders, and sparkling wines, a product $j$ was defined as an alcoholic beverage from a given quality segment, labeled with a given brand, produced by a given manufacturer, sold by a given retailer. For aperitifs and spirits, a product was defined according to the same characteristics plus the binary characteristic "alcohol-free" (7 brands for aperitifs, 3 brands for spirits). For still wines, a product $j$ was defined using only the subcategory, the manufacturer, and the retailer.

We now detail the construction of these characteristics. 
    \subsubsection{Quality segments}
    
    Each alcohol category is divided into quality segments making sub-categories:
    \begin{itemize}
    	\item Beers: Alcohol-free, Bock/Premium, Special, 
    	\item Ciders: Sweet, Raw,
    	\item Sparkling Wines: Champagne, other sparkling wines,
    	\item Still Wines: {Vin de table, Vin de pays, Vin d'appellation (with a designation of origin)},
    	\item Spirits: Rum, Whisky, Aniseed, Liquors, Other,
    	\item Aperitifs: Cocktails/Punches, Liquor wines, Natural sweet wines, Amer/Gentiane/Vermouth, Other.
    \end{itemize}
    
Vin de Table is France's lowest level of wine classification. They can result from the blending of wines from different geographical origins, made with very different grape varieties, produced with unregulated yields, and often made from imported grapes (often from Spain). They do not require a vintage date. Vin de Pays is a classification one level above Vin de Table. Grapes come from a well-defined geographical area and are produced in limited yields. Vin de Pays has recently been relabeled as IGP wines (IGP: Protected Geographic Indication). Vins d'Appellation is the highest classification. It corresponds to a name geographical areas, and producers are submitted to strict constraints regarding the yields, the use of grape varieties, how the grapes are gown and harvested and how the wine is made.

    \subsubsection{Brands, manufacturers and retailers}

     In 2014, in each market (except wines), some manufacturers had (and still have) large market shares and offered several brands. For modeling and inference, we decided to define an aggregate of small manufacturers that sold brands with tiny market shares and, therefore, a very small number of purchases in our 2014 data: in each market model, a manufacturer "Other" will sell a unique brand "Other". Larger manufacturers will also have an aggregate brand "Other". Table \ref{AppTab:manufacturers} lists the main manufacturers.

    We decided to define seven distribution channels: Leclerc/Galec, Intermarché, Auchan, Carrefour, Casino/EMC Distribution, an aggregate of hard discounters (11.56\%), and an aggregate of other shopping places (16.05\%). The first five are the main and largest retailing companies, representing 80$\%$ of food and drink sales in France. The aggregate of other shopping places is a heterogeneous collection of independent shops, cellars, direct-from-producers distribution channels, stores of minor retail companies.  
    
    In our data, about 95\% of alcohol purchases were made in 2014 in large retail outlets that were affiliated with a central purchasing agency. We observe very few purchases at independent cellars or direct purchases at cooperatives or farms. When we focus on wine drinkers, about 18\% of them shopped at least once in a place that was not a large retail outlet. This proportion increases to 25\% when we consider the upper median of wine drinkers (in quantity). For comparison, a 2019 consumer survey (N=1,200 alcohol consumers) found a proportion slightly higher than 30\%.\footnote{See \url{https://www.cavistesprofessionnels.fr/pro/etude-clients-cavistes-images-2019}}. Therefore, the Kantar Worldpanel data slightly cover the upper quality segment of the alcohol markets (luxury segment). We do not see this as a limit, as the economic and epidemiological burden of alcohol is likely to be concentrated in the everyday and premium segments. 

\begin{landscape}
\renewcommand{\arraystretch}{0.8}
\begin{table}[h!]
    \centering
    \caption{Main Manufacturers}\label{AppTab:manufacturers}
   \scriptsize
    \hspace*{-3.75cm}
    \begin{tabular}{c c c c c c c c}
        \hline \hline
         \textbf{Ciders} & \textbf{Beers} & \textbf{Aperitifs} & \textbf{Spirits}  & \makecell{\textbf{Still} \\ \textbf{wine}}  & \makecell{\textbf{Sparkling} \\ \textbf{wines}}  \\
            ETS MAEYAERT & KARLSBERG BRAUEREI KG WEBER & VINADEIS &RHUMS REUNION  & VINADEIS & LES GRANDS CHAIS DE FRANCE\\
            AGRIAL & AB INBEV & GROUPE CASTEL& DAMOISEAU FRERES & LES GRANDS CHAIS DE FRANCE  &GROUPE CASTEL\\
            LES CELLIERS ASSOCIES  & BRASSERIE DE SAINT OMER & UNICOOPS&BACARDI / MARTINI   & GROUPE CASTEL&LVMH\\
            DUJARDIN  & ABBAYE DE WESTMALLE & TERROIRS DISTILLERS&BOUGUET PAU   & BARON PHILIPPE DE ROTHSCHILD& COMPAGNIE FRANCAISE DES GRANDS VINS\\
                    &ABBAYE NOTRE-DAME DE SAINT-REMY   & CANTINE PELLEGRINO&GIFFARD &  UNION DES VIGNERONS DES COTES DU R& CENTRE VINICOLE CHAMPAGNE N FEUILL\\
                    &DIFCOM LA CAVE DES PERES & CROP ENERGIES&FAMILLE CAYARD & CHANTOVENT& CHARLES DE CAZANOVE\\
                   &BAVARIA BREWERY  &  J. GARCIA CARRION &&LISTEL& LANSON-BCC\\
                   &BRASSERIE CASTELAIN  & PICARD VINS ET SPIRITUEUX &LEJAY LAGOUTE &SOCIETE COOPERATIVE AGRICOLE WOLFB & CHAMPAGNE ALAIN THIENOT\\
                    &BRASSEURS DE GAYANT & FELIX SOLIS &LIXIR&UNION DES CAVES COOPERATIVES DE L' &LES CHAMPS RENIERS\\
                    &BRASSERIE LINDEMANS & BACARDI / MARTINI&MARIE BRIZARD   & MARIE BRIZARD &  LES PETITS FILS DE LA VEUVE AMBAL\\
                    &BRASSERIE DE BRABANDERE & BOUGUET PAU&MHD&&ACKERMAN\\
                  &DUCH EXPORT  & FAMILLE CAYARD&PERNOD RICARD   &&JAILLANCE\\
                    &GPE CARLSBERG  &  MARIE BRIZARD&ROTHSCHILD&&LAURENT PERRIER\\
                    &HEINEKEN & MHD &&& CHAMPAGNE JACQUART \\
                    &TSINGTAO  & PERNOD RICARD&&&EPI - COMPAGNIE CHAMPENOISE PIPER\\
                    &BRASSERIE LA CHOULETTE  & ROTHSCHILD&&&BARON PHILIPPE DE ROTHSCHILD \\
                    &BRASSERIE METEOR  & VRANKEN&&&LISTEL\\
                    &BRASSERIE SAINT SYLVESTRE &&&&FREIXENET\\
                    &DUVEL MOORTGAT &&& &SOCIETE COOPERATIVE AGRICOLE WOLFB\\
                    &BRASSERIE DU BOCQ  &&&&CHAMPAGNE ALFRED GRATIEN\\
                    &BRASSERIE DUYCK  &&&&FAMILLE MONMOUSSEAU\\

            &&&&& CHATEAU MONCONTOUR \\
            &&&&& ALLIANCE LOIRE  \\
            &&&&& BLMI \\
            &&&&& UJVR  \\
            &&&&& BARON DE HOEN  \\
            &&&&& CAVE DES PRODUCTEURS DE VINS DE MO \\
            &&&&& BACARDI / MARTINI \\
            &&&&& FAMILLE CAYARD \\
            &&&&& LEJAY LAGOUTE  \\
            &&&&& PERNOD RICARD  \\
            &&&&& VRANKEN  \\

 \hline\hline
    \end{tabular}
    \begin{center}
    		\begin{minipage}{1\textwidth}
    			{\footnotesize \textit{Notes}: This Table reports the names of the main manufacturers by alcohol category.}
    		\end{minipage}
    \end{center}
\end{table}
\end{landscape}

\subsection{Computation of key variables}\label{AppData:Alcohol}		
	
    \subsubsection{Alcohol contents}
    	
    KWP provides information on a number of observed product characteristics, such as brand, manufacturer, alcohol category, and alcohol degree $d$. However, information on alcohol degree was missing for about $9$\% of the references, representing $1$\% of purchase acts. We first imputed the missing values using information available in previous years of the panel for products with the same unique KWP-code, or from all other products with similar observed characteristics (using a weighted average). We also used additional information from producers' and retailers' websites. 
    
    The alcohol degree information was used to compute quantities of ethanol in grams or standard drinks, with the following formula:\footnote{For example, a standard wine bottle ($0.75$L) thus represents 0.8 x 12\% x 750 = 72 grams of ethanol, or 7.2 standard drinks.}
    \begin{equation}
    	e_{j}=0.8 \times d_{j} \times V_{j}*1000
    \end{equation}
    \noindent where $d_{j}$ is the degree and $V_{j}$ is the volume (in L) of one market unit of product $j$.
    
    \subsubsection{Formula for prices and consumptions}\label{AppData:KWPFormula}
    
    KWP provides for each purchase the quantity that was purchased $q$, with the corresponding expenditure $y$. KWP also provides correction factors that must be applied to quantities and expenditures to ensure that the purchase sample is representative of markets for each product category in each four-week period. By construction, the smallest time step is the period: one should not use these data to analyze purchase variations from one week to another. We use information on packaging (number of units per item and unit volume) to calculate the total volume of a given product in a given purchase act, taking into account bulk promotions.
    
    These household-purchase-level quantities and expenditures can be appropriately summed to compute period-specific househol-level quantities, ethanol purchases, and expenditures by alcohol category $a$, $Q_{t}^{ah}$, $E_{t}^{ah}$ and $Y_{t}^{ah}$, or by products $j$ $q_{jt}^{h}$, $e_{jt}^{h}$ and $y_{jt}^{h}$. Dividing expenditures by quantities at the product level yields unit prices.
    \bigskip	
    
    The price of a product $j$ in alcohol category $a$ at period $t$ was computed according to the following formula:
    \begin{equation}
    	p^{a}_{jt}=\frac{\sum_{h \in S1}\omega_{h,t}^{S1}y_{jt}^{h}}{\sum_{h \in S1}\omega_{h,t}^{S1}q_{jt}^{h}}
    \end{equation}
    \noindent where $\omega_{h,t}^{S1}$ is the sample weight for household $h$, $y_{jt}^{h}$ is the household total expenditure in period $t$ for product $j$ and $q_{jt}^{h}$ the corresponding purchased quantity. Note that the prices were constructed using sample S1, in order to use a maximum number of transaction prices.
    
    KWP also provides various formulas for computing averages of household consumption. 
    As households were not active in all 4-week periods, we had to correct household-level averages by the number of active periods $N_{h}^{ACT,S}=\sum_{t}\mathbf{1}\left\{ \omega_{h,t}^{S} \ne 0\right\}$. The yearly ethanol consumption at the household level is then given by:
    \begin{equation}
        \overline{E}^{h,S}=\sum_{t}\sum_{a}E_{t}^{ah}\frac{13}{N_{h}^{ACT,S}}
    \end{equation}
    This formula was used, inter alia, to calculate the average number of standard drinks per day per adult household member in 2013 (after division by $365 \times 10 \times$ the number of adult household members; see the statistics in Table \ref{AppData:S1S2S3S4_Var_HH_indiv}). 
    
    The average yearly quantities of ethanol were calculated at the population level by applying the following formula:
    \begin{equation}
    	\overline{E}^{S}=\frac{\sum_{h \in S}\left(\sum_{t}\left(\omega_{h,t}^{S}\sum_{a}E_{t}^{ah}\right)*\frac{13}{N_{h}^{ACT,S}}\right)}{H^{S}}
    \end{equation}

\subsection{Comparison with the French Family Budget Survey 2017}\label{AppData:CompareINSEE}
    
    Since the 2000s, economic studies analyzing markets for fast-moving consumer goods and evaluating policies aimed at regulating the consumption of such goods have mainly relied on scanner data. The Kantar WorldPanel (KWP) homescan data have three advantages over the \textit{Budgets de famille} (Family Budget Surveys, FBS) for the analysis of alcohol price regulations. First, they provide information on quantity, quality, and expenditure, while information on quantities is only available for 36\% of purchases in the 2017 FBS. Second, the alcohol categories of the FBS nomenclature are too aggregated to allow a precise study of the potential fiscal impact of reform scenarios targeting the alcohol content of beverages. Third, these panel scanner data follow purchases by the same households throughout the year, which limits the observation of zero consumption due to infrequent purchases \parencite{dubois_use_2022}. Fourth, they allow for very precise measurements of prices \parencite{ruhm_what_2012}. 
    
    Purchase scanner data are also less likely to be affected by bias due to under-reporting of alcohol quantities than are health data. As the survey does not specifically focus on the risks posed by alcohol, it does not make salient the stigma associated with excessive drinking. However, reporting requires more effort from respondents, raising questions about the quality of the data in terms of their representativeness and coverage of the population. 
    
     All of our analyzes use the sociodemographic sample weights provided by the KWP. These weights are determined using a margin calibration procedure that takes into account the socioprofessional and age categories of the reference person, the number of persons in the household combined with the age category of the reference person, the region of residence, and the standard of living of the household. The actual representativeness of the household panel and the quality of the collected scanner data can be questioned, especially compared to the FBS. \textcite{lecocq_alcohol_2023} compare the distribution of the sampling characteristics of households in the Kantar panel with that of households included in the 2017 FBS. They find that the KWP sample under-represents households where the reference person is aged between 50 and 64, as well as managers, intermediate occupations, and white-collar workers, and over-represents blue-collar workers and pensioners. Some of the differences between the two data sources can be explained by differences in the way the samples were built-up \parencite{zhen_understanding_2009}. Young, affluent and dual-income households are less represented in the scanner data because the survey requires a degree of diligence. Conversely, working-class retired households are over-represented, possibly because they have more free time and because active participation in the survey is rewarded with points that can be converted into vouchers.
    
     \textcite{lecocq_alcohol_2023} also compare the aggregated expenditure volumes for the categories of alcoholic beverages in the Classification of Individual Consumption by Purpose (COICOP) used in the FBS surveys. They show that the structure of expenditure observed in the 2014 KWP data is very similar to that calculated from the 2017 FBS data, with, for example, a total expenditure volume of 10.38 billion according to the former vs. 11.37 billion in the latter, with the difference attributed to higher expenditure in unit value among higher income households. Table \ref{AppTable:INSEE} reproduces their results. Finally, we should note a limitation common to both surveys. They do not allow for a precise identification of alcohol purchases for consumption away from home. This information is not available in the KWP data we have. It is included in the group "meals outside the home", which aggregates food and drinks, in the 2017 FBS.
\bigskip

\begin{table}[h!]
    \captionsetup{justification=centering}
    \caption{Aggregate expenditure, 2014 Kantar WorldPanel vs. 2017 Family Budget Survey }\label{AppTable:INSEE}
    \footnotesize
    \begin{tabular}{|c|c|c|c|c|}
    \hline  &
     \multicolumn{2}{|c|}{\textbf{Expenditure (€bn)}} & \multicolumn{2}{|c|}{\textbf{Budget share of alcohol expenditure (\%)}} \\ \hline
    \textbf{Survey} & \textbf{KWP 2014} & \textbf{FBS 2017}& \textbf{KWP 2014} & \textbf{FBS 2017}\\
    Spirits and liqueurs & 3.03 & 2.93 & 29.23 & 25.77 \\ \hline
    Wines and ciders & 4.04 & 5.08 & 38.95 & 44.68 \\ \hline
    Other wine-based aperitifs, champagne & 1.81 & 1.54 & 17.46 & 13.54 \\
    and other sparkling wines & & & & \\ \hline
    Beer and beer-based drinks & 1.49 & 1.82 & 14.37 & 16.01 \\ \hline
    Total alcoholic beverages & 10.38 & 11.37 & 100.00 & 100 \\ \hline
    \end{tabular}
    \begin{center}
    		\begin{minipage}{1\textwidth}
    			{\footnotesize \textbf{Notes}: Market size statistics by value calculated by applying the household sampling weights provided by each survey; for comparison, we have used the COICOP nomenclature from FBS 2017.\\
            \textit{Sources and coverage:} 2017 Family Budget Survey , INSEE; 2014 Kantar WorldPanel; full samples.}
    		\end{minipage}
    \end{center}
\end{table}
\FloatBarrier
\newpage

\subsection{Current tax system}\label{AppData:CurrentTax}

    Table \ref{Tab:CurrentTax} displays the excise taxes that were applied to alcohol products in 2014 and 2020 in France. It shows two important points. First, taxes on wines are lower than those on other alcohol, while spirits are heavily taxed. Second, in addition to wines, there are tax exemptions for products that are more likely to be produced in France or by French-owned companies, such as ciders, beers from small breweries, and spirits from overseas territories. In the study, we provide more precise statistics showing that excise taxes are partially decorrelated from the alcohol content of the products.
    
    \begin{table}[!h]
    	\centering
    	\caption{Excise taxes on alcohol products in 2024 and 2014}\label{Tab:CurrentTax}
    	\begin{tabular}{lllll}
    		\hline\hline
    		&      &    & 2024    & 2014    \\\hline
    		\multicolumn{3}{l}{Excise taxes}    &         &         \\
    		& \multicolumn{2}{l}{Droits de circulation}    &         &         \\
    		&      & Still wines (\euro/hl)      & 4.05    & 3.72    \\
    		&      & Sparkling wines (\euro/hl)       & 10.02    & 9.23    \\
    		&      & Ciders and alike (poirés, hydromel) (\euro/hl)    & 1.41    & 1.31    \\
    		&      &       &         &         \\\hline
    		& \multicolumn{2}{l}{Droits spécifiques}      &         &         \\
    		&      & Beer  $\le$ 2.8\% vol (\euro/°/hl)         & 3.98    & 3.66    \\
    		&      & Beer \textgreater 2.8\% vol + small brewery (\euro/°/hl)     & 3.98    & 3.66    \\
    		&      & Beer \textgreater 2.8\% vol + large brewery (\euro/°/hl)     & 7.96    & 7.33    \\\hline
    		&      &        &         &         \\
    		& \multicolumn{2}{l}{Droits de consommation}     &         &         \\
    		&      & Rum from DOM (\euro/hlap)      & 933.78  & 859.79  \\
    		
    		&      & Other alcohol, e.g., spirits (\euro/hlap)     & 1866.52 & 1718.61 \\
    		&      & Natural sweet / liquor wines (\euro/hl)   & 50.6   & 46.59   \\
    		&      & Other intermediary products (\euro/hl)    & 202.39  & 186.36  \\\hline
    		&      &     &         &         \\
    		\multicolumn{4}{l}{Social security tax (alcohol \textgreater 18\% vol or 0.5\% for beers)} &         \\
            &      & Vin doux naturels / de liqueur  (\euro/hl)   & 20.26   & 18.64   \\
    		&      & Other intermediary products (\euro/hl)   & 50.6   & 46.59   \\
    		&      & Beers, small brewery (\euro/hl)   & 50.6    & 1.47    \\
    		&      & Beers, large brewery (\euro/hl)   & 50.6   & 2.93 \\ 
            &      & Other alcohol  (\euro/hlap)   & 599.31   & 551.82   \\
                 \hline\hline 
    	\end{tabular}
    	\begin{center}
    		\begin{minipage}{0.8\textwidth}
    			{\footnotesize \textbf{Notes}: DOM: overseas territories ; hl = hectoliter; hlap = hectoliter of pure alcohol.}
    		\end{minipage}
    	\end{center}
    \end{table}
    \FloatBarrier
    \newpage
   
\subsection{Additional descriptive statistics }

\begin{table}[!h]
    \centering
    \caption{Purchase volumes (L) by household type and alcohol category}\label{Tab:DescStatHholdLiters}
    \footnotesize
    \begin{tabular}{l l l l l l l l l l l}
        \hline
          & Total &   & \multicolumn{4}{l}{Income category} &   & \multicolumn{3}{l}{Risk category} \\
          &   &   & High & Mid-high & Mid-low & Low &   & $\le 1$ & $]1,2]$ & $>2$ \\
        \hline\hline
        Yearly purchase volumes (L) & 76.19 &   & 62.51 & 72.95 & 82.66 & 80.29 &   & 29.54 & 101.63 & 234.17 \\
        \hline
        Shares (\%) &   &   &   &   &   &   &   &   &   &   \\
        Ciders & 2.22 &   & 2.54 & 2.15 & 2.20 & 2.13 &   & 4.77 & 2.59 & 0.84 \\
        Beers & 31.57 &   & 25.58 & 30.97 & 32.60 & 35.05 &   & 44.04 & 33.63 & 24.64 \\
        Aperitifs & 5.13 &   & 4.64 & 5.17 & 4.92 & 6.15 &   & 7.24 & 5.66 & 3.89 \\
        Spirits & 7.81 &   & 8.06 & 7.39 & 7.79 & 8.53 &   & 6.74 & 7.46 & 8.48 \\
        Still wines & 48.29 &   & 53.51 & 49.28 & 47.44 & 44.00 &   & 31.38 & 44.62 & 57.98 \\
        Sparkling wines & 4.98 &   & 5.67 & 5.04 & 5.05 & 4.14 &   & 5.83 & 6.04 & 4.17 \\
        \hline\hline
    \end{tabular}
    \begin{center}
			\begin{minipage}{\textwidth}
				{\footnotesize \textbf{Notes}: N=11,463 households (Sample S1). Statistics computed from purchase data using household and purchase sample weights. The four income categories are defined by KWP, based on household self-reported income and the number of OECD consumption units. The three risk categories are defined based on the average number of standard drinks per week per adult household member purchased by the household in 2013.}
			\end{minipage}
	\end{center}
\end{table}

\begin{table}[!h]
    \centering
    \caption{Purchase volumes (standard drinks) by household type and alcohol category}\label{Tab:DescStatHholdDrinks}
    \footnotesize
    \begin{tabular}{l l l l l l l l l l l}
    \hline\hline
          & Total &   & \multicolumn{4}{l}{Income category} &   & \multicolumn{3}{l}{Risk category} \\
          &   &   & High & Mid-high & Mid-low & Low &   & $\le$ 1 & $]1,2]$ & $>2$ \\
    \hline
    Weekly standard drinks & 13.6 &   & 11.7 & 13.0 & 14.7 & 14.1 &   & 4.5 & 17.8 & 45.4 \\
    \hline
    Share (\%) &   &   &   &   &   &   &   &   &   &   \\
    Ciders & 0.69 &   & 0.82 & 0.67 & 0.67 & 0.65 &   & 1.67 & 0.84 & 0.26 \\
    Beers & 13.17 &   & 11.17 & 13.10 & 13.53 & 14.15 &   & 20.93 & 14.50 & 9.71 \\
    Aperitifs & 5.51 &   & 4.75 & 5.70 & 5.51 & 5.85 &   & 8.30 & 6.25 & 4.17 \\
    Spirits & 25.72 &   & 25.09 & 24.20 & 25.93 & 28.90 &   & 24.06 & 24.99 & 26.61 \\
    Still wines & 49.87 &   & 52.74 & 51.17 & 49.23 & 46.30 &   & 38.18 & 47.19 & 55.30 \\
    Sparkling wines & 5.04 &   & 5.43 & 5.16 & 5.13 & 4.15 &   & 6.86 & 6.23 & 3.95 \\
    \hline\hline
    \end{tabular}
    \begin{center}
			\begin{minipage}{\textwidth}
				{\footnotesize \textbf{Notes}: N=11,463 households (Sample S1). Statistics computed from purchase data using household and purchase sample weights. Weekly standard drinks not adjusted for household size. The four income categories are defined by KWP, based on household self-reported income and the number of OECD consumption units. The three risk categories are defined based on the average number of standard drinks per week per adult household member purchased by the household in 2013.}
			\end{minipage}
	\end{center}
\end{table}
\FloatBarrier

\begin{landscape}
	
	\begin{table}[!htbp]
		\caption{Within-category market shares, prices and alcohol degree by sub-category}\label{Tab:DescStatClass}
		\centering
		\scriptsize
		\hspace*{-1.1cm}\begin{tabular}{l| l l l l l l}
			\hline \hline
			& \multirow{2}{*}{\quad \textbf{Beers}}  & \multirow{2}{*}{\quad \textbf{Ciders}} & \multirow{2}{*}{\quad \textbf{Sparkling wines}} & \multirow{2}{*}{\quad \textbf{Still wines}}  & \multirow{2}{*}{\quad \textbf{Aperitifs}} & \multirow{2}{*}{\quad \textbf{Spirits}}
			\\
			& & &  & & & \\
			\hline
			Market Shares (\%) & & & & & & \\
			\rowcolor{gray!10}&  Alcohol-free    &  Sweet    &  Champagnes    &  de Table    &  Cocktails and Punch    & Rum    \\
			& \quad 6.08 (2.35)  & \quad 36.51 (2.76) & \quad 27.94 (6.13) & \quad 18.93 (1.58) & \quad  18.00 (2.87) & \quad 11.39 (1.06) \\
			\rowcolor{gray!10}&  Bock and premium    &  Raw    &  Other sparkling wines    &  de Pays    & Liqueur wines      & Whisky    \\
			& \quad 44.48 (2.27) \quad  & \quad 63.49 (2.76) & \quad 72.06 (6.13) & 
			
			\quad 23.22 (2.96) & \quad 19.42 (2.36) & \quad 40.60 (2.16) \\
			\rowcolor{gray!10}&  Special    &       &      &  d'Appellation    &  Natural sweet wines     & Aniseed    \\
			& 
			\quad 48.68 (4.05) & & & \quad 57.85 (3.78) & \quad 14.76 (1.64) & \quad 26.72 (2.83) \\
			\rowcolor{gray!10}&       &       &      &       &  Amer, Gentiane, Vermouth     & Liquors    \\
			& & & & & \quad 26.18 (4.18) & \quad 12.66 (1.53)\\
			\rowcolor{gray!10}&       &       &      &       &   Other   & Other   \\
			& & & & & \quad 21.63 (5.45) & \quad 8.62 (1.02) \\
			\hdashline
			Prices (\euro/L) & & & & & & \\
			\rowcolor{gray!10}&  Alcohol-free    &  Sweet    &  Champagnes    &  de Table    &  Cocktails and Punch    & Rum    \\
			& \quad 1.37 (0.44) & \quad 2.66 (0.70) & \quad 25.95 (7.82) & \quad 2.52 (0.79) & \quad 5.90 (3.57) & \quad 16.13 (4.03) \\
			\rowcolor{gray!10}&  Bock and premium    &  Raw    &  Other sparkling wines    &  de Pays    &   Liquor wines    & Whisky    \\
			& \quad 2.64 (1.36) & \quad 2.78 (0.75) & \quad 5.99 (2.00) & \quad 3.16 (0.72) & \quad 9.58 (2.05) & \quad 21.80 (7.55) \\
			\rowcolor{gray!10}&  Special    &       &      &  d'Appellation    &    Natural sweet wines   & Aniseed    \\
			& \quad 3.21 (1.02) & & & \quad 5.49 (2.02) & \quad 6.02 (1.88) & \quad 12.74 (6.11)\\
			\rowcolor{gray!10}&       &       &      &       &  Amer, Gentiane, Vermouth     & Liquors    \\
			& & & & & \quad 7.08 (3.32) & \quad 14.89 (6.52) \\
			\rowcolor{gray!10}&       &       &      &       &   Other     & Other   \\
			& & & & & \quad 6.03 (4.02) & \quad 17.70 (4.06)\\
			\hdashline
			Alcohol degree (°) & & & & & & \\
			
			\rowcolor{gray!10}&  Alcohol-free    &  Sweet    &  Champagnes    &  de Table    &  Cocktails and Punch    & Rum    \\
			& \quad 0.71 (0.20) & \quad 2.00 (0.00) & \quad 12.00 (0.01) & \quad 12.00 (0.60) & \quad 8.40 (6.81) & \quad 44.92 (0.72) \\
			\rowcolor{gray!10}&  Bock and premium    &  Raw   &   Other sparkling wines    &  de Pays    &   Liquor wines    & Whisky    \\
			& \quad 4.58 (0.53) & \quad 4.47 (0.18) & \quad 11.38 (2.60) & \quad 12.00 (0.42) & \quad 18.34 (1.18) & \quad 40.00 (0.10) \\
			\rowcolor{gray!10}&  Special    &       &      &  d'Appellation    &    Natural sweet wines    & Aniseed    \\
			& \quad 7.04 (1.07) & & & \quad 12.03 (0.15) & \quad 15.56 (0.35) & \quad 32.44 (19.72)\\
			\rowcolor{gray!10}&       &       &      &       &   Amer, Gentiane, Vermouth    & Liquors    \\
			& & & & & \quad 12.79 (6.38) & \quad 21.21 (7.77) \\
			\rowcolor{gray!10}&       &       &      &       &     Other & Other    \\
 			& & & & & \quad 11.16 (6.33) & \quad 36.02 (5.66)\\
			\hdashline
			
			Tax level (\euro/L) - VAT excluded & & & & & & \\
			
			\rowcolor{gray!10}&  Alcohol-free    & Sweet    &  Champagnes    &  de Table    &  Cocktails and Punch    & Rum    \\
			& \quad 0.03 (0.01) & \quad 0.01 (0.00) & \quad 0.10 (0.12) & \quad 0.04 (0.00) & \quad 1.21 (1.53) & \quad 7.61 (1.77) \\
			\rowcolor{gray!10}&  Bock and premium    &  Raw    &   Other sparkling wines    &  de Pays    &   Liquor wines    & Whisky    \\
			& \quad 0.32 (0.07) & \quad 0.01 (0.00) & \quad 0.09 (0.02) & \quad  0.04 (0.00) & \quad 0.58 (0.07) & \quad 9.08 (0.02) \\
			\rowcolor{gray!10}&  Special    &       &      &  d'Appellation    &   Natural sweet wines    & Aniseed    \\
			& \quad 0.47 (0.11) & & & \quad  0.04 (0.00)& \quad 0.47 (0.00) & \quad 7.38 (4.44)\\
			\rowcolor{gray!10}&       &       &      &       &   Amer, Gentiane, Vermouth    & Liquors    \\
			& & & & & \quad 2.03 (1.07) & \quad 4.25 (2.07) \\
			\rowcolor{gray!10}&       &       &      &       &     Other  & Other   \\
			& & & & & \quad 1.24 (1.48) & \quad 8.11 (1.47)\\
			\hline \hline
		\end{tabular}
		\begin{center}
			\begin{minipage}{1.2\textwidth}
				{\footnotesize \textbf{Notes}: Market statistics computed from household purchase data of sample S2, using household and purchase acts sample weights. Standard deviation which describes the variability of market shares across periods, and the variability of prices, alcohol degrees and taxes across periods and product varieties.}
			\end{minipage}
		\end{center}
	\end{table}
	
\end{landscape}

\FloatBarrier

\subsection{Differences between the three samples}\label{App:sampledifferences}

One may be concerned that we do not calibrate all parts of our model on exactly the same datasets. The Appendix Table \ref{AppData:S1S2S3S4_Var_HH_indiv} shows that the different samples do not differ in terms of demographic characteristics. In Table \ref{AppData:SampleComp2} we also display statistics of household average consumption in 2014 in the sample used for the descriptive statistics (S1) and the two samples used for the model calibration (S2 and S3). The statistics for S2 are slightly higher than those for S1 and S3 because S2 excludes from S1 households that were not observed purchasing alcohol at least once in 2013. These S1 households were missing in S2 either because they were inactive in 2013, included in 2014 in the panel, or were truly abstinent in 2013. In the latter case, they are likely to have lower consumption levels than the average of S1 households. 

We have compared these statistics with other sources. Based on data from the national account statistics, an official report evaluated total ethanol consumption per adult older than 15 years to be 9584 g/year in 2014 (OFDT, 2020). In another official report (France Agrimer, 2019), the yearly average quantity of alcohol purchase is evaluated at 66L per household in 2017. Given our data and samples, it is not surprising that our estimates fall in between: first, we decided to focus on alcohol purchasers, and therefore we average purchases over a smaller population; second the national account data consider consumption for both at-home and away-from-home consumption (and nonresidents as well as residents).

\begin{table}[htbp]
	\caption{Household average alcohol consumption in 2014}\label{AppData:SampleComp2}
    \centering
    \medskip
    \footnotesize
	\begin{tabular}{lccccc}
		\hline\hline
		& Nb~HH  &  Week (L)& Yearly~(L)  & Standard drinks/week& Standard drinks/year\\ \hline 
		\textbf{Per~Household} & . & . & . & . & . \\ 
		S1 \& S3 & 11463 & 1.47& 76.19& 13.57& 709.35\\
		S2 & 9114 & 1.51& 78.29 & 14.24& 740.58\\
	 
		\textbf{Per~capita (>16 yo)} & . & . & . & . & . \\ 
		S1 \& S3 & 11463 & 0.84& 43.45 & 7.95& 413.15\\ 
		S2 & 9114 & 0.88& 45.72 & 8.36& 435.00\\ 
		\hline\hline 
	\end{tabular}
	\begin{center}
		\begin{minipage}{\textwidth}
			{\footnotesize \textbf{Notes}: All statistics are produced using the households sample-specific weights (average of all period weights over 2014) - see Appendix Section \ref{AppData:KWPFormula} for their computation.}
		\end{minipage}
	\end{center}
\end{table} 
\FloatBarrier

\pagebreak

    \FloatBarrier
    \pagebreak

    \section{Policy scenarios: details}\label{AppTaxDesign}
        In the tax reform scenarios, all current alcohol-specific taxes are replaced by a single volumetric excise tax. For a product $j$, the new excise tax per liter of beverage (in euro/L),  $T_{j}^{1}$, is defined as a function of $d_{j}$, the alcohol degree of the product.

Under a uniform tax per alcohol degree, $T_{j}^{1}$ is proportional to $d_{j}$: $T_{j}^{1}=t \times d_{j}$, where the constant $t$ can be varied by the policy-maker.

Under a progressive excise tax, $T_{j}^{1}$ is a convex increasing function of the alcohol degree, as the marginal tax rate increases at some thresholds. We consider a multi-tiered tax scheme with six intervals for the alcohol degree: [0;5[, [5;10[, [10;15[, [15;25[, [25;45[ and [45;100]. We assume that marginal taxation
is twice higher in the second interval than in the first, three times higher
in the third than in the first, etc. Table \ref{Tab:Tax} provides the rule for calculating the new excise tax.

\begin{table}[!h]
	\caption{New Excise Tax $T_{j}^{1}$}\label{Tab:Tax}
	\begin{center}
	\begin{tabular}
		[c]{ccl}
		\hline\hline
		$Uniform$ & $Progressive$ & \\\hline
		\multicolumn{1}{r}{$t\times d_{j}$} & \multicolumn{1}{l}{$\quad t\times d_{j}$} & if$\quad
		d_{j}\in\lbrack0;5[$\\
		\multicolumn{1}{r}{$t\times d_{j}$} & \multicolumn{1}{l}{$\quad t\times (2d_{j}-5)$} & if$\quad
		d_{j}\in\lbrack5;10[$\\
		\multicolumn{1}{r}{$t\times d_{j}$} & \multicolumn{1}{l}{$\quad t\times (3d_{j}-15)$} &
		if$\quad d_{j}\in\lbrack10;15[$\\
		\multicolumn{1}{r}{$t\times d_{j}$} & \multicolumn{1}{l}{$\quad t\times (4d_{j}-30)$} &
		if$\quad d_{j}\in\lbrack15;25[$\\
		\multicolumn{1}{r}{$t\times d_{j}$} & \multicolumn{1}{l}{$\quad t\times (5d_{j}-55)$} &
		if$\quad d_{j}\in\lbrack25;45[$\\
		\multicolumn{1}{r}{$t\times d_{j}$} & \multicolumn{1}{l}{$\quad t\times (6d_{j}-100)$} &
		if$\quad d_{j}\in\lbrack45;100[$ \\\hline\hline
	\end{tabular}
	\end{center}
\end{table}
\bigskip

For these two taxes, we calibrate the value of $t$ to
achieve one or another of the following two objectives under the assumption that there are no reactions on the demand and supply sides of the market. \textbf{This calibration exercise does not rely on our model}: it is a pure accounting calculation that uses the price and purchase data observed in 2014. 
\begin{enumerate}
	\item Fiscal neutrality: all alcohol tax revenues, about 3.463B\euro  (\textit{including}VAT revenues) are left unaltered by the reform.
	\item Neutrality for public finances: alcohol-specific tax revenues (\textit{excluding} VAT revenues) collected under the new tax scheme must cover the costs of alcohol for public finances (health cares and enforcement of the law, minus retirement pensions): 3.918B\euro as evaluated by an official report \cite{kopp_cout_2015}. A factor of 0.58 is applied to social cost to account for the fact that only 58\% of total consumption takes place at home according to the epidemiological data of the Nutrinet-Santé cohort.
\end{enumerate} 

The calibration results in a value of $t$ that is much lower (7.24 cts/degree/L) to achieve fiscal neutrality than to achieve neutrality for public finances (14.57 cts/degree/L). This is due to the large share of VAT in the current fiscal revenues, as ciders, beers and wines face little or no alcohol-specific taxes and represent a large share of purchase volume. Fiscal neutrality is partly achieved by maintaining the large contribution of VAT, while just equalizing the tax rates per grams of ethanol across alcohol categories. In contrast, neutrality for public finances is obtained by imposing a direct relationship between the social costs of alcohol and its specific taxes. The rationale here is that VAT revenues cannot be easily directed to alcohol-related expenditures (health cares, health education, information campaigns, etc.), while revenues from alcohol-specific taxes can.

Finally, we will simulate two scenarios for the minimum unit price policy (MUP). In a first scenario, the current tax system will be maintained, while it will be replaced by the progressive tax with fiscal neutrality in the second scenario. The minimum unit price will be set a 0.5\euro/standard drink, which is similar to the MUP implemented in Scotland. 
 
We end up with the following six scenarios:
\begin{enumerate}
	\item \textbf{Low uniform tax}: a volumetric tax, uniform per alcohol degree, replacing all other
alcohol-specific taxes with $t=0.0724$ (7.24 cts/degree/L): Fiscal neutrality,
	
	\item \textbf{High uniform tax}: a volumetric tax, uniform per alcohol degree, replacing all other
	alcohol-specific taxes with $t=0.1457$ (14.57 cts/degree/L): Neutrality for public finances,
	
	\item \textbf{Low progressive tax}: a volumetric tax, progressively increasing with alcohol degree, replacing all
	other alcohol-specific taxes with $t=0.0368$ (3.68 cts/degree/L): Fiscal neutrality,
	
	\item \textbf{High progressive tax}: a volumetric tax, progressively increasing with alcohol degree, replacing all
	other alcohol-specific taxes with $t=0.0674$ (6.74 cts/degree/L): Neutrality for public finances,
	
	\item \textbf{Minimum Unit Price}: The addition of a Minimum Unit Price set at 0.5\euro /standard drink to the current tax system
	
	\item \textbf{Minimum Unit Price + Low prog. tax}: Minimum Unit Price and a progressive Excise Tax with $x=0.0368$ (3.68 cts/degree/L) replacing all other alcohol-specific taxes: Fiscal neutrality.
\end{enumerate}

Each scenario will be simulated with and without the supply-side responses.
\FloatBarrier
\pagebreak

    \FloatBarrier
    \pagebreak

    \section{Expected utility of a consumption occasion, $u_{to}^{ah}$}\label{AppModel}
          
To identify the expected utility of a consumption occasion within a period, we conflate purchase acts with consumption occasions. An observed purchase act results from the choice of a product within a choice set that is made of all products $j$ available in the alcohol category, $j=1,...,J_t^a$, plus an outside option ($j=0$). Each purchase of a product $j$ observed in the data could be associated with one or more consumption occasions (depending on the number of purchased units) that are given the same utility. Thus, we assume that, within each alcohol category, variations in purchase volumes across purchase acts are not correlated with product quality. Purchase volumes vary randomly across purchase acts within a period, around an average that only depends on the purchase volume decided for the period in the first stage of decision, and the number of purchase acts in the category, which is exogenously determined. This rules out behaviors such as consumers systematically purchasing more or less units of a brand that has a higher consumption utility, as well as within-period optimal inventory management. 

Consider household $h$ decision to purchase a product $j$ in alcohol category $a$ at period $t$. On point-of-purchase, the anticipation of consumption occasions affects utility through the sum of the product-household deterministic effect, $\beta^{ah}Z_{j}^{ah}$, and a product-household-occasion specific shock $\varepsilon_{jto}^{ah}$. The utility derived from the purchase is $V_{jto}^{ah}=-\alpha^{ah}p_{jt}^{a}+\beta^{ah}Z_{j}^{a}+\varepsilon_{jto}^{ah}$, while the utility anticipated from consumption is $v_{jto}^{ah}=\beta^{ah}Z_{j}^{a}+\varepsilon_{0to}^{ah}$ (see equation \ref{Eq:RUM} in Section \ref{Sec:Demand_Model}).
 
 A consumption occasion corresponds to the consumption of one quantity unit. Here, and in the empirical implementation of the model, we assume that the specific shock $\varepsilon_{0to}^{ah}$ remains constant across all units of the same product purchased on the same shopping trip at the same point-of-purchase. This allows us to treat one purchase act as revealing the consumption utility of one consumption occasion.   

We introduce an outside option $j=0$ with utility $V_{0to}^{ah}=V_{0t}^{ah}+\varepsilon_{0to}^{ah}$ representing any other product in another alcohol category. We let $d_{to}^{ah}$ denote the decision of purchasing product $j$ (noted $d_{to}^{ah}=j$) in alcohol category $a$ for consumption occasion $o$. 

We note $u_{to}^{ah}$ the expected utility benefit from any
consumption occasion of product in
category $a$, conditional on the purchase of at least one product in
that category. It is equal (up to any strictly increasing transformation) to:

\begin{align}
u_{to}^{ah}  &  \sim\mathbb{E}\left[  \sum_{j\in J_{t}^{a}}\mathbf{1}\left[
d_{to}^{ah}=j\left\vert \exists l\in J_{t}^{a},V_{lto}^{ah}%
>V_{0to}^{ah}\right.  \right]  v_{jto}^{ah}\right] \nonumber\\
&  =\mathbb{E}\left[  \sum_{k\in J_{t}^{a}}\mathbf{1}\left[  d%
_{to}^{ah}=j\left\vert \exists l\in J_{t}^{a},V_{lto}^{ah}>U_{0to}%
^{ah}\right.  \right]  \left(  V_{jto}^{ah}+\alpha^{ah}p_{jt}^{a}\right)
\right] \nonumber\\
&  =\mathbb{E}\left[  Max_{j\in J_{t}^{a}}\left(  V_{jto}^{ah}\left\vert
\exists l\in J_{t}^{a},V_{lto}^{ah}>V_{0to}^{ah}\right.  \right)  \right]
+\mathbb{E}\left[  \sum_{j\in J_{t}^{a}}\mathbf{1}\left[  d_{to}%
^{ah}=j\left\vert \exists l\in J_{t}^{a},V_{lto}^{ah}>V_{0to}^{ah}\right.
\right]  \alpha^{ah}p_{jt}^{a}\right] \nonumber\\
&  =\ln\left(  \sum_{j\in J_{t}^{a}}exp\left(  V_{jt}^{ah}-V_{0t}^{ah}\right)
\right)  +C+\alpha^{ah}\sum_{j\in J_{t}^{a}}\pi_{jt}^{ah}p_{jt}^{a}%
\end{align}

\noindent where $C$ is a constant associated to the computation of the
first expectation term, and $\pi_{jt}^{ah}$ is the probability of purchasing
variety $j$ conditional on purchasing a product in category $a$:

\begin{equation}
\pi_{jt}^{ah}=\mathbb{E}\left[  \mathbf{1}\left[  d_{to}
^{ah}=j\left\vert \exists l\in J_{t}^{a},V_{lto}^{ah}>V_{0to}^{ah}\right.
\right]  \right]
\end{equation}

\noindent This probability does not depend on the consumption occasion $o$ as the latter affects purchase probabilities through the random error-term $\varepsilon_{0to}^{ah}$ only. The presence of the constant $C$ is related to the fact that the absolute
level of consumption utility cannot be defined. Some normalization is
therefore necessary. Here we use the counterfactual situation where category
$a$ would only be made of undifferentiated products, i.e., a single
homogeneous product. In comparison with this counterfactual situation, we let
$b_{t}^{ah}(\mathbf{p}_{t}^{a})$ be the conditional consumer surplus from
differentiation across products \emph{within} category $a$:
\begin{equation}\label{Eq:quality_index}
b_{t}^{ah}(\mathbf{p}_{t}^{a})=\frac{1}{\alpha^{ah}}\left[  \ln\left(
{\displaystyle\sum\limits_{k\in J_{t}^{a}}}\exp(V_{jt}^{ah}-V_{\star t}
^{ah})\right)  +\alpha^{ah}{\sum\limits_{k\in J_{t}^{a}}}\pi_{jt}^{ah}\left(
p_{jt}^{a}-p_{\star t}^{a}\right)  \right]
\end{equation}
where $V_{\star t}^{ah}$ is the (deterministic) utility of a homogeneous
product in category $a$ when there is no differentiation across products $j$, and $p_{\star t}
^{a}$ is the price of this product. This derivation uses
the fact that if the products were homogeneous in quality, then they would have the same prices
and market shares: ${\textstyle\sum_{j\in J_{t}^{a}}}\pi_{\star t}^{ah}p_{\star t}^{a}=p_{\star t}^{a}=
{\textstyle\sum_{j\in J_{t}^{a}}}\pi_{jt}^{ah}p_{jt}^{a}$

The quality index $b_{t}^{ah}(\mathbf{p}_{t}^{a})$ is a measure of the
consumer welfare gains from having access to differentiated products in market
$a$. This surplus is zero when the products are not differentiated. Using
surplus as a measure of utility is a way to ensure that utility flows are
normalized across alcohol categories. The motivation here is that the utility flows are identified by using the purchase data for each alcohol category and fitting
random coefficient logit models for each market separately. However, these models identify the random utility function up to a scale parameter. Since the models are estimated separately for each category $a$, we must normalize the scale. A natural scale normalization is to take a money-metric measure of utility, by using the estimated marginal utility of expenditure on each market $\alpha^{ah}$.
Additionally, our normalization for quality differentiation naturally implies that the utility of purchasing products in $a$ if they were not differentiated
would simply be equal to one
quantity unit of alcohol from category $a$, as we would have $b_{t}^{ah}(\mathbf{p}_{t}^{a})=0$.

With these two normalizations, the utility of a consumption occasion is $u_{to}^{ah}=(1+b_{t}^{ah}(\mathbf{p}_{t}%
^{a}))$
\FloatBarrier
\pagebreak

    \FloatBarrier
    \pagebreak

    \section{Habit formation and stockpiling effects}\label{AppStockpiling}
        

Household purchases may show some dependence between periods due to (1) short-term habit formation and (2) stockpiling during sale periods (\cite{hendel_measuring_2006}). Here we apply the test procedure implemented in \textcite{griffith_price_2020} to test for these forms of intertemporal dependence. The results suggest that these forms of dynamics are not of primary importance once household preference heterogeneity is taken into account. Our results and conclusions are therefore very
similar to theirs.

We test for the presence of habit formation by running two regressions. In the
first regression, the dependent variable is a dummy variable equal to one if
the household purchases alcohol in a given period; in the second regression,
the dependent variable is the number of standard
drinks that the household purchased conditional on purchasing. We consider periods of either one
week or four weeks. We regress these variables on the number of standard drinks
purchased by the household in each of the past eight periods, and period
dummies. We estimate each regression with and without household fixed-effects.
As shown in Table \ref{Tab:Dependence1}, we obtain a moderate relationship between past and
present behaviors in Ordinary Least Squares regressions with household and time fixed-effects. For example, considering that the period length is a week, purchasing one more standard drink per adult two weeks ago ($t-2$) is associated with (i) an increase in the probability of purchasing alcohol of 0.12 percentage points, for an average of 17.9\% and (ii) an increase of 0.04 standard drinks in conditional purchase volumes for an average of 21.6 standard drinks. The results are of the same order of magnitude when we take a period of four weeks. Finally, it is interesting to note that, in the conditional quantity regressions, the coefficients on lagged quantities drop when we control for fixed-effects. This shows that long-term habit formation is much more important than short-term habit formation. We account for this heterogeneity by conditioning all of our empirical analyses on the household average purchase volumes observed in 2013.

We also assess the existence of a state dependence arising from consumer
stockpiling during sale periods; if short-run price reductions lead to an
increase in alcohol purchases, which are then stored rather than immediately
consumed, then the own-price elasticities of demand would be overestimated
(see \textcite{hendel_measuring_2006}). To test this effect, we follow one of the
suggestions in \textcite{hendel_sales_2006}. We assume that each household has a
constant consumption level equal to the average number of standard drinks
purchased per period. We calculate this level for each household and combine
it with its purchases per period to calculate its inventory at the beginning
of each period. We then regress (i) the probability of buying in a given
period and (ii) the number of standard drinks purchased (conditional on
purchasing a positive amount) on this inventory variable, period effects
(which control for price changes, promotions, advertising, etc.), and household
fixed-effects. \textcite{hendel_sales_2006} argue that if there is stockpiling, then
a high inventory should be associated with a lower probability of purchase or
a smaller quantity purchased (conditional on purchase). Whether the period
is one or four weeks, we find in contrast a very weak positive
relationship between the inventory variable on the one hand and the
probability and quantity of alcohol purchased on the other (see Table \ref{tab:Dependence2}).

\begin{landscape}
    \begin{table}[]
        \centering
        \caption{Dependence of current purchase decisions on past purchases}
        \label{Tab:Dependence1}
        \footnotesize
        \begin{tabular}[c]{llccccccccc}
            \hline\hline
            &  & \multicolumn{4}{c}{1 week} &  & \multicolumn{4}{c}{4 weeks}%
            \\\cline{3-6}\cline{8-11}
            &  & Purchase? & Purchase? & Quantity & Quantity &  & Purchase? & Purchase? &
            Quantity & Quantity\\\hline
            \multicolumn{2}{l}{Past quantity} & \multicolumn{1}{l}{} &
            \multicolumn{1}{l}{} & \multicolumn{1}{l}{} & \multicolumn{1}{l}{} &
            \multicolumn{1}{l}{} & \multicolumn{1}{l}{} & \multicolumn{1}{l}{} &
            \multicolumn{1}{l}{} & \multicolumn{1}{l}{}\\
            & $t-1$ & \multicolumn{1}{l}{\ 0.0019***} & \multicolumn{1}{l}{\ 0.0009***} &
            \multicolumn{1}{l}{\ 0.0736***} & \multicolumn{1}{l}{-0.0382***} &
            \multicolumn{1}{l}{} & \multicolumn{1}{l}{\ 0.0015***} &
            \multicolumn{1}{l}{\ 0.0008***} & \multicolumn{1}{l}{\ 0.2346***} &
            \multicolumn{1}{l}{\ 0.0191}\\
            &  & \multicolumn{1}{l}{(0.0001)} & \multicolumn{1}{l}{(0.0001)} &
            \multicolumn{1}{l}{(0.0075)} & \multicolumn{1}{l}{(0.0071)} &
            \multicolumn{1}{l}{} & \multicolumn{1}{l}{(0.0001)} &
            \multicolumn{1}{l}{(0.0000)} & \multicolumn{1}{l}{(0.0141)} &
            \multicolumn{1}{l}{(0.0139)}\\
            & $t-2$ & \multicolumn{1}{l}{\ 0.0023***} & \multicolumn{1}{l}{\ 0.0012***} &
            \multicolumn{1}{l}{\ 0.1510***} & \multicolumn{1}{l}{\ 0.0363**} &
            \multicolumn{1}{l}{} & \multicolumn{1}{l}{\ 0.0010***} &
            \multicolumn{1}{l}{\ 0.0004***} & \multicolumn{1}{l}{\ 0.1626***} &
            \multicolumn{1}{l}{-0.0100}\\
            &  & \multicolumn{1}{l}{(0.0001)} & \multicolumn{1}{l}{(0.0001)} &
            \multicolumn{1}{l}{(0.0164)} & \multicolumn{1}{l}{(0.0176)} &
            \multicolumn{1}{l}{} & \multicolumn{1}{l}{(0.0001)} &
            \multicolumn{1}{l}{(0.000)} & \multicolumn{1}{l}{(0.0093)} &
            \multicolumn{1}{l}{(0.0122)}\\
            & $t-3$ & \multicolumn{1}{l}{\ 0.0018***} & \multicolumn{1}{l}{\ 0.0008***} &
            \multicolumn{1}{l}{\ 0.1113***} & \multicolumn{1}{l}{\ 0.0088} &
            \multicolumn{1}{l}{} & \multicolumn{1}{l}{\ 0.0007***} &
            \multicolumn{1}{l}{\ 0.0003***} & \multicolumn{1}{l}{\ 0.1187***} &
            \multicolumn{1}{l}{-0.0229*}\\
            &  & \multicolumn{1}{l}{(0.0001)} & \multicolumn{1}{l}{(0.0000)} &
            \multicolumn{1}{l}{(0.0084)} & \multicolumn{1}{l}{(0.0069)} &
            \multicolumn{1}{l}{} & \multicolumn{1}{l}{(0.0000)} &
            \multicolumn{1}{l}{(0.0000)} & \multicolumn{1}{l}{(0.0074)} &
            \multicolumn{1}{l}{(0.0108)}\\
            & $t-4$ & \multicolumn{1}{l}{\ 0.0018***} & \multicolumn{1}{l}{\ 0.0008***} &
            \multicolumn{1}{l}{\ 0.1120***} & \multicolumn{1}{l}{\ 0.0140*} &
            \multicolumn{1}{l}{} & \multicolumn{1}{l}{\ 0.0005***} &
            \multicolumn{1}{l}{\ 0.0002***} & \multicolumn{1}{l}{\ 0.0849***} &
            \multicolumn{1}{l}{-0.0338***}\\
            &  & \multicolumn{1}{l}{(0.0001)} & \multicolumn{1}{l}{(0.0000)} &
            \multicolumn{1}{l}{(0.0092)} & \multicolumn{1}{l}{(0.0085)} &
            \multicolumn{1}{l}{} & \multicolumn{1}{l}{(0.0000)} &
            \multicolumn{1}{l}{(0.0000)} & \multicolumn{1}{l}{(0.0088)} &
            \multicolumn{1}{l}{(0.0103)}\\
            & $t-5$ & \multicolumn{1}{l}{\ 0.0016***} & \multicolumn{1}{l}{\ 0.0006***} &
            \multicolumn{1}{l}{\ 0.1076***} & \multicolumn{1}{l}{\ 0.0143*} &
            \multicolumn{1}{l}{} & \multicolumn{1}{l}{\ 0.0004***} &
            \multicolumn{1}{l}{\ 0.0001**} & \multicolumn{1}{l}{\ 0.0775***} &
            \multicolumn{1}{l}{-0.0375***}\\
            &  & \multicolumn{1}{l}{(0.0001)} & \multicolumn{1}{l}{(0.0000)} &
            \multicolumn{1}{l}{(0.0078)} & \multicolumn{1}{l}{(0.0075)} &
            \multicolumn{1}{l}{} & \multicolumn{1}{l}{(0.0000)} &
            \multicolumn{1}{l}{(0.0000)} & \multicolumn{1}{l}{(0.0076)} &
            \multicolumn{1}{l}{(0.0092)}\\
            & $t-6$ & \multicolumn{1}{l}{\ 0.0015***} & \multicolumn{1}{l}{\ 0.0005***} &
            \multicolumn{1}{l}{\ 0.0959***} & \multicolumn{1}{l}{\ 0.0057} &
            \multicolumn{1}{l}{} & \multicolumn{1}{l}{\ 0.0003***} &
            \multicolumn{1}{l}{\ 0.0000} & \multicolumn{1}{l}{\ 0.0758***} &
            \multicolumn{1}{l}{-0.0367***}\\
            &  & \multicolumn{1}{l}{(0.0001)} & \multicolumn{1}{l}{(0.0000)} &
            \multicolumn{1}{l}{(0.0089)} & \multicolumn{1}{l}{(0.0078)} &
            \multicolumn{1}{l}{} & \multicolumn{1}{l}{(0.0000)} &
            \multicolumn{1}{l}{(0.0000)} & \multicolumn{1}{l}{(0.0099)} &
            \multicolumn{1}{l}{(0.0125)}\\
            & $t-7$ & \multicolumn{1}{l}{\ 0.0015***} & \multicolumn{1}{l}{\ 0.0004***} &
            \multicolumn{1}{l}{\ 0.1024***} & \multicolumn{1}{l}{\ 0.0073} &
            \multicolumn{1}{l}{} & \multicolumn{1}{l}{\ 0.0003***} &
            \multicolumn{1}{l}{\ 0.0000} & \multicolumn{1}{l}{\ 0.0663***} &
            \multicolumn{1}{l}{-0.0462***}\\
            &  & \multicolumn{1}{l}{(0.0001)} & \multicolumn{1}{l}{(0.0000)} &
            \multicolumn{1}{l}{(0.0079)} & \multicolumn{1}{l}{(0.0078)} &
            \multicolumn{1}{l}{} & \multicolumn{1}{l}{(0.0000)} &
            \multicolumn{1}{l}{(0.0000)} & \multicolumn{1}{l}{(0.0078)} &
            \multicolumn{1}{l}{(0.0099)}\\
            & $t-8$ & \multicolumn{1}{l}{\ 0.0015***} & \multicolumn{1}{l}{\ 0.0005***} &
            \multicolumn{1}{l}{\ 0.0929***} & \multicolumn{1}{l}{\ 0.0003} &
            \multicolumn{1}{l}{} & \multicolumn{1}{l}{\ 0.0002***} &
            \multicolumn{1}{l}{-0.0000} & \multicolumn{1}{l}{\ 0.0927***} &
            \multicolumn{1}{l}{-0.0342***}\\
            &  & \multicolumn{1}{l}{(0.0001)} & \multicolumn{1}{l}{(0.0000)} &
            \multicolumn{1}{l}{(0.0070)} & \multicolumn{1}{l}{(0.0072)} &
            \multicolumn{1}{l}{} & \multicolumn{1}{l}{(0.0001)} &
            \multicolumn{1}{l}{(0.0000)} & \multicolumn{1}{l}{(0.0086)} &
            \multicolumn{1}{l}{(0.0104)}\\\hline
            \multicolumn{2}{l}{Mean of dep. var.} & \multicolumn{1}{l}{\ 0.1791} &
            \multicolumn{1}{l}{\ 0.1791} & \multicolumn{1}{l}{21.5831} &
            \multicolumn{1}{l}{21.5831} & \multicolumn{1}{l}{} &
            \multicolumn{1}{l}{\ 0.3929} & \multicolumn{1}{l}{\ 0.3929} &
            \multicolumn{1}{l}{39.3574} & \multicolumn{1}{l}{39.3574}\\
            \multicolumn{2}{l}{Period effects?} & Yes & Yes & Yes & Yes &  & Yes & Yes &
            Yes & Yes\\
            \multicolumn{2}{l}{Household fixed effects?} & No & Yes & No & Yes &  & No &
            Yes & No & Yes\\\hline\hline
        \end{tabular}
        \begin{center}
            \begin{minipage}\textwidth
            \noindent{\footnotesize \textbf{Notes}: The dependent variable in columns "Purchase?"
            is a dummy variable equal to one if the household purchased alcohol in the
            period. The dependent variable in columns "Quantity" is the number of standard
            drinks purchased per adult in that period, conditional on making a non-zero
            purchase. The table shows the estimated coefficients on the number of standard
            drinks purchased per adult in the preceding 8 periods. Standard errors are in
            parentheses. ***, ** and * indicate a significant effect at the 1\%, 5\% and
            10\% levels.}
        \end{minipage}
        \end{center}
   \end{table}   
\end{landscape}

\begin{table}[]
    \caption{Dependence of current purchase decisions on steady-state inventory level}
    \label{tab:Dependence2}
    \footnotesize
    \centering
    \begin{tabular}[c]{lllcll}\hline\hline
        & \multicolumn{2}{c}{1 week} &  & \multicolumn{2}{c}{4 weeks}\\\cline{2-3}%
        \cline{5-6}
        & Purchase? & Quantity & \multicolumn{1}{l}{} & Purchase? & Quantity\\\hline
        Inventory & \ 0.0003*** & \ 0.0393*** & \multicolumn{1}{l}{} & \ 0.0007*** &
        \ 0.1223***\\
        & (0.0000) & (0.0039) & \multicolumn{1}{l}{} & (0.0000) & (0.0094)\\\hline
        Mean of dep. var. & \ 0.1791 & 21.5831 & \multicolumn{1}{l}{} & \ 0.3929 &
        39.3574\\
        Period effects? & \multicolumn{1}{c}{Yes} & \multicolumn{1}{c}{Yes} &  &
        \multicolumn{1}{c}{Yes} & \multicolumn{1}{c}{Yes}\\
        Household fixed effects? & \multicolumn{1}{c}{Yes} & \multicolumn{1}{c}{Yes} &
        & \multicolumn{1}{c}{Yes} & \multicolumn{1}{c}{Yes}\\\hline\hline
    \end{tabular}
    \begin{center}
        \begin{minipage}\textwidth
            \noindent{\footnotesize \textbf{Notes}: The dependent variable in columns "Purchase?"
            is a dummy variable equal to one if the household purchased alcohol in the
            period. The dependent variable in columns "Quantity" is the number of standard
            drinks purchased per adult in that period, conditional on making a non-zero
            purchase. The table shows the estimated coefficients on a variable for the
            household's alcohol inventory. This is calculated by assuming that the
            household has a fixed level of consumption (equal to its mean purchases per
            period). Standard errors are in parentheses. ***, ** and * indicate a
            significant effect at the 1\%, 5\% and 10\% levels.}
        \end{minipage}
    \end{center}
\end{table}

\FloatBarrier

\pagebreak
\newpage
\cleardoublepage

    \FloatBarrier
    \newpage
    
    \newpage
    \section{Calibration: details}\label{AppInference}
        
\subsection{Market models: details}

    \subsubsection{Demand for quality: specification}\label{AppInference:fixedeffects}

Consistent with equation \ref{Eq:RUM} in Section \ref{Sec:Demand_Model}, household's $h$ utility of purchasing product $j$ at time $t$ for consumption occasion $o$ in alcohol market $a$ is:

\begin{equation}
	V^{ah}_{jto}=V^{ah}_{jt}+\varepsilon^{ah}_{jto}
\end{equation}

\noindent where $V^{ah}_{jt}$ denotes the deterministic component of the utility function. It is a function of product attributes and households' characteristics, The error term $\varepsilon^{ah}_{jto}$ is specific to the household, the product, and the (anticipated) consumption occasions. It captures the effect of unobserved variables such as occasion-specific effects (e.g., the anticipation of a party), retailer promotions, variations in the product positioning on shelves etc.

The deterministic component of utility is further specified as:

\begin{equation}
	V^{ah}_{jt}=-\alpha^{ah} p^{a}_{jt}+\beta^{ah}Z^{a}_{jt}
\end{equation}

\noindent with $\alpha^{ah}= \alpha^{a}+\delta^{a} D^{h} + \sigma^{a}\zeta^{a}_h$, where $\alpha^{a}$ is the mean marginal disutility for the category and $D_{h}$ is a vector of discrete demographic characteristics ($\zeta^{a}$ is a category-specific vector of coefficients), $\sigma^{a}$ represents the unobserved heterogeneity in the disutility of price and $\zeta_h^{a}$ is a random term that captures the unobserved taste of household $h$ for alcohol category $a$. We assume that $\zeta_h^{a}$ follows a standard normal distribution.

Table \ref{Tab:DescStatClass} shows that there is a huge difference between the average price of champagnes (26\euro/L) and that of other sparkling wines (6\euro/L), while in the other alcohol categories, the average prices are more similar between the different subcategories. Hence, for the Sparkling Wine category, we decided to further distinguish the average disutility of price between the two subcategories by adding a term capturing the specific mean marginal disutility of the price for champaigns. For sparkling wines, the disutility of the price is specified as $\alpha^{ah}+\delta^{Champagne}$, but we only use the estimated $\alpha^{ah}$ in the calculation of the quality index. 

We allow the product-specific household utility $\beta^{ah}Z^{a}_{jt}$ to be a set of fixed-effects that capture the average utility of specific product characteristics (having in mind that a product is a brand-retailer pair) - see Table \ref{tab:vectorfixedeffect}: retailer fixed-effects $\beta_{r(j)}$; brand fixed-effects $\beta_{b(j)}$; alcohol-free fixed-effect $\beta_{Alc(j)}$; sub-category fixed-effects interacted with demographic fixed-effects $\beta_{c(j)} \times D^{h} $ where the demographics are the household income category ($D^{Income_{h}}$: 4 classes), the age category of the main shopper ($D^{Age_{h}}$: 3 categories), and alcohol habit as measured by the average purchase volumes per household member in 2013 ($D^{Habit_{h}}$: 3 categories). The selection was made based on specification testings by eliminating progressively sets of fixed-effects that were not significant. 

\FloatBarrier
\renewcommand{\arraystretch}{1.5}
\begin{table}[H]
    \caption{Vector of fixed effects per alcohol market}
    \label{tab:vectorfixedeffect}
    \centering
    \medskip
    \small
    \begin{tabular}{l| c}
        \hline \hline
        Alcohol category \textit{a} & Product-specific household utility: $\beta^{ah}Z^{a}_{jt}$ \\
        \hline
        Ciders & $\beta_{r(j)}+\beta_{c(j)} \times D^{Age_{h}}+\beta_{b(j)}$\\
        
        Beers & $\beta_{r(j)}+\beta_{b(j)} \times \beta_{Alc(j)}$ \\
        
        Aperitif & $\beta_{r(j)}+\beta_{c(j)} \times D^{Habit_{h}}+\beta_{b(j)}+ \beta_{Alc(j)}$ \\
        
        Spirits & $\beta_{r(j)}+\beta_{c(j)} \times D^{Income_{h}} + \beta_{Alc(j)}$ \\
        
        Still wines & $\beta_{r(j)}+\beta_{c(j)} \times D^{Income_{h}}$ \\
        
        Sparkling Wines & $\beta_{r(j)}+\beta_{c(j)}\times  D^{Age_{h}}+\beta_{b(j)}$ \\
        
        \hline \hline
    \end{tabular} 
     \begin{center}
        \begin{minipage}{0.7\textwidth}   
        \small \textbf{Notes:} $D^{Income_{h}}$: low-, mid-low, mid-high, high-income household (Kantar WorldPanel classification based on self-reported household income and income composition); $D^{Age_{h}}$: age of main shopper <35 years-old, 35-55 years-old, 55+ years-old; $D^{Habit_{h}}$: average household purchases  in 2013 <1, 1-2,2+ standard drinks/adult/week.
        \end{minipage}
    \end{center}
\end{table}

\FloatBarrier

Assuming that $\varepsilon^{ah}_{jto}$ is independently and identically
distributed like an extreme value type I distribution, we can write the individual probability that household $h$ buys product $j$ of the alcohol market $a$ at period $t$ for any consumption occasion, conditional on some value $\zeta^{a}_{h}$ of the unobserved household taste shifter:
\begin{equation}
	d^{a}_{jt}(\zeta_{h}, D_{h})= \frac{\exp (V^{ah}_{jt})}{%
		1+\sum_{k=1}^{J^{a}_{t}}\exp (V^{ah}_{kt})}
	\label{Ind_Share}
\end{equation}
\noindent with $J^{a}_{t}$ denoting the set of alternatives from market \textit{a} in period \textit{t}. Hence, the probability of purchasing product $j$ of the alcohol market $a$ at period $t$ is: \begin{equation}
	s^{ah}_{jt}=\int_{\zeta} d^{a}_{jt}(\zeta^{a}_{h}, D_{h})dP_{\zeta }(\zeta^{a}_{h} )
	\label{Share}
\end{equation}
\noindent where $dP_{\zeta }$ is the standard gaussian distribution. The market share of product $j$ of the alcohol market $a$ at period $t$ is:
\begin{equation}
	s^{a}_{jt}=\int_{D} \int_{\zeta} d^{a}_{jt}(\zeta^{a}_{h}, D_{h})dP_{\zeta }(\zeta^{a}_{h} )dP_{D}(D_{h})
	\label{Share}
\end{equation}
\noindent where $P_{D}$ is the distribution of household demographic characteristics observed from the data.
\bigskip

\subsubsection{Price endogeneity}\label{App:firststage}

In deriving the expression of household $h$'s purchase probability \ref{Ind_Share}, we have assumed that the explanatory variables, especially prices, are independent of the error disturbances $\varepsilon^{ah}_{jto}$. Although we do not let prices vary across household-related characteristics such as their place of residence, the assumption of price exogeneity may not hold if prices depend on time-varying product-specific characteristics such as the display on shelves or the advertising efforts by firms.

In practice, we can split up the household error term in two components: 
\begin{equation}
	\varepsilon^{ah}_{jto}=\xi^{a}_{jt}+e^{ah}_{jto}
\end{equation} 
\noindent where $e^{ah}_{jto}$ is a household-occasion specific taste shock varying across products and occasions and over time, while $\xi^{a}_{jt}$ captures the influence of time-varying product-specific characteristics. While $e^{ah}_{jto}$ is not observed by the supply-side, the unobserved product characteristics captured in $\xi^{a}_{jt}$ impact household purchase decisions and they are likely related to the pricing policies of firms, making prices endogenous. 

To solve this endogeneity issue, we apply a two-stage residual inclusion approach as in \textcite{petrin_control_2010}. We first regress the prices on a set of instrumental variables $W_{jt}$ as well as the set of product specific fixed-effects $\left\lbrace \beta_{r(j)},\beta_{b(j)},\beta_{c(j)},\beta_{Alc(j)} \right\rbrace$ of the baseline utility function:
\begin{equation*}
	p^{a}_{jt}=\psi^{a}W^{a}_{jt} +\beta_{r(j)}+\beta_{b(j)}+\beta_{c(j)}+\beta_{Alc(j)}+\eta^{a}_{jt}\label{eq:IVprice}
\end{equation*}
\noindent The estimated error term $\widehat{\eta}^{a}_{jt}$ of the first stage is a proxy measure of the impact of unobserved time-varying product characteristics on prices. We include this residual in the indirect utility $V^{ah}_{jt}$ in order to control for the impact of these unobserved product characteristics (second stage):
\begin{equation*}
	V^{ah}_{jt}=- \alpha^{ah} p^{a}_{jt}+\beta^{a}_{jt}Z^{a}_{jt} +\rho^{a} \widehat{\eta}^{a}_{jt}
\end{equation*}
\noindent where $\rho^{a}$ will be the estimated parameter associated with the estimated error term of the first stage. This approach also shows that we must choose instruments that induce significant variations in prices, otherwise $\widehat{\eta}^{a}_{jt}$ could be collinear with prices and $Z^{a}_{jt}$. The choice of instruments is discussed in the main text.

Table \ref{Tab:firststage} displays the estimation results of the instrumental equation (\ref{eq:IVprice}) in which product prices are instrumented on a set of instrumental variables (IVs) to construct the function controlling for price endogeneity in the demand models. Various sets of IVs were tested (see the main text) and we selected the variables that were significant to avoid weak IV issues. The F-test statistics reported at the bottom of the table are above the recommended threshold of 10 \parencite{stock_survey_2002}. The signs of the coefficients reflect various mechanisms, such as the market power of manufacturers or retailers, or incentives for price differentiation. Typically, a manufacturer may increase its prices if it offers many products (more market power), but will tend to set lower prices when other manufacturers also propose many products (more competition).  

\FloatBarrier
\begin{landscape}
	\begin{center}
		\begin{table}[!h]
			\caption{Price Regressions}
			\label{Tab:firststage}
			\def\sym#1{\ifmmode^{#1}\else\(^{#1}\)\fi}
			\centering
			\medskip
			\footnotesize
			\begin{tabular}{l| c| c| c| c| c| c}
				\hline \hline
				& \multirow{2}{*}{\textbf{Ciders}} & \multirow{2}{*}{\textbf{Beers}}  & \multirow{2}{*}{\textbf{Aperitifs}} & \multirow{2}{*}{\textbf{Spirits}} & \multirow{2}{*}{\textbf{Still wines}} & \multirow{2}{*}{\textbf{Sparkling wines}}  \\
				& & & & & & \\
				\hline
			Instrumental Variables (IV) & &&&&&\\
			\quad Taxes &  & & 1.226 (0.048)*** &1.744 (0.062)*** & & \\
			\quad Nb of competing products$^{1}$ & & 0.005 (0.000)*** & 0.105 (0.021)***  & 0.129 (0.023)*** &   &   \\
			\quad Nb of competing products & &  -0.010 (0.003)*** & -0.113 (0.021)***& -0.079 (0.021)*** &     & -0.022 (0.012) *    \\
			\quad  \quad offered by the other firms &  &&   & &  & \\  
			
			\quad Mean price of other products && && &&\\
			\makecell{\quad Mean price of other products in \\
            same retailer at other periods} &-7.9594 (1.332) & & & & -7.684 (1.514) *** ***& -3.298 (0.777) ***  \\
			\makecell{\quad Mean price of same brand products \\ in other periods} & -9.930 (1.475) ***& &  && 0.685 (0.019) ***   & -6.210 (0.379) ***  \\
			\hline
			Fixed Effects & &&&&&\\
			\quad Retailers    & Yes & Yes & Yes & Yes & Yes & Yes \\
			\quad Sub-category    & Yes & Yes & Yes & Yes & Yes  & Yes \\
			\quad Brands    & Yes & Yes & Yes & Yes & No & Yes \\
			\quad Brands x Alcohol free   &    No  &    Yes  &    No  &    No  &    No  &    No  \\
			\quad Alcohol-free & No & No & Yes & Yes & No & No \\
			\hline
			F-test IV   &  53.37***  &   28.24*** &   216.71***  & 275.15***  &   660.28***     &   106.06***         \\
			Adj. $R^2$ &   0.98  &  0.96 &  0.93 & 0.97 &   0.94    &   0.96               \\
			N  &   774&    4,100  &    3,119   &   4,417 &   2,147 &   1,927         \\
			\hline \hline
			\end{tabular}
			\begin{center}
				\begin{minipage}{1.5\textwidth}
					{\scriptsize \textbf{Notes}: Standard errors in parentheses. $* p < 0.10$, $** p < 0.05$, $*** p < 0.01$. Number of competing products defined as the number of other products within a given classification of alcoholic beverage, offered at the same period by the manufacturer itself and by the other manufacturers. N: number of price-period observations.}
				\end{minipage}
			\end{center}
		\end{table}
	\end{center}
\end{landscape}

\subsubsection{Estimation}\label{App:maxlik}

The model is maximized by applying a maximum simulated log-likelihood procedure, whereby we sample the standard normal distribution by applying the sparse grid methodology of \textcite{heiss_likelihood_2008}. This approach allows approximating the likelihood function without suffering from the dimensional problem of the large number of draws that are required to minimize the estimation error of the simulated likelihood function.
\FloatBarrier

 \subsubsection{Demand for quality: elasticity formula}\label{App:elasticity}
    
At the market level, the own- and cross-price elasticities
of product market shares $s^{a}_{jt}$ were derived using the following formula:
\begin{equation}
	\frac{\partial s^{a}_{jt}}{\partial p^{a}_{kt}}\frac{p^{a}_{kt}}{s^{a}_{jt}}=\left \{
	\begin{array}{cl}
		\frac{p^{a}_{jt}}{s^{a}_{jt}}\displaystyle \int_{D} \int_{\zeta} \alpha^{ah}d^{ah}_{jt}(1-d^{ah}_{jt})dP_{\zeta }(\zeta^{a}_{h} )dP_{D}(D_{h}) 
		& \mbox{if}\;j=k \\
		\frac{p^{a}_{kt}}{s^{a}_{jt}} \displaystyle \int_{D} \int_{\zeta} \alpha^{ah}d^{ah}_{jt}d^{ah}_{kt} dP_{\zeta }(\zeta^{a}_{h} )dP_{D}(D_{h}) & 
		\mbox{otherwise}
	\end{array}
	\right.  \label{elasti}
\end{equation}

Household-specific level price elasticities can also be easily calculated, as well as elasticities by population subgroups or by product subcategories. Table \ref{Tab:Elastclassif} reports such elasticities. 

\begin{landscape}
    \renewcommand{\arraystretch}{1.5}
	\begin{table}
		\caption{Average estimated own-price elasticities by sub-category}
		\label{Tab:Elastclassif}
		\centering
		\medskip
		\footnotesize
	\begin{tabular}{l| l l l l l l}
		\hline \hline
		& \quad \textbf{Ciders}& \quad \textbf{Beers} & \quad \textbf{Aperitifs} & \quad \textbf{Spirits} & \quad \textbf{Still wines} & \quad \textbf{Sparkling wines}     \\
		\hline
		
		By sub-category &  Sweet &  Alcohol-free & Cocktails and Punch   & Rum    &  de Table     &  Champagne          \\
		& -3.59 (0.84) &  -4.74 (1.73)&  -2.93 (1.52) &  -3.45 (0.23) & -5.64 (3.61) & -2.78 (1.14)      \\

		&  Raw&  Bock and premium  &   Liquor wines  & Whisky &  de Pays   &  Other sparkling wines                \\
		& -3.48 (0.90) &  -5.29 (1.68) & -4.55 (0.73) &  -3.68 (0.17) & -6.60 (2.43)& -2.73 (1.26)    \\
		
		& &  Special &   Natural sweet wines & Aniseed & d'Appellation    &               \\
		& &  -5.44 (2.08)& -3.05 (0.78)&  -2.89 (1.05)& -4.95 (1.82)     &    \\
		
		&       &       &  Amer, gentiane and vermouth    &  Liquors     &       &     \\
		& & &-3.46 (1.33) &-3.18 (0.51) &  &  \\
		
		&       &       &   Other aperitifs   & Other spirits       &       &    \\
		& & &-2.96 (1.58) &-3.56 (0.20) &  &   \\
			& & & & & &\\
		\textbf{By category} & \textbf{-3.40 (0.84)}& \textbf{-5.02 (1.60)} & \textbf{-3.04 (1.20)} & \textbf{-3.55 (0.20)}& \textbf{-4.43 (2.15)}   & \textbf{-2.71 (1.09)}   \\
		\hline \hline
	\end{tabular}
		\begin{center}
			\begin{minipage}{1.2\textwidth}
				{\scriptsize \textbf{Notes}: in parenthesis, the standard deviation of estimated elasticities across periods and products in the same subcategory or category.}
			\end{minipage}
		\end{center}
	\end{table}
\end{landscape}
\FloatBarrier

\subsection{Demand for quantity}

    \subsubsection{Specification}\label{App:quantity_spe}

The demand for quantity model is estimated on pseudo-panel data that are constructed by aggregating data at the level of sociodemographic clusters indexed by $c$.
The demand functions are derived from a Quadratic Almost Ideal Demand System (QUAIDS). The budget share $w_{t}^{ac}$ on
alcohol category $a$ for cluster $c$ at period $t$ varies with log total
expenditure $\ln Y_{t}^{c}=\ln\left( \sum_{a=1}^{A}Y_{t}^{ac}\right)=\ln\left(\sum_{a=1}^{A}\mathbb{P}_{t}^{ac}\mathbb{Q}
_{t}^{ac}\right)$ and the vector of log price indices, ln$\mathbf{P}_{t}^{c}=(\ln
\mathbb{P}_{t}^{1c},...,\ln\mathbb{P}_{t}^{Ac})^{\prime}$, according to the following econometric specification:

\begin{equation}
	w_{t}^{ac}=\frac{Y_{t}^{ac}}{Y_{t}^{c}}=\kappa^{ac}+\mathbf{\gamma}^{a\prime}
	\ln\mathbf{P}_{t}^{c}+\chi^{a}(\ln Y_{t}^{c}-G_{1}(\ln\mathbf{P}^{c}_{t},\mathbf{\theta
	}))+\lambda^{a}\frac{(\ln Y_{t}^{c}-G_{1}(\ln\mathbf{P}_{t}^{c},\mathbf{\theta}))^{2}
	}{G_{2}(\ln\mathbf{P}^{c}_{t},\mathbf{\theta)}}+\nu_{t}^{ac} \label{Eq:budget}
\end{equation}

with the following non-linear price aggregators

\begin{align*}
	G_{1}(\ln\mathbf{P}_{t}^{c},\Theta^{Q}) &  =\kappa_{0}^{c}+\kappa^{c\prime}\ln\mathbf{P}_{t}^{c}+\frac{1}{2}\ln\mathbf{P}_{t}^{c\prime}\Gamma\ln\mathbf{P}_{t}^{c}\\
	G_{2}(\ln\mathbf{P}_{t}^{c},\Theta^{Q}) &  =\exp(\mathbf{\chi}^{\prime}\ln\mathbf{P}_{t}^{c})
\end{align*}

\noindent where $\kappa^{c}=(\kappa^{1c},...,\kappa^{Ac})^{\prime}$, $\chi=(\chi^{1},...,\chi^{A})^{\prime}$, $\Gamma=(\mathbf{\gamma}%
^{1},...,\mathbf{\gamma}^{A})^{\prime}$, $\Theta^{Q}$ is the set of all
parameters, and $\nu_{t}^{ac}$ is an error term. The parameter $\kappa_{0}^{c}$ in the
price aggregator is unidentified and can be set to 0 or to any other fixed
value. The heterogeneity of households in clusters enters the demand system through
the $\kappa^{ac}$'s, which are modelled as linear combinations of the
sociodemographic variables $D^{c}$ observed in the data (including a constant):

\begin{equation}
	\kappa^{ac}=K^{a\prime}D^{ct}
\end{equation}

\subsubsection{Construction of the price indices}\label{App:construction_price_indices}

Consider first a household $h$ that purchases a positive quantity
in category $a$. Following equation (\ref{eq:Adjusted_price}), the price index is:
for each category $a$:
\begin{equation}
	\mathbb{P}_{t}^{ah}=\frac{Y_{t}^{ah}}{Q_{t}^{ah}(1+b_{t}^{ah}(\mathbf{p}_{t}^{a}))} \label{Eq:PriceIndex}
\end{equation}
\noindent Total expenditure $Y_{t}^{ah}$ will be constructed by deflating observed expenditure by the monthly CPI. The quantity 
$Q_{t}^{ah}$ will be the observed alcohol intake of $a$.

 The quality-adjustment term $(1+b_{t}^{ah}(\mathbf{p}_{t}^{a}))$ is less straightforward to compute, because its formula includes the marginal utility of expenditure $\alpha^{ah}$ (cf. Equation \ref{Eq:quality_index}):
\begin{equation}
	b_{t}^{ah}(\mathbf{p}_{t}^{a})=\frac{\ln\left(  {\displaystyle\sum\limits_{j\in J_{t}^{a}}}\exp(V_{jt}^{ah}-V_{\star t}^{ah})\right)  +\alpha^{ah}{\displaystyle\sum\limits_{j\in J_{t}^{a}}}\pi_{jt}^{ah}\left( p_{jt}^{a}-p_{\star t}^{a}\right)  }
	{\alpha^{ah}}
\end{equation}
\noindent where $\pi^{ah}_{jt}$ is the conditional purchase probability:
\begin{equation}
    \pi^{ah}_{jt}=d_{jt}^{a}(\zeta^{ah},D^h)/(1-d^{a}_{0t}(\zeta^{ah},D^h))=\frac{\exp (V^{ah}_{jt})}{
		\sum_{k=1}^{J^{a}_{t}}\exp (V^{ah}_{kt})}
\end{equation} 
\noindent $V_{\star t}^{ah}$ and price $p_{\star t}^{a}$ are \enquote{reference} purchase utility and prices that we specify as the weighted average of purchase utilities
${V}_{jt}^{ah}$ and prices $p_{jt}^{a}$ with weights equal to the
conditional purchase probabilities $\pi^{ah}_{jt}$.
\begin{align}
    V_{\star t}^{ah} = \sum_j \pi^{ah}_{jt} {V}_{jt}^{ah} \\
    p_{\star t}^{a} = \sum_j \mathbb{E}_h(\pi^{ah}_{jt}) {p}_{jt}^{a}
\end{align}

\bigskip
In Step 1's random utility model, the marginal utility of expenditure is specified as a random variable: $\alpha^{ah}= \alpha^{a} + \delta^{a} D_{h} + \sigma^{a}\zeta^{a}_h$, where $\alpha^{a}$ is the average
marginal utility of expenditure in category $a$ and $\zeta^{a}_h$ is the household-specific taste shock. As the latter is not observed, we cannot directly compute $b_{t}^{ah}(\mathbf{p}_{t}^{a})$. However, we can calculate its expectation conditional on all available information on household preferences $\Omega_{t}^{ah}$. This information includes, in particular, its observed choices in the period. Hence, we apply Bayes rule to compute:
\begin{equation}
	\mathbb{B}_{t}^{ah}=\mathbb{E}\left[b_{t}^{ah}(\mathbf{p}_{t}^{a})\left|\Omega_{t}^{ah}\right.\right]=\int_{\zeta}b_{t}^{ah}(\mathbf{p}_{t}^{a})dP_{\zeta\left|\Omega_{t}^{ah}\right. }(\zeta_{h} )  \label{Eq:IndexQualPred}
\end{equation}
\noindent where the conditional distribution $P(\zeta|\Omega)$ is derived using the Bayes rule. This conditional expectation is calculated by integrating out over the simulated distribution of $\zeta^{a}_h$ using Halton draws and the same $\zeta^{a}_h$ for all purchase acts of
the same household. 

The price index is finally constructed as:
	\begin{equation}
		\mathbb{P}_{t}^{ah}=\frac{Y_{t}^{ah}}{Q_{t}^{ah}(1+\mathbb{B}_{t}^{ah})} \label{Eq:PriceIndex}
	\end{equation}
\bigskip

\subsubsection{Construction of the pseudo-panel}\label{App:pseudopanel}

The model estimation uses sample S3 (see definition in Section \ref{App:Subsamples}), as it maximizes the number of households that we can aggregate to form the clusters. This leaves us with 120 clusters observed over 26 time periods (13 periods in 2014), 3120 observations (cluster-periods), half of them being used for estimating the demand in 2014. Table \ref{Tab:Clusters} reports the median, minimum and maximum numbers of households and purchases aggregated in each cluster, and by year and period for purchases.

\begin{table}[!htbp]
	\caption{Characteristics of clusters}\label{Tab:Clusters}
	\centering
	\medskip
	\begin{tabular}[c]{llccc}\hline\hline
		\multicolumn{2}{l}{} & Median & Min & Max\\\hline
		\multicolumn{2}{l}{Number of households per clusters} & \multicolumn{1}{r}{212} &
		\multicolumn{1}{r}{3} & \multicolumn{1}{r}{1389}\\
		\multicolumn{2}{l}{Number of purchases per clusters} & \multicolumn{1}{r}{} & \multicolumn{1}{r}{}
		& \multicolumn{1}{r}{}\\
		& Over the year & \multicolumn{1}{r}{8,105} & \multicolumn{1}{r}{28} &
		\multicolumn{1}{r}{38,079}\\
		& Minimum in a period & \multicolumn{1}{r}{243} & \multicolumn{1}{r}{1} &
		\multicolumn{1}{r}{1,262}\\
		& Maximum in a period & \multicolumn{1}{r}{431} & \multicolumn{1}{r}{8} &
		\multicolumn{1}{r}{1,753}\\\hline\hline
	\end{tabular}
	\begin{center}
		\begin{minipage}{0.5\textwidth}
			{\footnotesize \textbf{Notes}: All statistics are computed over $N=120$ clusters. The first line reports statistics on the median, minimum and maximum number of households in clusters. The second line reports the median, minimum and maximum number of purchases observed in each cluster over a year. The third and fourth lines reports the same statistics for the minimum number of purchases and the maximum number of purchases that can be observed in a period for each cluster.}
		\end{minipage}
	\end{center}
\end{table}

For each alcohol category $a$, using the QUAIDS specification for budget shares (\ref{Eq:budget}), we estimate a demand function:
\begin{equation}
	\mathbb{Q}_{t}^{ac}(\mathbb{P}_{t}^{1c},...,\mathbb{P}_{t}^{Ac},Y_{t}^{c},D^{c};\Theta^Q)
\end{equation}

\noindent where $D_{c}$ is a vector of average characteristics of households in
cluster $c$,
$Y_{t}^{c}$ is the average alcohol expenditure, $\Theta^Q$ is the vector of parameters to
be estimated, and the price indices represent average expenditures for one
unit of utility expressed in money-metric as in \ref{Eq:PriceIndex}.

We apply the following formula for computing these variables:
\begin{align}
	Y_{t}^{c} &=\sum_{a=1}^{A}Y_{t}^{ac} \\
	Y_{t}^{ac}&=\displaystyle\sum\limits_{h\in c \cap S3}\frac{\omega_{ht}^{S3}}{\sum_{h\in c}\omega_{ht}^{S3}}Y_{t}^{ah} \\
	D^{c}&=\displaystyle\sum\limits_{h\in c \cap S3}\frac{\omega_{ht}^{S3}}{\sum_{h\in c}\omega_{ht}^{S3}}D^{h}\\
	\mathbb{P}_{t}^{ac}&=\frac{Y_{t}^{ac}}{\displaystyle\sum\limits_{h\in c \cap S3}\frac{\omega_{ht}^{S3}}{\sum_{h\in c \cap S3}\omega_{ht}^{S3}}Q_{t}^{ah}(1+\mathbb{B}_{t}^{ah})}\\
	\mathbb{Q}_{t}^{ac}&=\frac{Y_{t}^{ac}}{	\mathbb{P}_{t}^{ac}}
\end{align}
where $\omega_{ht}^{S3}$ is the number of French households represented by household $h$ (see the weight normalization in Section \ref{AppData:weightnorm}).  

In our preferred specification, we include as socio-demographic shifters $D^{ct}$ the proportion of households in each
modality of the variables used in the clustering, period dummies, as well as
the proportion of households living in the 8 administrative regions (ZEAT) defined by INSEE (the French National Statistics Office) and in town with less than 20,000 inhabitants. These variables will thus control (adjust) for a variety of effects: position in the life-cycle, income, household structure and spatial variations in tastes such as  rural/urban differences. We also control for seasonality effects through the inclusion of period fixed-effects.

\bigskip
 
\subsubsection{Demand for quantity: elasticity formula}

After having estimated the demand system, we derive the budget
and price elasticities. Let $a$ and $k$ denote two alcohol categories, differentiating (\ref{Eq:budget})
 with respect to $Y_{t}^{c}$ and $\mathbb{P}_{t}^{ac}$ yields, respectively:
 
\begin{align*}
	\mu^{ac}  &  =\chi^{a}+2\lambda^{a}\frac{(\ln Y_{t}^{c}-G_{1}(\ln\mathbf{P}
		_{t}^{c},\Theta^Q))}{G_{2}(\ln\mathbf{P}_{t}^{c},\Theta^Q)},\\
	\mu^{ack}  &  =\gamma^{ak}-\mu^{a}(\mathbf{\alpha}^{ak}+\mathbf{\gamma}%
	^{k\prime}\ln\mathbf{P}_{t}^{c})-\lambda^{a}\chi^{a'}\frac{(\ln Y_{t}^{c}
		-G_{1}(\ln\mathbf{P}_{t}^{c},\Theta^Q))^{2}}{b(\ln\mathbf{P}_{t}^{c}
		,\Theta^Q)}
\end{align*}
Budget elasticities are then given by:
\begin{equation}
	Elas_{a}^{c}=\mu^{a}/w^{ac}+1 \label{Eq:ElastBudget}
\end{equation}
\noindent Uncompensated price elasticities for the \emph{adjusted} quantities $\mathbb{Q}$ are:
\begin{equation}
	Elas_{ak}^{uc}=\mu^{ak}/w^{ac}-\delta^{ak} \label{Eq:UncompElast}
\end{equation}
\noindent where $\delta^{ak}$ is the Kronecker
delta and $w^{ac}$ is the budget share for cluster $c$. Compensated price elasticities are given by:
\begin{equation}
	Elas_{ak}^{c}=Elas_{ak}^{uc}+Elas_{a}^{c}w^{kc} \label{Eq:CompElast}
\end{equation}.

The elasticities reported in the main text are computed setting all variables to their average sample value.

\subsection{Impacts on consumer welfare}\label{App:EVformula}

We use equivalent variations to estimate the impact of counterfactual price policies on consumer welfare. Importantly, we are able to account for consumer adjustments in quality and quantity, as the quality-adjusted prices take into account changes in market prices and consumers' substitutions between products within a market.

Equivalent variation (EV) is the adjustment in income required \textit{before}
the change has occurred to reach the final level of utility (i.e., the level
that would occur if the change happened) at the original prices. It must
satisfy the following equalities:
\[
V^{1}=V(\mathbf{P}^{0},Y^{0}+EV)=V(\mathbf{P}^{1},Y^{1})
\]
where $V^{1}$ is the utility levels of the household after the policy, and $\mathbf{P}^{0}$ and $\mathbf{P}^{1}$ are two levels of quality-adjusted prices. 

Then,
EV is given by
\begin{align*}
c(\mathbf{P}^{0},V(\mathbf{P}^{0},Y^{0}+EV))  &  =c(\mathbf{P}^{0},V^{1})\\
Y^{0}+EV  &  =c(\mathbf{P}^{0},V^{1})\\
EV  &  =c(\mathbf{P}^{0},V^{1})-c(\mathbf{P}^{0},V^{0})
\end{align*}
EV can be decomposed as
\begin{align*}
EV  &  =\left[  c(\mathbf{P}^{1},V^{1})-c(\mathbf{P}^{0},V^{0})\right]
+\left[  c(\mathbf{P}^{0},V^{1})-c(\mathbf{P}^{1},V^{1})\right] \\
&  =YV+EV^{\ast}
\end{align*}
where $YV$ is the variation in expenditure, and $EV^{\ast}$ computed as follows. Since $Y^{1}=\mathbf{P}^{1}\mathbf{Q}
^{1}=c(\mathbf{P}^{1},V^{1})$

\begin{itemize}
\item use $\mathbf{P}^{1}$ and $Y^{1}=c(\mathbf{P}^{1},V^{1})$ to compute the
indirect utility in 1, $V^{1}$, using the formula
\[
\ln V^{1}=\left\{  \left[  \frac{\ln Y^{1}-\ln a(\ln\mathbf{P}^{1}
,\mathbf{\theta})}{b(\ln\mathbf{P}^{1},\mathbf{\theta})}\right]  ^{-1}
+c(\ln\mathbf{P}^{1},\mathbf{\theta})\right\}  ^{-1}
\]

\item and use $\mathbf{P}^{0}$ and $V^{1}$ to compute the cost $c(\mathbf{P}
^{0},V^{1})=\widetilde{Y}^{0}$, using the reverse formula
\[
\ln\widetilde{Y}^{0}=\ln a(\ln\mathbf{P}^{0},\mathbf{\theta})+b(\ln
\mathbf{P}^{0},\mathbf{\theta})\left[  \frac{\ln V^{1}}{1-c(\ln\mathbf{P}
^{0},\mathbf{\theta})\ln V^{1}}\right]  
\]

\end{itemize}

\FloatBarrier
\pagebreak

    \FloatBarrier
    \pagebreak

    \section{Simulation: details}\label{AppSimulation}
        The simulations are based on the estimated marginal costs from the supply model
as well as on the other estimated structural demand parameters. We first show how we use the supply-side model to simulate the impact of new price regulations on market equilibrium outcomes.

    \subsection{Step 1: Markets simulations}

 When a reform of price regulations is implemented, the producer faces new constraints taking the form of either a change in alcohol-specific excise taxes levied at the producer level or a minimum unit price policy. 
\bigskip

For a change in excise tax from $\mathbf{T}^{0}$ to $\mathbf{T}^{1}$, the new vector of equilibrium prices $\mathbf{p}_{t}^{1}$ is the solution of the following system of equations derived from first-order conditions (\ref{eq1}) and (\ref{eq2}) for each separate market $a$ (we omit this superscript):

\begin{equation*}
	\sum\nolimits_{k \in G_{f}}\left(\frac{p^1_{kt}}{1+\tau}-\widehat{C}_{kt}-T_{k}^{1}\right) \frac{\partial s_{kt}(\mathbf{p}^1_{t})}{\partial p^1_{j}}+s_{jt}(\mathbf{p}^1_{t})+\sum \nolimits_{k \in \widetilde{S}_{r}}\left( \frac{p^1_{kt}}{1+\tau}-\widehat{C}_{kt}-T_{k}^{1}\right) \frac{\partial s_{kt}(\mathbf{p}^1_{t})}{\partial p^1_{jt}}  =0 , \forall j\in G_{fr}, \forall f, \forall r
\end{equation*}
\begin{equation*}
	\sum\nolimits_{k \in \widetilde{S}_{r}}\left( \frac{p^1_{kt}}{1+\tau}-\widehat{C}_{kt}-T_{k}^{1}\right) \frac{\partial s_{kt}(\mathbf{p}^1_{t})}{\partial p^1_{jt}}+s_{jt}(\mathbf{p}^1_{t})  =0, \forall j\in \widetilde{S}_{r}, \forall r.
\end{equation*}
\bigskip

This system of first-order conditions can be rewritten compactly with matrix notation, and can be inverted such that the vector of new equilibrium margins $\mathbf{m}^{1}$ (with elements $m_{kt}^{1}$ equal to $\frac{p_{kt}^{1}}{1+\tau}-\widehat{C}_{kt}-T_{k}^{1}$) can be expressed as a function $\mathcal{M}$ of the demand parameters and the new equilibrium prices $\mathbf{p}^{1}$:
\begin{equation}
	\mathbf{m}^{1}=\mathcal{M}\left(\mathbf{p}^{1}\right)
\end{equation}

By construction, this vector of new equilibrium prices must satisfy the following condition:
\begin{equation}
	\mathbf{p}^{1}-\mathbf{m}^{1}=\mathbf{\widehat{C}}+\mathbf{T}^{1}
\end{equation}

This suggests finding the equilibrium prices that minimize the following criterion:
\begin{equation}
	\mathbf{p}^{1}=argmin_{\mathbf{p}}\left\Vert \mathbf{p}-\mathcal{M}\left(\mathbf{p}\right)-\mathbf{\widehat{C}}-\mathbf{T}^{1}\right \Vert
	\label{eq-simul}
\end{equation}%
\noindent where $\left \Vert .\right \Vert $ is the Euclidean norm. In practice, we can implement an iterative algorithm, whereby at each iteration $s$ and for a vector of prices $\mathbf{p}(s)$, we have:
\begin{equation}
	\mathbf{p}(s+1)=\mathcal{M}\left(\mathbf{p}(s)\right)-\mathbf{\widehat{C}}-\mathbf{T}^{1} \label{Eq:Iter}
\end{equation} 
\noindent The algorithm will be initialized with $\mathbf{p}^{0}$ equal to the actual prices, which means that the focus is on the fixed-point solution $\mathbf{p}(\infty)$ that is consistent with the initial market situation. 

The pass-through of the tax onto consumer prices will be computed as the ratio of the price difference on the cost difference:
\begin{equation}
	PT_{jt} = \frac{p_{jt}^{1} - p_{jt}^{0}}{C_{jt}^{1} - C_{jt}^{0}} 
\end{equation}
\bigskip

Finally, variations in profits will be calculated for each firm, notably allowing us to distinguish the economic impact by size of manufacturers and by alcohol category.
\bigskip

\textbf{Simulating a new market equilibrium after a Minimum Unit Price policy}
\medskip

Let $\mathbf{\underline{p}}^{1}$ be the vector of minimum prices imposed by a MUP policy, and $\mathcal{P}$ the set of prices that meet the MUP constraint.
For this policy, the new equilibrium prices will be estimated by solving the following program:
\begin{equation}
	\mathbf{p}^{1}=argmin_{\mathbf{p} \in \mathcal{P}}\left\Vert \mathbf{p}-\mathcal{M}\left(\mathbf{p}\right)-\mathbf{\widehat{C}}-\mathbf{T}^{1}\right \Vert
	\label{eq-simul2}
\end{equation}%
We will use a modified version of the iterative algorithm \ref{Eq:Iter}:
\begin{equation}
	\forall j, p_{j}(s+1)=max\left\lbrace\underline{p}_{j};  \mathcal{M}_{j}\left(\mathbf{p}(s)\right)-\widehat{C}_{j}-T_{j}^{1}
	\right\rbrace 
\end{equation}

        \subsection{Step 2: Simulating changes in household purchases of pure alcohol}

Denote $r=0,1$ the superscript indicating whether a variable $X$ is
constructed using the actual or new price vector, and $\Delta X=(X^{1}%
-X^{0})/X^{0}$ the \emph{relative} variation in $X$ generated by the change from
actual to new prices. 

The relative variation in total alcohol (ethanol) purchases $E^{h}$ by household $h$ is a weighted sum of category-specific variations $\Delta E^{ah}$:
\begin{align}
	\Delta E^{h} &=\sum_{a}\Delta E^{ah}*\frac{E^{ah0}}{\sum_{k}E^{kh0}}\\
\end{align}

The impact of the price change on  $\Delta
E^{ah}$, is a combination of a quantity effect and a \textquotedblleft
quality\textquotedblright\ effect, as shown in the following formula:
\begin{equation}
	\Delta E^{ah}=\Delta Q^{ah}(1+\Delta \Psi^{ah})+\Delta \Psi^{ah} \label{Eq:Ethanol_a}
\end{equation}

\noindent where $\Delta Q^{ah}$ and $\Delta \Psi^{ah}$ are the \emph{relative} variations in,
respectively, the aggregated \emph{physical} quantity (in Liter) and the
average alcohol-content of purchases in category $a$ by household $h$.

We assume that households who did not purchase any variety in a category
throughout 2014 will continue not to purchase in this category regardless of the price level: households who did not
consume before the policy will not consume after the policy, so that for them
we have $\Delta E^{ah}=0$.

For the other households, the relative variation in quantity is bounded as
follows:
\begin{equation}
	\Delta Q^{ah}=Max\left\lbrace \frac{Q^{ah1}-Q^{ah0}}{Q^{ah0}};-1\right\rbrace 
\end{equation}
\bigskip

Simulation Step 1 provides us with $\mathbf{p}^{ar}$, the vector of new market prices. The aggregate price index for category $a$ and household $h$ can be expressed
as the average of variety prices, $p^{ar}_{j}$ ($\forall j\in J^{a}$), weighted by
the corresponding conditional purchase probabilities, $\pi_{j}^{ahr}(p^{ar})$ (the probability that $h$ purchases $j$ in $a$ conditional on purchasing any product in $a$). Formally
\begin{equation}
	\mathbb{P}^{ahr}=\frac{1}{(1+\mathbb{B}^{ahr})}\sum_{j \in
		J^{a}}\frac{\pi_{j}^{ahr}}{\sum_{k \in
			J^{a}}\pi_{k}^{ahr}}p^{r}_{j}
\end{equation}
\noindent where we replace the average unit value $Y^{ah}_{t}/Q^{ah}_{t}$ in formula (\ref{eq:Adjusted_price}) by the weighted average of product prices at $r$, and we replace $b^{ah}(\mathbf{p}^{ar})$ by $\mathbb{B}^{ahr}$, the
expected quality obtained by household $h$ from a purchase occasion in
category $a$ with products priced at $\mathbf{p}^{ar}$. $\mathbb{B}^{ahr}$ and $\pi_{j}^{ahr}$ are calculated by integrating over the simulated distribution of unobserved preferences, as we did previously to evaluate equation (\ref{Eq:IndexQualPred}).
\bigskip

The transmission of the changes in variety price to the aggregated price indices
is then measured by $\Delta\mathbf{P}^{h}=(\Delta\mathbb{P}^{1h}%
,...,\Delta\mathbb{P}^{Ah})$, and its impact on the quantity index for category
$a$, household $h$, by
\begin{equation}
	\Delta\mathbb{Q}^{ah}=\mathbf{Elas}_{P}^{ac(h)}\mathbf{\times}\Delta\mathbf{P}%
	^{h}+Elas_{Y}^{ac(h)}\times(\mathbf{Elas}_{PY}^{c(h)}\times\Delta\mathbf{P}^{h})
\end{equation}

\noindent where $c(h)$ is the cluster to which $h$ belongs, $\mathbf{Elas}_{P}^{ac(h)}$ is the vector of uncompensated direct and cross-price
elasticities for alcohol category $a$, calculated according to formula (\ref{Eq:UncompElast}), $\mathbf{Elas}_{PY}^{c(h)}$ is the elasticity vector of total expenditures to price
variations on total expenditures, as estimated in the first-stage instrumental
equation of total expenditure for the QUAIDS, and $Elas_{Y}^{ac(h)}$ is the budget elasticity of $a$.

Finally, knowing $\mathbb{Q}^{ah0}=Q^{ah0}(1+\mathbb{B}^{ah0})$ and $\Delta\mathbb{Q}^{ah}$,
we will calculate $\mathbb{Q}^{ah1}=(1+\Delta\mathbb{Q}^{ah })\mathbb{Q}^{ah0}$. This will give $Q^{ah1}=\mathbb{Q}^{ah1}/(1+\mathbb{B}^{ah1})$, as well as variety quantities
$q_{j}^{ah1}$, since $q_{j}^{ahr}=\pi_{j}^{ahr}(p^{ar})Q^{ahr}$.
\bigskip

In addition to the relative variation in aggregate quantity of
$a$, the change in the price of product varieties may induce substitutions
between varieties, which may also alter the average alcohol-content of the
purchases in that category. Let $\psi_{j}^{a}$ be the alcohol-content of variety $j$
(in grams per liter) in group $a$, then the relative variation in the
household-specific average alcohol-content of purchases in category $j$ can be
written as:
\begin{equation}
	\Delta \Psi^{ah}=\sum_{j \in J^{a}}\psi_{j}^{a}\Delta\pi_{j}^{ah}
\end{equation}
Having calculated $\Delta \Psi^{ah}$ and $\Delta \Psi^{ah}$, we will have the outcome of interest, $\Delta E^{ah}$, the relative variation in alcohol purchase from category $a$ (equation \ref{Eq:Ethanol_a}).

\FloatBarrier
    \pagebreak

    \section{Additional results \& sensitivity analysis}\label{AppResults}
        \subsection{Prices in the QUAIDS model: sensitivity analysis}\label{AppResults}

Table \ref{AppTab:SensitivityQuaids} below presents the results of the estimation of the QUAIDS model for two alternative specifications (2 and 3). These results are commented in the main text. To save space, we do not report the cross-price elasticities.
	\begin{table}[h]
        \caption{Average budget and own-price elasticities: sensitivity analysis}\label{AppTab:SensitivityQuaids}
        \centering
        \begin{tabular}[c]{llccc}
            \hline\hline
            &  & {\small Specification 1} & {\small Specification 2} & {\small Specification 3}\\\hline
            &  & {\small Baseline} & {\small +income/habit-period FE} & {\small Laspeyres price indices} \\\hline
            \multicolumn{5}{l}{{\small Budget}}\\
            & {\small Ciders} & \multicolumn{1}{l}{{\small \ 0.503 (0.302)*}} &
            \multicolumn{1}{l}{{\small \ 0.894 (0.308)***}} &
            \multicolumn{1}{l}{{\small \ 1.145 (0.044)***}}\\
            & {\small Beers} & \multicolumn{1}{l}{{\small \ 1.210 (0.112)***}} &
            \multicolumn{1}{l}{{\small \ 1.131 (0.108)***}} &
            \multicolumn{1}{l}{{\small \ 0.951 (0.016)***}}\\
            & {\small Aperitifs} & \multicolumn{1}{l}{{\small \ 1.017 (0.154)***}} &
            \multicolumn{1}{l}{{\small \ 0.943 (0.152)***}} &
            \multicolumn{1}{l}{{\small \ 1.011 (0.023)***}}\\
            & {\small Spirits} & \multicolumn{1}{l}{{\small \ 1.434 (0.083)***}} &
            \multicolumn{1}{l}{{\small \ 1.480 (0.081)***}} &
            \multicolumn{1}{l}{{\small \ 0.976 (0.013)***}}\\
            & {\small Still wines} & \multicolumn{1}{l}{{\small \ 0.349 (0.126)***}} &
            \multicolumn{1}{l}{{\small \ 0.352 (0.125)***}} &
            \multicolumn{1}{l}{{\small \ 1.134 (0.021)***}}\\
            & {\small Sparkling wines} & \multicolumn{1}{l}{{\small \ 0.385 (0.187)**}} &
            \multicolumn{1}{l}{{\small \ 0.387 (0.182)**}} &
            \multicolumn{1}{l}{{\small \ 0.943 (0.035)***}}\\\hline
            \multicolumn{5}{l}{Un{\small compensated own-price}}\\
            & {\small Ciders} & \multicolumn{1}{l}{{\small -0.504 (0.085)***}} &
            \multicolumn{1}{l}{{\small -0.524 (0.085)***}} &
            \multicolumn{1}{l}{{\small -0.639 (0.024)***}}\\
            & {\small Beers} & \multicolumn{1}{l}{{\small -0.828 (0.047)***}} &
            \multicolumn{1}{l}{{\small -0.758 (0.045)***}} &
            \multicolumn{1}{l}{{\small -0.742 (0.012)***}}\\
            & {\small Aperitifs} & \multicolumn{1}{l}{{\small -0.531 (0.045)***}} &
            \multicolumn{1}{l}{{\small -0.533 (0.04)***}} &
            \multicolumn{1}{l}{{\small -0.686 (0.015)***}}\\
            & {\small Spirits} & \multicolumn{1}{l}{{\small -0.881 (0.030)***}} &
            \multicolumn{1}{l}{{\small -0.888 (0.030)***}} &
            \multicolumn{1}{l}{{\small -0.818 (0.010)***}}\\
            & {\small Still wines} & \multicolumn{1}{l}{{\small -0.396 (0.046)***}} &
            \multicolumn{1}{l}{{\small -0.389 (0.046)***}} &
            \multicolumn{1}{l}{{\small -0.747 (0.013)***}}\\
            & {\small Sparkling wines} & \multicolumn{1}{l}{{\small -0.161 (0.033)***}} &
            \multicolumn{1}{l}{{\small -0.159 (0.032)***}} &
            \multicolumn{1}{l}{{\small -0.686 (0.014)***}}\\\hline\hline       
        \end{tabular}
        \begin{center}
        \begin{minipage}{1\textwidth}
            \noindent{\scriptsize \textbf{Notes}: Kantar Worldpanel 2013-2014; average elasticities where each cluster observation is weighted by the proportion of French households represented by the cluster; standard errors in parentheses; ***,** and * significant at the 1\%. 5\% and 10\% level; the first six lines display budget elasticities. i.e. the \% change in the quantity for a given alcohol category when total alcohol expenditures increase by 1\%; the last six lines display the \% change in the quantity for each alcohol category when their own
            price increases by 1\%; Specification 1 is our preferred specification used for producing the main results. Specifications 2 adds to specification 1 interactions between time fixed effects and income categories and interactions between time fixed effects and alcohol habit categories; In specification 3 we replace the quality-adjusted price indices by cluster-specific Laspeyres indices.}
        \end{minipage}
		\end{center}
	\end{table}

\FloatBarrier
\subsection{Key simulation results under QUAIDS estimates obtained with Laspeyres price indices}

\begin{center}
    \begin{table}[h!]
        \caption{Key simulation results: sensitivity analysis}
    	\label{AppTab:Laspeyresresults}
    	\centering
    	\medskip
    	\small
        \begin{tabular}{l c c c c c }
            \hline\hline
             \textbf{Prices in QUAIDS:} & \multicolumn{2}{c}{\textbf{Quality-adjusted prices}} &  & \multicolumn{2}{c}{\textbf{Laspeyres price indices}} \\
            Impacts on & \textit{Pure alcohol} & \textit{Purchase volume} & & \textit{Pure alcohol}  & \textit{Purchase volume} \\
            \hline
            Low uniform tax & +14.0\% & +1.7\% &  & +4.2\% & -4.5\% \\
            High uniform tax & -5.9\% & -9.3\% &  & -19.0\% & -22.2\% \\
            Low progressive tax & +6.8\% & +1.3\% &  & +2.2\% & -2.3\% \\
            High progressive tax & -10.3\% & -6.8\% &  & -18.8\% & -15.9\% \\
            Minimum Unit Price & -15.0\% & -13.5\% &  & -23.1\% & -22.0\% \\
            MUP + Low prog. Tax & -14.7\% & -13.9\% &  & -26.1\% & -25.6\% \\
            \hline\hline
        \end{tabular}
        \begin{center}
            \begin{minipage}{0.8\textwidth}
                {\scriptsize \textbf{Notes}: Policy simulation results when the QUAIDS is estimated with quality-adjusted prices and when it is estimated with Laspeyres price indices computed using cluster-specific average market shares and product prices.}
            \end{minipage}
        \end{center}
    \end{table}
\end{center}

\FloatBarrier

\subsection{Additional results for still wines by quality segment}

\begin{center}
    \begin{table}[h!]
        \caption{Impacts of Minimum Price on the market for still wines}
    	\label{AppTab:wineresults}
    	\centering
    	\medskip
    	\small
        \begin{tabular}{l c c c c c c c}
            \hline\hline
             & \multicolumn{3}{c}{\textbf{Price (\euro/L)}} && \multicolumn{3}{c}{\textbf{Volume (L/hhold)}}\\
            & Baseline & \multicolumn{2}{c}{Variation} && Baseline & \multicolumn{2}{c}{Variation} \\
            & \euro/L &  \multicolumn{2}{c}{(\% of baseline)} & & L/hhold & \multicolumn{2}{c}{(\% of baseline)} \\
            Supply-side reaction & & Without & With &  & & Without & With \\
            \hline
            \multicolumn{8}{l}{\textit{Segmentation by quality label}}\\
            Table & 2.1  & +132.6 & +136.0 && 8.9  & -95.5 & -96.2 \\
            Pays & 2.7 & +80.4  & +82.2 && 12.3 & -87.6 & -89.0 \\
            Appellation & 4.8 & +8.3  & +10.6  && 16.3  & +69.7 & +65.4  \\
            \hline
            \multicolumn{8}{l}{\textit{Segmentation by initial price}} \\
            $\le$ 3\euro/L & 2.3  & +107.8   & +108.0 && 19.3 & -94.1 & -94.6 \\
            $\in ]3;5]$\euro/L & 4.1 & +19.1  & + 20.3 && 9.6 & -4.5 & -11.3 \\
            $\ge5$\euro/L & 5.3  & +0.8   & +3.3 && 8.6 & +124.4  & +122.4  \\
            \hline\hline
        \end{tabular}
        \begin{center}
            \begin{minipage}{0.8\textwidth}
                {\scriptsize \textit{Notes}: Impacts of minimum price policy on unit prices and volumes purchased by households - sample average.}
            \end{minipage}
        \end{center}
    \end{table}
\end{center}

\FloatBarrier

\subsection{MUP: decomposition of variations in tax revenues and firms profits}

Let $Q^{a}$ be the size of market $a$, $s_j$ the market share of product $j$, and $(p_j-c_j)$, the unit margin, then the relative variation in profit can be approximately decomposed as (taking the log): 

$\frac{\Delta \Pi_j}{\Pi_j} \approx \underbrace{\frac{\Delta Q^{a}}{Q^{a}}}_{Quantity}+ \underbrace{\frac{\Delta s_{j}}{s_j}}_{Quality}+ \underbrace{\frac{\Delta p_j}{p_j - c_j}}_{Price}$

\begin{center}
    \begin{table}[h!]
        \caption{MUP impacts on tax revenues and profits}
    	\label{AppTab:taxprofitsresults}
    	\centering
    	\medskip
    	\small

        \begin{tabular}{ll cccccccc} 
        \hline\hline
        & & Ciders & Beers & Aperitifs & Spirits & Still wines & Sparkling wines & &Total \\
        \hline
        \multicolumn{10}{l}{Tax revenues} \\ 
        Total & & +3.2\% & -6.8\% & -3.2\% & -7.6\% & +15.8\% & +2.0\% & & -2.3\% \\
         & Alcohol taxes & +3.0\% & -12.5\% & -8.2\% & -9.9\% & -22.9\% & -3.8\% & & -10.3\% \\
         & VAT  & +3.2\% & -1.2\% & +2.5\% & -0.4\% & +18.2\% & +2.3\% & & +7.0\% \\
        \hline
        \multicolumn{10}{l}{Profits} \\
        Total && +3.4\% & +1.1\% & +3.4\% & +6.9\% & +2.3\% & +1.8\% & & +3.7\% \\
        & Quantity & +3.0\% & -10.6\% & -9.0\% & -8.5\% & -30.4\% & -4.3\% & & -14.1\% \\
        & Quality & -0.0\% & -15.1\% & -8.6\% & -5.8\% & -139.7\% & +1.3\% & & -43.6\% \\
        & Price & +0.5\% & +26.8\% & +21.1\% & +21.2\% & +172.5\% & +4.8\% & & +61.5\% \\
        \multicolumn{10}{l}{\textit{Small firms}}\\
        Total && +3.4\% & +7.6\% & +6.7\% & +5.8\% & +39.3\% & +2.7\% & & +24.5\% \\
        & Quantity & +3.0\% & -11.3\% & -9.3\% & -8.4\% & -41.4\% & -4.3\% & & -26.6\% \\
        & Quality & +0.4\% & +6.0\% & +5.0\% & -2.2\% & -44.4\% & +5.8\% & &-24.1\% \\
        & Price & -0.0\% & +12.9\% & +11.0\% & +16.4\% & +125.2\% & +1.2\% & & +75.3\% \\
        \multicolumn{10}{l}{\textit{Large firms}}\\
        Total && +3.4\% & +1.0\% & +2.5\% & +7.0\% & -38.7\% & +1.3\% & & -3.5\% \\
        & Quantity & +3.0\% & -10.6\% & -9.0\% & -8.5\% & -18.2\% & -4.3\% & & -9.8\% \\
        & Quality & -0.1\% & -15.7\% & -12.6\% & -6.1\% & -245.5\% & -1.2\% & &-50.4\% \\
        & Price & +0.5\% & +27.2\% & +24.0\% & +21.6\% & +225.0\% & +6.7\% & & +56.7\% \\
        \hline\hline
        \end{tabular}
         \begin{center}
            \begin{minipage}{1\textwidth}
                {\scriptsize \textit{Notes}: \% pts impacts of minimum price policy on tax revenues and firms profits. The latter are defined as the total marginal profits made by manufacturers and retailers on a product. Large firms correspond to national brands produced by large manufacturers and to products under private labels. Small firms correspond to the aggregate of small manufacturers. For each alcohol market, the quantity effects relate to the variation in the size of the market, the quality effects to the reshuffling of market shares, and the price effect to the change in product specific-margins.  }
            \end{minipage}
        \end{center}
    \end{table}
\end{center}

\FloatBarrier
\pagebreak

    \FloatBarrier
    \pagebreak
    \newpage

    \singlespacing
    \printbibliography	
    \end{refsection}

\end{document}